\pdfoutput=1
\documentclass[aps,prd,10pt,preprintnumbers,nofootinbib,superscriptaddress,showpacs]{revtex4-2}
\usepackage{graphicx}
\usepackage{bm}
\usepackage{amsmath,amssymb}
\usepackage[colorlinks=true,linkcolor=blue,citecolor=blue,urlcolor=blue]{hyperref}
\usepackage{multirow}
\usepackage[section] {easy-todo}

\newcommand{\hc}{\mathrm{h.c.}}
\newcommand{\tr}{\mathrm{tr}}
\newcommand{\ts}{\hskip 0.08em} 
\newcommand{\tts}{\hskip 0.04em}
\newcommand{\ahat}{\hat{a}}
\newcommand{\bhat}{\hat{b}}
\newcommand{\chat}{\hat{c}}
\newcommand{\dhat}{\hat{d}}
\newcommand{\ehat}{\hat{e}}

\allowdisplaybreaks

\begin{document}

\preprint{KEK--TH--2543}
\preprint{KU--PH--036}

\title{Flavor physics in SU(5) GUT with scalar fields in the 45 representation}

\author{Toru Goto}
\affiliation{Theory Center, IPNS, KEK, Tsukuba 305-0801, Japan}
\author{Satoshi Mishima}
\affiliation{Theory Center, IPNS, KEK, Tsukuba 305-0801, Japan}
\affiliation{Department of Liberal Arts, Saitama Medical University, Moroyama, Saitama 350-0495, Japan}
\author{Tetsuo Shindou}
\affiliation{Division of Liberal-Arts, Kogakuin University, 2665-1 Nakano-machi, Hachioji, Tokyo, 192-0015, Japan}

\begin{abstract}
We study a realistic SU(5) grand unified model, where a $\mathbf{45}$
representation of scalar fields is added to the Georgi-Glashow model 
in order to realize the gauge coupling unification and the masses and
mixing of quarks and leptons. 
The gauge coupling unification together with constraints from proton
decay implies mass splittings in scalar representations. 
We assume that an SU(2) triplet component of the $\mathbf{45}$ scalar, 
which is called $S_3$ leptoquark, has a TeV-scale mass, and color-sextet
and color-octet ones have masses of the order of $10^6$ GeV. 
We calculate one-loop beta functions for Yukawa couplings in the model,  
and derive the low-energy values of the $S_3$ Yukawa couplings which are consistent
with the grand unification.
We provide predictions for lepton-flavor violation and
lepton-flavor-universality violation induced by the $S_3$ leptoquark, 
and find that current and future experiments have a chance to find a
footprint of our SU(5) model.  
\end{abstract}

\pacs{12.10.Dm, 14.80.Sv, 13.20.He, 13.35.Dx}

\maketitle

\section{Introduction}

The idea of Grand Unification is an attractive candidate for the fundamental theory behind 
the present understanding of particle physics
described by the Standard Model (SM)~\cite{Pati:1973rp,Pati:1974yy,Georgi:1974sy,Georgi:1974yf,Georgi:1974my,Fritzsch:1974nn}. 
Although the SM has been established as a successful effective model
at the electroweak (EW) scale by the discovery of the Higgs boson,
reaching a deeper understanding of nature is a desire of particle physicists. 
Interestingly, some properties of the SM suggest the existence of a Grand Unified Theory (GUT) as a high-energy theory beyond the SM. 
For example, the renormalization group (RG) runnings of the gauge couplings in the SM 
show a unification tendency at a high scale~\cite{Georgi:1974yf}, and 
the charge quantization of the SM fermions suggests the unification of matter.
Once the SM gauge groups for electromagnetic, weak, and strong interactions are unified to 
a GUT gauge symmetry group, quarks and leptons are consequently unified in a single or a few representations of the GUT group. 
Various groups, such as SU(5), SO(10), E$_6$, \textit{etc.}, have been
considered as the GUT gauge symmetry group~\cite{Langacker:1980js}. 

The SU(5) is the minimal simple group which contains the SM gauge
groups
SU(3)$_{\text{C}}\times$SU(2)$_{\text{L}}\times$U(1)$_{\text{Y}}$. 
The minimal version of the GUT model based on the SU(5) symmetry,
called the minimal SU(5) GUT, was originally proposed by Georgi and Glashow~\cite{Georgi:1974sy}. 
In the minimal SU(5) GUT model, the right-handed down quarks 
and the left-handed lepton doublets are embedded in $\mathbf{\bar{5}}$
representations of SU(5), and the left-handed quark doublets, the right-handed up quarks, 
and the right-handed charged leptons are embedded in $\mathbf{10}$ representations. 
The SM Higgs doublet is embedded in a $\mathbf{5}$ representation of scalars. 
In addition, the minimal model also contains a $\mathbf{24}$ 
representation of scalars, 
which breaks the SU(5) gauge symmetry to the SM ones. 

Although the concept of the minimal SU(5) GUT is beautiful, 
there are two serious issues that have to be solved to construct a
more realistic model. 
First, the three gauge couplings are not unified at a
high-energy scale only with the RG runnings in the SM. 
In the minimal SU(5) model, there is a grand desert between the EW and
the GUT scales, where there is no new contribution to the RG runnings. 
Second, the measured values of the masses of the charged leptons and the down-type
quarks cannot be accommodated with the minimal SU(5) GUT, where they
originate from a common Yukawa interaction in the GUT Lagrangian.

The first issue on the gauge coupling unification can be overcome by introducing extra fields 
in the grand desert, since such fields modify the RG runnings of the gauge couplings. 
A famous example of this direction is the supersymmetric SU(5) GUT model, 
in which the superpartners of the SM particles as well as the second
Higgs doublet are introduced and the gauge coupling unification occurs
at the scale of the order of $10^{16}$ GeV~\cite{Langacker:1990jh,Ellis:1990wk,Amaldi:1991cn,Langacker:1991an,Giunti:1991ta}. 
In nonsupersymmetric SU(5) GUT models, a single or a few particles in an
extra representation of SU(5) are predicted to lie in the grand desert in order to realize 
the gauge coupling unification~\cite{Babu:1984vx,Murayama:1991ah,Giveon:1991zm,Dorsner:2005fq,Dorsner:2005ii,Dorsner:2006dj,Dorsner:2006hw,Bajc:2006ia,FileviezPerez:2007bcw,Dorsner:2007fy,FileviezPerez:2008afb,Dorsner:2009mq,FileviezPerez:2016sal,Cox:2016epl,Dorsner:2017wwn,Becirevic:2018afm,Schwichtenberg:2018cka,Haba:2018vvu}. 

The second issue on the masses of the charged leptons and the
down-type quarks can be resolved by introducing a $\mathbf{45}$
representation of scalar fields to the minimal SU(5) GUT.  
Similar to the $\mathbf{5}$ scalar, the $\mathbf{45}$ scalar couples
with the $\mathbf{10}$ and the $\mathbf{\bar{5}}$ fermions since 
$\mathbf{10}\otimes\mathbf{\bar{5}}=\mathbf{5}\oplus\mathbf{45}$. 
This coupling makes modifications in the relation between the charged-lepton
and the down-type-quark Yukawa matrices at the GUT scale through 
the Georgi-Jarlskog mechanism~\cite{Georgi:1979df}. 

We combine the above two ideas on the extensions of the minimal SU(5)
GUT, and construct a concrete example of a realistic SU(5) GUT model, 
where the gauge coupling unification and the correct fermion masses
are realized simultaneously. 
This kind of model having the $\mathbf{45}$ scalar can be found, for example, in
Refs.~\cite{Babu:1984vx,Giveon:1991zm,Dorsner:2006dj,FileviezPerez:2007bcw,Dorsner:2007fy,FileviezPerez:2008afb,Dorsner:2009mq,FileviezPerez:2016sal,Dorsner:2017wwn,Becirevic:2018afm}. 
In the current study, we introduce the $\mathbf{45}$ scalar to reproduce the charged-lepton and
the down-type-quark Yukawa matrices correctly, and make 
an SU(2) triplet component of the $\mathbf{45}$ scalar light enough to
achieve the gauge coupling unification~\cite{Dorsner:2006dj}. 
This triplet scalar is called $S_3^*$. 
The Yukawa interactions between the $\mathbf{45}$ scalar and the $\mathbf{10}$ and the $\mathbf{\bar{5}}$ fermions are given by $\mathbf{10}\cdot\mathbf{10}\cdot\mathbf{45}$ and $\mathbf{10}\cdot\mathbf{\bar{5}}\cdot\mathbf{\overline{45}}$, where we omit the former by hand to suppress baryon-number-violating interactions mediated by the light $S_3^*$. 
In addition to the $S_3^*$, we assume that color-sextet and color-octet 
components in the $\mathbf{24}$ and the $\mathbf{45}$ scalars have masses of the order of $10^6$ GeV in order to 
avoid too rapid proton decay mediated by the GUT gauge
bosons.\footnote{For example, one can increase the GUT scale to evade the 
constraint from the proton decay by making the
$(\mathbf{8},\mathbf{2},1/2)$ scalar in 
the $\mathbf{45}$ representation light~\cite{Babu:1984vx,Dorsner:2006dj,Haba:2018vvu}.}
In this case, the SU(2) triplet scalar has a mass of
$\mathcal{O}(10^3-10^6\,\mathrm{GeV})$, and the GUT scale is of
$\mathcal{O}(10^{16}-10^{17}\, \mathrm{GeV})$. 
We do not consider any mechanisms
to generate the mass splittings in the GUT multiplets  
and 
to forbid the $\mathbf{10}\cdot\mathbf{10}\cdot\mathbf{45}$ interactions, 
which are beyond the scope of the current work. 
Moreover, we do not specify the origin of the nonzero neutrino masses, which are studied in the framework of the SU(5) GUT with the $\mathbf{45}$ scalar, for example, in Refs.~\cite{FileviezPerez:2007bcw,Dorsner:2007fy,FileviezPerez:2016sal,Dorsner:2017wwn,Rahat:2018sgs,Perez:2019aqq,Perez:2020nqq}. 

This triplet scalar $S_3^*$ carries the SM gauge quantum numbers 
$(\mathbf{3},\mathbf{3},-1/3)$, and has Yukawa couplings to a lepton
and a quark. The conjugate state of $S_3^*$, having 
$(\mathbf{\bar{3}},\mathbf{3},1/3)$, is often called $S_3$
leptoquark~\cite{Buchmuller:1986zs,Dorsner:2016wpm}. 
If the mass of the $S_3$ leptoquark lies at the TeV scale, $S_3$ 
can affect various flavor observables. 
Unlike the phenomenological models where the $S_3$ leptoquark is
introduced by hand as, for instance, in Refs.~\cite{Sakaki:2013bfa,Hiller:2014yaa,Alok:2017jgr,Dorsner:2017ufx,DiLuzio:2017fdq,Fajfer:2018bfj,Alda:2018mfy,Mandal:2019gff,DiLuzio:2019jyq,Angelescu:2021lln,Crivellin:2021egp,Kosnik:2021wyp}, 
the flavor structure of the
Yukawa couplings associated with $S_3$ is constrained by the measured
values of the charged-lepton and the down-type quark masses. 
It provides peculiar correlations in the flavor observables. 
We study the impact of the $S_3$ leptoquark at the TeV scale 
in our model on the phenomenology of flavor observables, such as leptonic
and semileptonic $B$ decays, $B_s-\bar{B}_s$ mixing, 
$\Upsilon(nS)$ decays, tau-lepton decays, and $Z\to\mu^\mp\tau^\pm$ decay. 
We show that 
Belle II with 50 ab$^{-1}$ and LHCb with 300 fb$^{-1}$
have a chance to find a footprint of our SU(5) GUT model. 

This paper is organized as follows.
In Sec.~\ref{sec:model}, we introduce an SU(5) GUT model with a $\mathbf{45}$ scalar, and 
explain how it solves the issues in the minimal SU(5) GUT. 
In Sec.~\ref{sec:pheno}, we present and discuss phenomenological
implications of our model. 
Section~\ref{sec:summary} contains our summary and conclusions. 
Some technical details are given in the Appendixes.

\section{Model}
\label{sec:model}
\subsection{Lagrangian}
\label{subsec:lagrangian}

We consider an SU(5) GUT model, where the SM fermions reside in
$\mathbf{10}$ and $\mathbf{\bar{5}}$ representations of SU(5), denoted by
$\Psi_{10\ts i}$ and $\Psi_{\bar{5}\tts i}$ with $i=1,2,3$ being the
generation index, and the scalar sector is composed of one $\mathbf{24}$,
one $\mathbf{5}$, and one $\mathbf{45}$-dimensional scalar representation,
denoted by $\Sigma$, $\Phi_5$, and $\Phi_{45}$, respectively.  
The SU(5)-symmetric renormalizable Lagrangian is given by 
\begin{align}
\mathcal{L}
  &=
  - \frac{1}{4}\ts(V^{\mu\nu})^{B}_{\ \,A}(V_{\mu\nu})^{A}_{\ \,B}
  + i\ts(\bar\Psi_{10\ts i})_{AB}\tts\gamma^\mu D_\mu(\Psi_{10\ts i})^{AB}
  + i\ts(\bar\Psi_{\bar{5}\tts i})^{A}\tts\gamma^\mu D_\mu(\Psi_{\bar{5}\tts i})_{A}
  + \big[D^\mu\Sigma^{B}_{\ \,A}\big]
    \big[D_\mu\Sigma^{A}_{\ \,B}\big]
  \nonumber\\
  &\hspace{4mm}
  + \big[D^\mu(\Phi_{5}^\dagger)_{A}\big] 
    \big[D_\mu(\Phi_{5})^{A}\big]
  + \big[D^\mu(\Phi_{45}^\dagger)^{C}_{AB}\big] 
    \big[D_\mu(\Phi_{45})_{C}^{AB}\big]
  + \mathcal{L}_Y
  - V(\Sigma,\Phi_5,\Phi_{45})\,,
  \label{eq:SU5Lag}
\end{align}
where $V_{\mu\nu}$ is the field strength tensor of the SU(5) gauge bosons, 
$A,B,C=1,\ldots,5$ are SU(5) indices, and 
$\mathcal{L}_Y$ and $V(\Sigma,\Phi_5,\Phi_{45})$ represent the Yukawa
interactions and the scalar potential, respectively.
The summation over repeated indices is implied. 
Here the fields $\Psi_{10\ts i}$, $\Sigma$, and $\Phi_{45}$ satisfy the
following relations:
\begin{align}
(\Psi_{10\ts i})^{AB} = -(\Psi_{10\ts i})^{BA}\,,\quad
(\Sigma^{B}_{\ \,A})^* = \Sigma^{A}_{\ \,B}\,,\quad
\Sigma^{A}_{\ \,A}=0\,,\quad
(\Phi_{45})^{AB}_C = -(\Phi_{45})^{BA}_C\,,\quad
(\Phi_{45})^{AB}_A = 0\,. 
\end{align}
In general the Yukawa term $\mathcal{L}_Y$ in Eq.~\eqref{eq:SU5Lag} consists of
the four interactions: 
\begin{align}
-\mathcal{L}_Y
  &=
  \frac{1}{8}\ts (Y_5^U)_{ij}\epsilon_{ABCDE}\ts
    (\Psi_{10\ts i})^{AB}(\Phi_5)^C(\Psi_{10\ts j})^{DE}
  + (Y_5^D)_{ij}
    (\Psi_{10\ts i})^{AB}(\Phi_{5}^\dagger)_A(\Psi_{\bar{5}\tts j})_B
  \nonumber\\
  &\hspace{4mm}
  + \frac{1}{4}(Y_{45}^U)_{ij}\epsilon_{ABCDE}\ts
    (\Psi_{10\ts i})^{AB}(\Phi_{45})^{CD}_F(\Psi_{10\ts j})^{EF}
  + \frac{1}{2}\ts (Y_{45}^D)_{ij}
    (\Psi_{10\ts i})^{AB}(\Phi_{45}^\dagger)_{AB}^C(\Psi_{\bar{5}\tts j})_C
  + \mathrm{h.c.}\,,
  \label{eq:SU5Yukawa}
\end{align}
where the totally antisymmetric tensor is defined as 
$\epsilon_{12345}=1$, and $Y_{5}^U$ and $Y_{45}^U$ are symmetric and
antisymmetric matrices in the generation space, respectively: 
\begin{align}
(Y_{5}^U)_{ij}=(Y_{5}^U)_{ji},\qquad 
(Y_{45}^U)_{ij}=-(Y_{45}^U)_{ji}.
\end{align}
The explicit expression for the scalar potential
$V(\Sigma,\Phi_5,\Phi_{45})$ is given in
Appendix~\ref{sec:scalarMasses}.

The SU(5) gauge symmetry is assumed to be broken down to
the SM gauge symmetry
SU(3)$_{\text{C}}\times$SU(2)$_{\text{L}}\times$U(1)$_{\text{Y}}$ by
the vacuum expectation value (VEV) of a SM-singlet scalar field in
$\Sigma$: 
$\langle\Sigma\rangle=v_{24}\tts\mathrm{diag}(2,2,2,-3,-3)$. 
The field $\Sigma$ is decomposed around the VEV as 
\begin{align}
\Sigma^{A}_{\ \,B}
=
\begin{pmatrix}
(\Sigma_8)^{\ahat}_{\ \,\bhat}
  + 2\ts \bigg(v_{24} 
    - \dfrac{1}{2\sqrt{15}}\,\Sigma_{1}\bigg)\ts 
    \delta^{\tts\ahat}_{\ \,\bhat}
&
\dfrac{1}{\sqrt{2}}\ts (\Sigma_G)^{\ahat}_{\ \,\beta}
\\[3mm]
\dfrac{1}{\sqrt{2}}\ts (\Sigma_G^*)^{\alpha}_{\ \,\bhat}
&
(\Sigma_3)^{\alpha}_{\ \,\beta}
  - 3\ts \bigg(v_{24} 
    - \dfrac{1}{2\sqrt{15}}\,\Sigma_{1}\bigg)\ts
     \delta^{\tts\alpha}_{\ \,\beta}
\end{pmatrix},
\label{eq:Sigma}
\end{align}
where $\hat{a},\hat{b}=1,2,3$ and $\alpha,\beta=1,2$ are
SU(3) and SU(2) indices, respectively. 
The spontaneous breaking of SU(5) typically provides 
the masses of the scalars $\Sigma_1$, $\Sigma_3$, and $\Sigma_8$ of
the order of $v_{24}$, while $\Sigma_G^{(*)}$
corresponds to the massless would-be Nambu-Goldstone boson, which gives
masses to the gauge bosons associated with the broken symmetries. 
These massive vector bosons are called $X$ bosons.

\subsection{Fermions}
\label{subsec:fermion}

The SM fermions $q_{Li}$, $u_{Ri}^c$, $d_{Ri}^{\ts c}$, $\ell_{Li}$, and $e_{Ri}^{\tts c}$ 
are embedded into the $\mathbf{10}$ and $\mathbf{\bar{5}}$ representations as 
\begin{align}
(\Psi_{10\ts i})^{AB}
=
\frac{1}{\sqrt{2}}
\begin{pmatrix}
\epsilon^{\hat{a}\hat{b}\hat{c}}\ts (V_{QU})_i{}^k\tts
u^{c}_{Rk\hat{c}} &
q_{Li}^{\tts \hat{a}\beta}
\\[1mm]
-q_{Li}^{\tts \hat{b}\alpha} &
\epsilon^{\alpha\beta}\tts (V_{QE})_i{}^k\tts
e_{Rk}^{\tts c}
\end{pmatrix},
\qquad
(\Psi_{\bar{5}\tts i})_{A} =
\begin{pmatrix}
d_{Ri\hat{a}}^{\ts c} &
\epsilon_{\alpha\beta}\tts (V_{DL})_i{}^k\ts
\ell_{Lk}^{\ts \beta} 
\end{pmatrix},
\label{eq:10and5bar}
\end{align}
where $i,k$ are the generation indices, and 
the totally antisymmetric tensors are defined as 
$\epsilon^{12}=\epsilon_{12}=1$ and
$\epsilon^{123}=\epsilon_{123}=1$. 
Without loss of generality, one can rotate the basis of
$\Psi_{10}^{}$ and $\Psi_{\bar{5}}^{}$ as  
\begin{align}
\Psi_{10}\to U_{10}\ts \Psi_{10}\,,\qquad
\Psi_{\bar{5}}\to U_5\ts \Psi_{\bar{5}}\,,
\label{eq:U10U5}
\end{align}
where $U_{10}$ and $U_5$ are arbitrary unitary matrices in the generation space. 
By using the degrees of freedom associated with the unitary rotations, 
we can take the basis where the up-type quarks and the charged leptons are in their mass eigenstates: 
\begin{align}
q_{Li}=
\begin{pmatrix}
\hat{u}_{Li}
\\[0.5mm]
(V_{\text{CKM}})_{i}{}^{j}\ts \hat{d}_{Lj}
\end{pmatrix},\qquad 
u_{Ri}=\hat{u}_{Ri}\,,\qquad 
d_{Ri}=\hat{d}_{Ri}\,,\qquad
\ell_{Li}=
\begin{pmatrix}
\hat{\nu}_{Li}
\\
\hat{e}_{Li}
\end{pmatrix},\qquad
e_{Ri}=\hat{e}_{Ri}\,,
\label{eq:SMfermions}
\end{align}
where the mass eigenstates are denoted with a hat, and 
$V_{\text{CKM}}$ is the Cabibbo-Kobayashi-Maskawa (CKM) matrix
in the Particle Data Group (PDG) phase convention~\cite{Chau:1984fp,ParticleDataGroup:2020ssz}. 
Analogous to the CKM matrix that represents a mismatch of the bases in $q_L$, the unitary matrices $V_{QU}$, $V_{QE}$, and $V_{DL}$
are introduced in $\Psi_{10}$ and $\Psi_{\bar{5}}$ as in Eq.~\eqref{eq:10and5bar}.

\subsection{Scalar spectrum and gauge coupling unification}
\label{subsec:scalar}

The scalar $\Phi_5$ is decomposed to the so-called color triplet
Higgs $S_1^{(5)*}$ and the SU(2)$_{\text{L}}$ doublet $H^{(5)}$: 
\begin{align}
(\Phi_{5})^A
&=
\begin{pmatrix}
S_1^{(5)*\ahat} 
\\[0.5mm]
H^{(5)\alpha}  
\end{pmatrix},
\label{eq:Phi5}
\end{align}
while the $\Phi_{45}$ consists of the scalars 
$\tilde{S}_1$, $R_2^*$, $S_3^*$, $S_6^*$, $S_8$, 
$H^{(45)}$, and $S_1^{(45)*}$ as 
\begin{align}
(\Phi_{45})^{\ahat\bhat}_{\chat}
  &=
  \frac{1}{\sqrt{2}}\ts\epsilon^{\ahat\bhat\dhat}
  \bigg[
  (\eta_{a})_{\chat\dhat}\ts S^{*a}_{6} 
  - \frac{1}{2}\ts\epsilon_{\chat\dhat\ehat}\ts S_1^{(45)*\ehat}
  \bigg]\,,
  &
(\Phi_{45})^{\ahat\bhat}_{\gamma}
  &=
  \frac{1}{\sqrt{2}}\ts
  \epsilon^{\ahat\bhat\dhat} R_{2\dhat\gamma}^*\,,
  \nonumber\\
(\Phi_{45})^{\ahat\beta}_{\chat}
  &=
  \frac{1}{\sqrt{2}}\ts
  \bigg[
  \frac{1}{\sqrt{2}}(\lambda_a)^{\ahat}{}_{\chat}\ts S_{8}^{\tts a\beta}
  + \frac{1}{2\sqrt{3}}\ts\delta^{\ahat}_{\chat}\ts H^{(45)\beta}
  \bigg]\,,
  &
(\Phi_{45})^{\alpha\beta}_{\chat}
  &=
  \frac{1}{\sqrt{2}}\ts
  \epsilon^{\alpha\beta}\tilde{S}_{1\chat}\,,
  \nonumber\\
(\Phi_{45})^{\alpha\bhat}_{\gamma}
  &=
  \frac{1}{\sqrt{2}}\ts
  \bigg(
  \frac{1}{\sqrt{2}}(\sigma_a)^{\alpha}_{\ \,\gamma}\ts S_{3}^{*\bhat}
  - 
  \frac{1}{2}\ts\delta^{\alpha}_{\gamma}\ts S_{1}^{(45)*\bhat}
  \bigg)\,,
  &
(\Phi_{45})^{\alpha\beta}_{\gamma}
  &=
  - \frac{\sqrt{3}}{2\sqrt{2}}\ts
  \epsilon^{\alpha\beta}\epsilon_{\gamma\delta} H^{(45)\delta},
\label{eq:Phi45}
\end{align}
where $\sigma_a\, (a=1,2,3)$ are the Pauli matrices, 
$\lambda_a\, (a=1,2,\ldots,8)$ the Gell-Mann matrices, and 
$\eta_a\, (a=1,2,\ldots,6)$ the symmetric matrices defined by 
\begin{align}
\{\eta_1,\eta_2,\eta_3,\eta_4,\eta_5,\eta_6\}
  = \left\{
  \begin{pmatrix}
  1&0&0\\ 0&0&0\\ 0&0&0 
  \end{pmatrix},\,
  \begin{pmatrix}
  0&0&0\\ 0&1&0\\ 0&0&0 
  \end{pmatrix},\,
  \begin{pmatrix}
  0&0&0\\ 0&0&0\\ 0&0&1 
  \end{pmatrix},\,
  \frac{1}{\sqrt{2}}\!
  \begin{pmatrix}
  0&1&0\\ 1&0&0\\ 0&0&0 
  \end{pmatrix},\,
  \frac{1}{\sqrt{2}}\!
  \begin{pmatrix}
  0&0&1\\ 0&0&0\\ 1&0&0
  \end{pmatrix},\,
  \frac{1}{\sqrt{2}}\!
  \begin{pmatrix}
  0&0&0\\ 0&0&1\\ 0&1&0
  \end{pmatrix}
  \right\}.
\end{align}
The decompositions of $\Sigma$, $\Phi_5$, and $\Phi_{45}$ are summarized 
in Table~\ref{tab:scalars}. 
\begin{table}[t]
\centering
\caption{
  The decomposition of the scalar fields $\Sigma$, $\Phi_{5}$, and
  $\Phi_{45}$ under the SM gauge groups. \label{tab:scalars}} 
\renewcommand\arraystretch{1.3}
\vspace{1mm}
\begin{tabular}{c|c|c|c|c|c||c|c|c|c|c|c}
\hline\hline
  Field & SU(5) & Field
  & SU(3)$_{\text{C}}$
  & SU(2)$_{\text{L}}$
  & U(1)$_{\text{Y}}$
  & Field & SU(5) & Field
  & SU(3)$_{\text{C}}$
  & SU(2)$_{\text{L}}$
  & U(1)$_{\text{Y}}$
  \\ \hline \hline
  && $\Sigma_{1}$ & $\mathbf{1}$ & $\mathbf{1}$ & $0$
  &&& 
  $\tilde{S}_1$ & $\mathbf{\bar{3}}$ & $\mathbf{1}$ & $4/3$
  \\ 
  \cline{3-6}\cline{9-12}
  && $\Sigma_{3}$ & $\mathbf{1}$ & $\mathbf{3}$ & $0$
  &&& 
  $R_2^*$ & $\mathbf{\bar{3}}$ & $\mathbf{\bar{2}}$ & $-7/6$
  \\ \cline{3-6}\cline{9-12}
  $\Sigma$ & $\mathbf{24} $
  & $\Sigma_{G}$ & $\mathbf{3}$ & $\mathbf{\bar{2}}$ & $-5/6$
  &&& 
  $S_3^{*}$ & $\mathbf{3}$ & $\mathbf{3}$ & $-1/3$
  \\ \cline{3-6}\cline{9-12}
  && $\Sigma_{G}^*$ & $\mathbf{\bar{3}}$ & $\mathbf{2}$ & $5/6$
  & $\Phi_{45}$ & $\mathbf{45}$
  & 
  $S_6^{*}$ & $\mathbf{\bar{6}}$ & $\mathbf{1}$ & $-1/3$
  \\ \cline{3-6}\cline{9-12}
  && $\Sigma_{8}^*$ & $\mathbf{8}$ & $\mathbf{1}$ & $0$
  &&& 
  $S_8^{}$ & $\mathbf{8}$ & $\mathbf{2}$ & $1/2$
  \\ \cline{1-6}\cline{9-12}
  \multirow{2}{*}{$\Phi_5$} & 
  \multirow{2}{*}{$\mathbf{5}$} 
  & $H^{(5)}$ & $\mathbf{1}$ & $\mathbf{2}$ & $1/2$
  &&& 
  $H^{(45)}$ & $\mathbf{1}$ & $\mathbf{2}$ & $1/2$
  \\ \cline{3-6}\cline{9-12}
  && $S_1^{(5)*}$ & $\mathbf{3}$ & $\mathbf{1}$ & $-1/3$
  &&& 
  $S_1^{(45)*}$ & $\mathbf{3}$ & $\mathbf{1}$ & $-1/3$
  \\ \hline\hline
\end{tabular}
\end{table}
Here the scalar $H^{(45)}$ ($S_1^{(45)}$) has the same quantum numbers 
under the SM gauge group as $H^{(5)}$ ($S_1^{(5)}$). 
Therefore, they can mix with each other, and we define the mass
eigenstates $H$, $H'$, $H_C$, and $S_1$ by introducing the mixing
angles $\theta_H$ and $\theta_S$ and the phases $\delta_H$ and
$\delta_S$: 
\begin{align}
  \begin{pmatrix}
  H\\ H^{\prime}
  \end{pmatrix}
  = 
  \begin{pmatrix}
  c_H & e^{-i\delta_H}s_H\\
  -e^{i\delta_H}s_H & c_H 
  \end{pmatrix}
  \begin{pmatrix}
  H^{(5)}\\ H^{(45)}
  \end{pmatrix},
  \qquad
  \begin{pmatrix}
  H_C\\ S_1
  \end{pmatrix}
  = 
  \begin{pmatrix}
  c_S & e^{-i\delta_S} s_S\\
  -e^{i\delta_S}s_S & c_S 
  \end{pmatrix}
  \begin{pmatrix}
  S_1^{(5)}\\ S_1^{(45)}
  \end{pmatrix},
\label{eq:mixing}
\end{align}
where 
$c_H=\cos\theta_H$, $s_H=\sin\theta_H$,
$c_S=\cos\theta_S$, and $s_S=\sin\theta_S$.
The presence of the two doublet scalars allows us to explain the
masses of the down-type quarks and the charged leptons
simultaneously.

Owing to the symmetry breaking of SU(5) to the SM gauge groups, mass splitting may occur among the scalar fields embedded in the SU(5) multiplets. 
At least one SU(2)$_{\text{L}}$-doublet scalar has to be light to break the EW symmetry spontaneously below the TeV scale.\footnote{The EW symmetry breaking 
can also be driven by the VEV of $\Sigma_3$ below the TeV scale. 
However, we assume that $\Sigma_3$ does not develop a VEV since it causes a dangerous contribution to the $\rho$ parameter.}
We assume that the scalar $H$ is light and corresponds to the SM Higgs doublet. 

It is well-known that the SM gauge couplings do not unify only by naive RG running in the SM.
The mass splitting of the SU(5) scalar multiplets can improve the situation. 
We consider the scenario where some of the scalar fields, in addition to $H$, are much lighter than others. 
At the energy scale above the mass of an additional light scalar, the scalar contributes to the RG running of the gauge couplings. 
The gauge coupling unification is realized if an appropriate set of light scalars is considered.
We define $\alpha_3(\mu)$, $\alpha_2(\mu)$, and $\alpha_1(\mu)$ as 
\begin{align}
\alpha_3(\mu)
= \alpha_s(\mu) = \frac{g_s(\mu)^2}{4\pi}\,,
\qquad
\alpha_2(\mu)
= \frac{g(\mu)^2}{4\pi}\,,
\qquad
\alpha_1(\mu)
= \frac{5}{3}\ts\frac{g'(\mu)^2}{4\pi}\,,
\end{align}
where $g_s$, $g$, and $g'$ are the gauge couplings 
of SU(3)$_{\text{C}}$, SU(2)$_{\text{L}}$, and U(1)$_{\text{Y}}$, 
respectively, and $\mu$ is the renormalization scale. 
Our analysis assumes that the SM gauge couplings are unified at the scale $M_X$, 
\textit{i.e.}, $\alpha_3(M_X)=\alpha_2(M_X)=\alpha_1(M_X)\equiv \alpha_X(M_X)$, 
and all the scalar masses are not heavier than $M_X$. 
Then, above the $M_X$ scale, all the scalars contribute to the running
as complete SU(5) multiplets so that the coupling unification holds
above $M_X$. We also make an ansatz that the mass of the $X$ boson is equal to the unification scale $M_X$.

Solving the renormalization group equations (RGEs) in Appendix~\ref{sec:RGE} with the unification assumption, 
we get the three relations, 
\begin{align}
  \alpha_X^{-1}(M_X)=&\ \alpha_3^{-1}(m_Z)-\left(\frac{B_{g_s}^{\mathrm{SM}}}{2\pi}\log\frac{M_X}{m_Z} + 
  \sum_{\phi}\frac{B_{g_s}^{\phi}}{2\pi}\log\frac{M_X}{m_{\phi}}\right),\nonumber\\
  \alpha_X^{-1}(M_X)=&\ \alpha_2^{-1}(m_Z)-\left(\frac{B_{g}^{\mathrm{SM}}}{2\pi}\log\frac{M_X}{m_Z} + 
  \sum_{\phi}\frac{B_{g}^{\phi}}{2\pi}\log\frac{M_X}{m_{\phi}}\right),\nonumber\\
  \alpha_X^{-1}(M_X)=&\ \alpha_1^{-1}(m_Z)-\frac{3}{5}\left(\frac{B_{g'}^{\mathrm{SM}}}{2\pi}\log\frac{M_X}{m_Z} + 
  \sum_{\phi}\frac{B_{g'}^{\phi}}{2\pi}\log\frac{M_X}{m_{\phi}}\right),
\end{align}
where $m_Z$ is the $Z$-boson mass, $\phi$ is summed over all the relevant scalars, 
and the coefficients $B_{g_i}^{\mathrm{SM}}$ and $B_{g_i}^{\phi}$ are given in Table~\ref{Table:beta_gi}.  
Eliminating $\alpha_X^{-1}(M_X)$~\cite{Hisano:1992jj}, 
one can get two independent equations,
\begin{align}
  &
  \frac{2}{5}\log\frac{m_{H_C}}{m_Z}
  +\frac{2}{5}\log\frac{m_{S_1}}{m_{H'}}
  +\frac{7}{5}\log\frac{m_{\tilde{S}_1}}{m_{S_3}}
  +\frac{4}{5}\log\frac{m_{R_2}}{m_{S_3}}
  +\frac{9}{5}\log\frac{m_{S_6}}{m_{S_3}}
  +\frac{4}{5}\log\frac{m_{S_8}}{m_{S_3}}
  +\log\frac{m_{\Sigma_8}}{m_{\Sigma_3}}
  \nonumber\\
  &\phantom{Space}=
  2\pi\left[-2\ts\alpha_3^{-1}(m_Z)+3\ts\alpha_2^{-1}(m_Z)-\alpha_1^{-1}(m_Z)\right]
  \simeq 79.8
  \,,
  \\
  & 
  44\log\frac{M_X}{m_Z}
  +6\log\frac{m_{S_3}}{m_{R_2}}
  +\log\frac{m_{S_6}}{m_{\tilde{S}_1}}
  +4\log\frac{m_{S_8}}{m_{\tilde{S}_1}}
  +\log\frac{m_{\Sigma_3}m_{\Sigma_8}}{M_X^2}
  \nonumber\\
  &\phantom{Space}=
  2\pi\left[-2\ts\alpha_3^{-1}(m_Z)-3\ts\alpha_2^{-1}(m_Z)+5\ts\alpha_1^{-1}(m_Z)\right]
  \simeq 1193
  \,,
\end{align}
where the gauge couplings at $\mu=m_Z$ 
are evaluated for six active quark flavors~\cite{Arason:1991ic} 
with the input values shown in Table~\ref{tab:inputs}. 
\begin{table}[t]
\centering
\caption{
  Input values for the $Z$-boson mass $m_Z$,
  the gauge couplings $\alpha_s(m_Z)$ and $\alpha^{-1}(m_Z)$,
  the weak mixing angle $\sin^2\theta_W(m_Z)$, 
  the quark masses, and the CKM parameters $s_{ij}$ and $\delta$,
  taken from Ref.~\cite{ParticleDataGroup:2020ssz}.
  Other parameters, 
  such as the pole masses of the charged leptons $m_e$, $m_\mu$, and $m_\tau$, 
  are also taken from Ref.~\cite{ParticleDataGroup:2020ssz}. 
  }
\label{tab:inputs}
\begin{tabular}{ll|ll|ll|ll}
\hline\hline
Parameter & Value &
Parameter & Value &
Parameter & Value &
Parameter & Value
\\
\hline\hline
$m_Z$& 
$91.1876$ GeV
&
$m_u(2\,\mathrm{GeV})$&
$0.00216$ GeV
&
$m_d(2\,\mathrm{GeV})$&
$0.00467$ GeV
&
$s_{12}$ &
$0.22650$
\\
$\alpha_s(m_Z)$ &
$0.1179$
&
$m_c(m_c)$&
$1.27$ GeV
&
$m_s(2\,\mathrm{GeV})$&
$0.093$ GeV
&
$s_{13}$ &
$0.00361$
\\
$\alpha^{-1}(m_Z)$ &
$127.952$
&
$m_t^{\mathrm{pole}}$&
$172.76$ GeV
&
$m_b(m_b)$&
$4.18$ GeV
&
$s_{23}$ &
$0.04053$
\\
$\sin^2\theta_W(m_Z)$ &
$0.23121$
&
&&
&&
$\delta$ &
$1.196$ rad
\\
\hline\hline
\end{tabular}
\end{table}

As a general property, the $S_3$ contribution improves the gauge coupling unification~\cite{Dorsner:2006dj}. 
In the case that only the Higgs boson and $S_3$ are lighter than $M_X$ in the scalar sector, 
the gauge coupling unification occurs at 
$M_X\sim \mathcal{O}(10^{14}\,\mathrm{GeV})$
with
$m_{S_3}\sim \mathcal{O}(10^{8}\,\mathrm{GeV})$. 
However, $M_X$ is severely constrained by proton decay search
experiments, since contributions from the GUT gauge-boson exchange generate 
the dimension-six operators relevant to the proton decay. 
Then the proton lifetime is naively expected as~\cite{Langacker:1980js,Nath:2006ut}
\begin{align}
  \tau_p\sim \frac{M_X^4}{\alpha_X^2m_p^5}\,,
\end{align}
where $m_p$ is the mass of proton, 
and one finds a naive lower bound as 
\begin{align}
  M_X>~5\times 10^{15}~\text{GeV}\,,
\end{align}
by using the experimental lower limit on the lifetime $\tau(p\to \pi^0e^+)>2.4\times 10^{34}$ years~\cite{Super-Kamiokande:2020wjk} and 
$\alpha_X^{2}\sim\mathcal{O}(10^{-3})$.

In order to avoid the rapid proton decay by making $M_X$ much heavier, 
we assume that $S_6$, $S_8$, and $\Sigma_8$, in addition to $S_3$,
are lighter than the unification scale $M_X$. 
With their contributions to the RGEs of the gauge couplings, $M_X$ can
be significantly heavier with keeping the coupling unification. 
Therefore, we consider a scenario where the masses of $S_6$, $S_8$,
and $\Sigma_8$ are below $M_X$. 
For simplicity, we assume that the other scalar components, except for
the SM-like Higgs doublet $H$, are as heavy as $M_X$.

Let us explain in more detail the masses of the other scalars embedded
in the GUT representations. 
The mass parameter $m_H^2$ associated with the SM-like Higgs doublet
$H$ is of the order of the weak scale according to the LHC
measurements, while the other scalars associated
with the $\mathbf{5}$ and $\mathbf{45}$ representations obey the mass 
relation given in Eq.~\eqref{eq:massRelation}. 
We simply choose that $\Sigma_1$, $\Sigma_3$, $H'$, $H_C$, $S_1$, and
$\tilde{S}_1$ have a common mass $M_X$. 
As a consequence, the mass of $R_2$ is determined as  $m_{R_2}^2\approx (2+4s_H^2)M_X^2/3$, 
where $s_H$ is the sine of the mixing angle defined in Eq.~\eqref{eq:mixing}.
For $s_H^2<1/4$, $m_{R_2}$ is lighter than $M_X$.

\begin{figure}[t]
  \centering
  \begin{tabular}{c@{\hskip 10mm}c}
  \includegraphics[height=7cm]{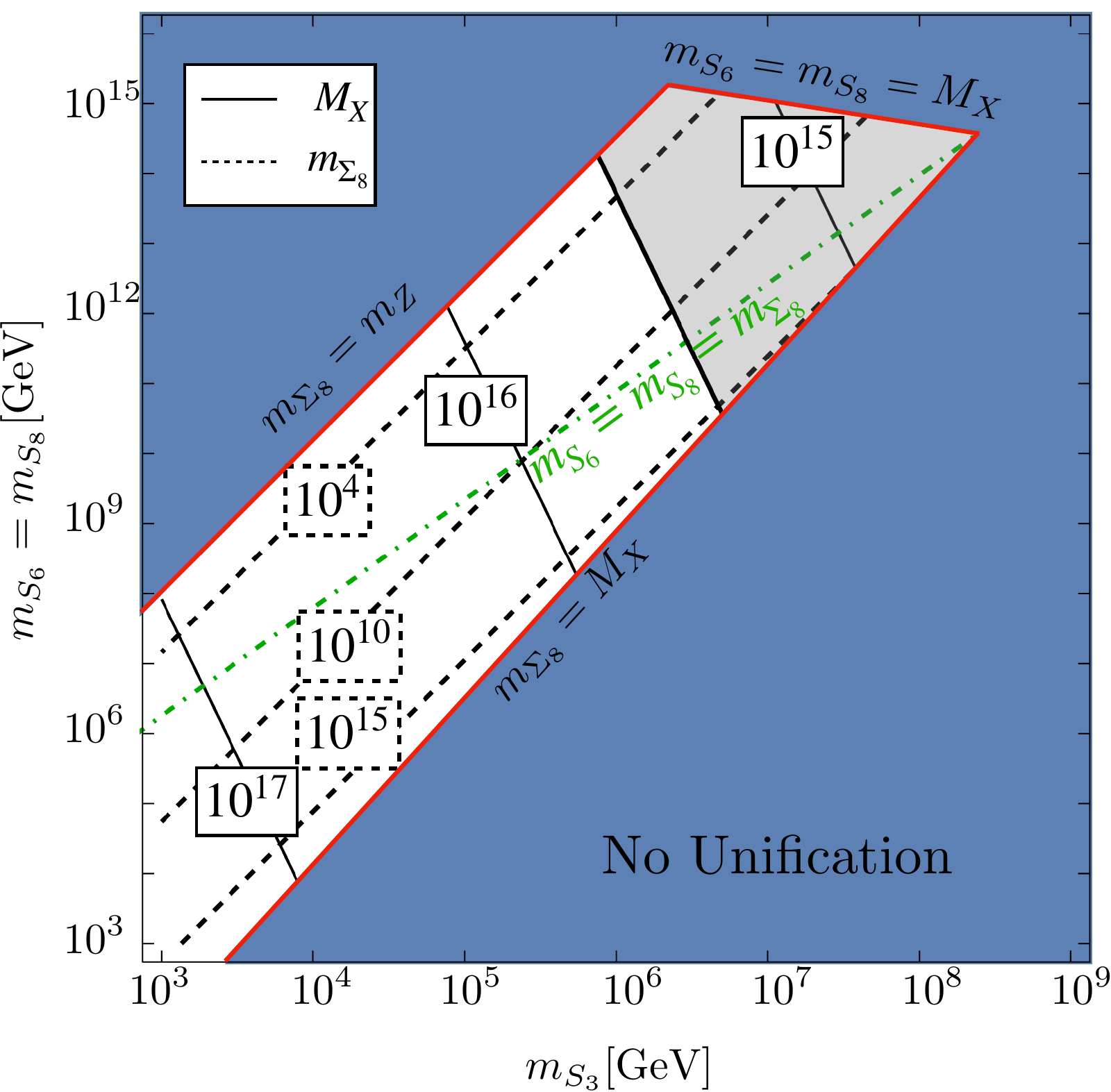}&
  \includegraphics[height=7cm]{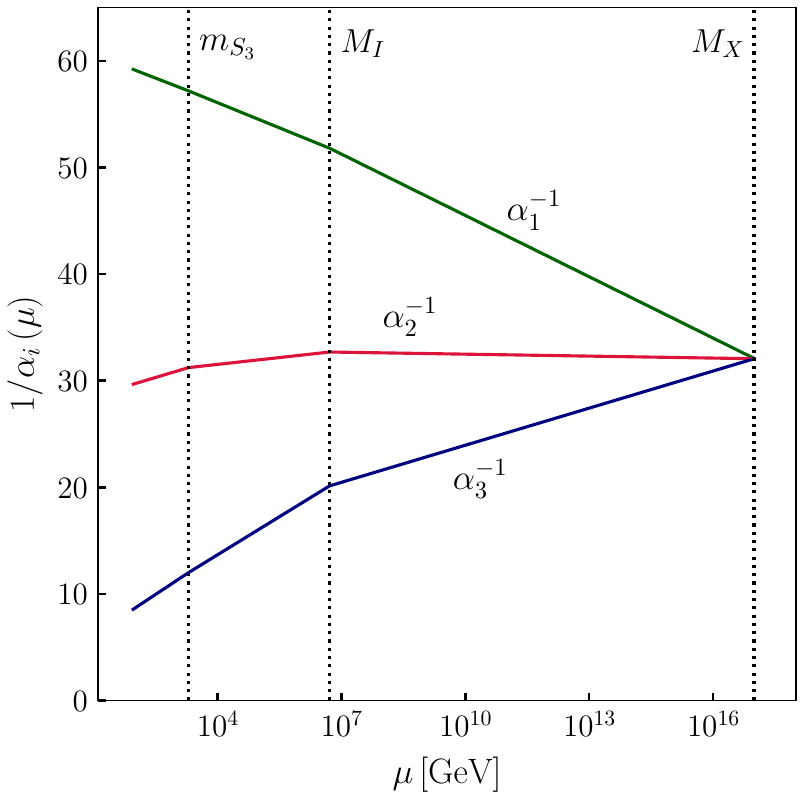}
  \\
  (a)&(b)
  \end{tabular}
\caption{
  (a) The contours of $M_X$(solid) and $m_{\Sigma_8}$ (dashed) 
  in the unit of GeV
  for realizing the coupling unification in the plane of 
  $m_{S_3}$ and $m_{S_6}=m_{S_8}$ for $\cot\theta_H=50$. 
  In the blue shaded region, the gauge coupling unification does not
  occur by RG running. The light gray region is disfavored by the 
  proton decay experiments because $M_X$ is too small. 
  The green dot-dashed line corresponds to the case with $m_{S_6}=m_{S_8}=m_{\Sigma_8}$.
  (b) RG runnings of the gauge couplings for $m_{S_3}=2$ TeV,
  $m_{S_6}=m_{S_8}=m_{\Sigma_8}\equiv M_I =5.2\times 10^{6}$ GeV, and $\cot\theta_H=50$. }
\label{fig:GCU}
\end{figure}

In Fig.~\ref{fig:GCU}(a), the contours of $M_X$ and $m_{\Sigma_8}$ are
shown in the parameter space of $m_{S_3}$ and $m_{S_6}=m_{S_8}$.
The gauge coupling unification favors rather light $S_3$, which can be as light as a TeV scale. 
The light gray regions are for $M_X<3\times 10^{15}$ GeV, which is disfavored by the proton decay search as mentioned above. 
For example, if we take $m_{S_3}=2$ TeV, $m_{S_6}=m_{S_8}=m_{\Sigma_8}=5.2\times 10^{6}$ GeV, and $\cot\theta_H=50$, 
the gauge coupling unification is realized at $M_X=9.7\times 10^{16}$ GeV as shown in Fig.~\ref{fig:GCU}(b).

In the phenomenological analysis, we use a benchmark scenario with the following mass spectrum:
\begin{itemize}
\item The masses of the quarks and leptons, the SM gauge bosons, and the SM-like Higgs boson $H$ are set to be 
consistent with their measurements; 
\item $S_3$ has a TeV-scale mass: $m_{S_3}\sim \mathcal{O}(10^3\,\mathrm{GeV})$;
\item $S_6$, $S_8$, and $\Sigma_{8}$ have intermediate masses, and we set them to an 
identical scale, \textit{i.e.},
$m_{S_6}=m_{S_8}=m_{\Sigma_8}\equiv M_I\sim \mathcal{O}(10^6\,\mathrm{GeV})$; 
\item The other particles including the $X$ bosons have masses of the order of the GUT scale $M_X$.
\end{itemize}

\subsection{Yukawa couplings}
\label{sec:Yukawa}

Below the GUT scale, the Yukawa
interactions with the scalars $H$, $S_3$, $S_6$, and $S_8$ are
given by 
\begin{align}
-\mathcal{L}_Y 
  &=
  (Y_U)_{ij}\epsilon_{\alpha\beta}\ts
    \bar{u}_{R\ahat i}\ts H^{\alpha}q^{\ahat\beta}_{Lj}
  + (Y_D)_{ij}\ts
    \bar{d}_{R\ahat i}\ts H_{\alpha}^*\ts q^{\ahat\alpha}_{Lj}
  + (Y_E)_{ij}\ts
    \bar{e}_{Ri}\ts H_{\alpha}^*\, \ell^{\ts\alpha}_{Lj}
    + \frac{(Y_3^{QQ})_{ij}}{2}\ts
    \epsilon_{\ahat\hat{b}\hat{c}}\ts\epsilon_{\alpha\beta}\ts
    \bar{q}_{Li}^{c\ts\ahat\alpha}(\sigma^a)^{\beta}{}_{\gamma}\ts
    S_3^{* a\bhat}q_{Lj}^{\hat{c}\ts\gamma}
  \nonumber\\
  &\hspace{5mm}
  + (Y_3^{QL})_{ij}\epsilon_{\alpha\beta}\ts
    \bar{q}_{Li}^{\ts c\ts\ahat\gamma}(\sigma_a)^{\alpha}{}_{\gamma}\ts
    S_{3\ahat}^a\ts\ell_{Lj}^{\ts\beta}
    + \frac{(Y_6^{QQ})_{ij}}{2}\ts\epsilon_{\alpha\beta}\ts
    \bar{q}_{Li}^{\ts c\ts\ahat\alpha} (\eta^A)_{\ahat\hat{b}}\ts 
    S_{6}^{A*}\tts q_{Lj}^{\hat{b}\beta}
    + (Y_6^{DU})_{ij}\ts
    \bar{d}_{R\ahat i}\ts (\eta^A)^{\ahat\hat{b}}
    S_{6}^{A}\ts u_{R\hat{b} j}^c
  \nonumber\\
  &\hspace{5mm}
    + (Y_8^{UQ})_{ij}\epsilon_{\alpha\beta}\ts
    \bar{u}_{R\ahat i}^{}\ts (\lambda^A)^{\ahat}{}_{\hat{b}}\ts
    S_{8}^{A \alpha }\ts q_{Lj}^{\hat{b}\beta}
    + (Y_8^{DQ})_{ij}\ts
    \bar{d}_{R\ahat i}\ts (\lambda^A)^{\ahat}{}_{\hat{b}}\ts
    S_{8\alpha}^{A*}\ts q_{Lj}^{\hat{b}\alpha}
  + \hc\,,    
\label{eq:LYukawaEFT}
\end{align}
where the first three terms lead to the fermion mass terms after the
Higgs field $H$ acquires a VEV at the EW scale. 
The $\mathbf{45}$ scalar plays an essential role in reproducing the
masses of the SM fermions. 
If the $\mathbf{45}$ scalar is absent, the Yukawa matrices must obey a
condition $Y_E = V_{QE}^T\ts Y_D^T \ts V_{DL}$ at the
GUT scale. 
This condition conflicts with the low-energy values of the masses of 
the down-type quarks and the charged leptons. 
In the current model, this problem is solved by the
presence of the Yukawa coupling $Y_{45}^D$. 
Because the SM-like Higgs field $H$ is a mixture of $H^{(5)}$ and
$H^{(45)}$ as in Eq.~\eqref{eq:mixing}, the Yukawa matrices $Y_U$,
$Y_D$, and $Y_E$ are given at the GUT scale by 
\begin{align}
Y_U =&
  -\frac{1}{2}\ts V_{QU}^{\ts T}
  \bigg( c_H\ts Y_{5}^U 
  + \sqrt{\frac{2}{3}}\,e^{i\delta_H}s_H\ts Y_{45}^{U} \bigg)^{\! T},
  \nonumber\\
Y_D &= 
  - \frac{1}{\sqrt{2}}
  \bigg( c_H\ts Y_{5}^{D}
  - \frac{1}{2\sqrt{6}}\,e^{-i\delta_H}s_H\ts Y_{45}^{D} \bigg)^{\! T},
  \nonumber\\ 
Y_E &= 
  - \frac{1}{\sqrt{2}}\ts V_{QE}^T
  \bigg( c_H\ts Y_{5}^D
  + \frac{\sqrt{3}}{2\sqrt{2}}\,e^{-i\delta_H}s_H\ts Y_{45}^{D}\bigg)
  V_{DL}\,,
\label{eq:YUYDYE}
\end{align}
which can lead to realistic Yukawa matrices at the low energy. 
Moreover, the GUT-scale matching conditions for the other couplings 
in Eq.~\eqref{eq:LYukawaEFT} read as
\begin{align}
Y_3^{QQ} &= 
  \frac{1}{2}\ts Y_{45}^{U}\,,
&
Y_6^{QQ} &=
  - \frac{1}{\sqrt{2}}\ts Y_{45}^{U}\,,
&
Y_8^{UQ} &=
  - \frac{1}{2}\ts V_{QU}^{\ts T}\ts Y_{45}^{U}\,,
\nonumber\\
Y_3^{QL} &=
  - \frac{1}{2\sqrt{2}}\ts Y_{45}^{D}\ts V_{DL}\,,
&
Y_6^{DU} &=
  \frac{1}{2}\ts (Y_{45}^{D})^{T}\tts  V_{QU}\,,
&
Y_8^{DQ} &=
  \frac{1}{2\sqrt{2}}\ts (Y_{45}^{D})^{T}.
\label{eq:Y3QLY6DUY8DQ}
\end{align}
The scalar $S_3$ couples to a quark and a lepton simultaneously and
thus is a leptoquark. 
The RGEs for these couplings are given in Appendix~\ref{sec:RGE}.

Let us count the physical degrees of freedom in the Yukawa sector. 
In the general case, there are four Yukawa matrices $Y_{5}^{U}$, $Y_{45}^{U}$,
$Y_5^{D}$, and $Y_{45}^{D}$ in the GUT Lagrangian. 
Since $Y_{5}^{U}$ and $Y_{45}^{U}$ are symmetric and antisymmetric
matrices, respectively, the four matrices contain $54$ parameters in
total. 
By the redefinitions of the fermion fields by $U_{10}$ and $U_{5}$ in
Eq.~\eqref{eq:U10U5}, $18$ degrees of freedom out of the $54$
can be eliminated. 
Thus there remain 36 physical parameters in the Yukawa matrices. 
Taking the basis where the up-type quarks and the charged
leptons are their mass eigenstates, the Yukawa matrices $Y_{5}^{U}$, $Y_{45}^{U}$,
$Y_5^{D}$, and $Y_{45}^{D}$ are written as 
\begin{align}
Y_{5}^U
  &=
  -\frac{1}{c_H}
  \Big(\tts V_{QU}^*\ts \hat{Y}_U 
  + \hat{Y}_U\ts V_{QU}^\dagger \Big)\,,
&Y_{45}^U
  &=
  \frac{\sqrt{3}}{\sqrt{2}\,e^{i\delta_H}s_H}
  \Big(\tts V_{QU}^*\ts \hat{Y}_U 
  - \hat{Y}_U\ts V_{QU}^\dagger \Big)\,,
  \nonumber\\ 
Y_{5}^{D}
  &= 
  - \frac{1}{2\sqrt{2}\ts c_H}
  \Big( 3\ts V_{\text{CKM}}^*\tts \hat{Y}_D
  + V_{QE}^*\ts \hat{Y}_E\tts V_{DL}^{\dagger}
  \Big)\,,
&Y_{45}^{D}
  &=
  \frac{\sqrt{3}}{e^{-i\delta_H}s_H}
  \Big(
  V_{\text{CKM}}^{*}\tts \hat{Y}_D
  - V_{QE}^*\ts \hat{Y}_E\tts V_{DL}^{\dagger}
  \Big)\,,
\label{eq:YukawaMatching2}
\end{align}
where $\hat{Y}_U$, $\hat{Y}_D$, and $\hat{Y}_E$ represent diagonal
matrices in the mass basis. 
It is noted that an overall phase in $V_{QU}$ and 
three phases in $V_{QE}$ (and/or $V_{DL}$) can be removed by U(1)$_{B}$, U(1)$_{e}$, U(1)$_{\mu}$, and U(1)$_{\tau}$ transformations. 
The right-hand sides of Eq.~\eqref{eq:YukawaMatching2} then contain nine eigenvalues in $\hat{Y}_U$, $\hat{Y}_D$, and $\hat{Y}_E$,  three mixing angles and one phase in $V_{\text{CKM}}$, eight parameters in $V_{QU}$, and fifteen ones in $V_{QE}$ and $V_{DL}$.

In general, the scalar $S_3$ can have two types of 
Yukawa couplings, $Y_3^{QQ}$ and $Y_3^{QL}$, and 
the combination of these couplings 
leads to baryon-number-violating dimension-six operators,
which cause too fast proton decay.
For example, the bound from $p\to\pi^0e^+$ is estimated as 
\begin{equation}
  \big|(Y_3^{QQ})_{12}(Y_3^{QL})_{11}(V_{\mathrm{CKM}}^{})_2{}^1\big|
  \lesssim 10^{-25}\left(\frac{m_{S_3}}{2~\mathrm{TeV}}\right)^2. 
  \label{eq:Y3QQY3QLconstraint}
\end{equation}
Because $(Y_3^{QL})_{11}\sim y_d/s_H$ with $y_d$ being the Yukawa coupling for down quark, 
this condition implies a strong upper bound on $(Y_{45}^{U})_{12}$: 
\begin{equation}
  \big|(Y_{45}^U)_{12}\big|\lesssim 10^{-20}\left(\frac{m_{S_3}}{2~\mathrm{TeV}}\right)^2
  s_H
\end{equation}
Other components in $Y_{45}^U$ also have to be highly suppressed to avoid the constraints from the proton decay. 
As explained in Appendix~\ref{sec:RGE}, the coupling 
$Y_3^{QQ}$ in Eq.~(\ref{eq:LYukawaEFT}) is forbidden 
in the whole range of the renormalization scale 
by an accidental global symmetry $U(1)_B\times U(1)_L$ if 
$Y_3^{QQ}$ is once set to be zero at the GUT scale. 
Therefore, in the following, 
we make an ansatz that $Y_{45}^U=0$ at the GUT scale.

We here show a parametrization for the mixing matrices 
$V_{QU}$, $V_{QE}$, and $V_{DL}$. According to the matching condition in Eq.~\eqref{eq:YukawaMatching2}, 
the ansatz $Y_{45}^U=0$ at the GUT scale requires that 
$V_{QU}$ should be a diagonal phase matrix: 
\begin{align}
V_{QU} =
\begin{pmatrix}
  1&0&0\\
  0&e^{\ts i\alpha^{QU}_2}&0\\
  0&0&e^{\ts i\alpha^{QU}_3}
\end{pmatrix}.
\end{align}
The other two matrices $V_{DL}$ and $V_{QE}$ can be 
parametrized as 
\begin{align}
V_{QE} =
V_{\text{CKM}}
\begin{pmatrix}
  1&0&0\\
  0&e^{\ts i\alpha^{QE}_2}&0\\
  0&0&e^{\ts i\alpha^{QE}_3}
\end{pmatrix}
\hat{V}_{QE}
\,,\qquad
V_{DL} =
\begin{pmatrix}
  1&0&0\\
  0&e^{\ts i\alpha^{DL}_2}&0\\
  0&0&e^{\ts i\alpha^{DL}_3}
\end{pmatrix}
\hat{V}_{DL}
\begin{pmatrix}
  e^{\ts i\beta^{DL}_1}&0&0\\
  0&e^{\ts i\beta^{DL}_2}&0\\
  0&0&e^{\ts i\beta^{DL}_3}
\end{pmatrix},
\label{eq:VQEVDL}
\end{align}
where $\hat{V}_{QE}$ and $\hat{V}_{DL}$ are 
the $3\times 3$ unitary matrices parametrized by three angles and one
phase as the CKM matrix, and $V_{\text{CKM}}$ is extracted in $V_{QE}$.

We define the coupling $\bar{Y}_3^{QL}$ in the
mass basis of the down-type quarks and the charged leptons:
\begin{align}
\bar{Y}_3^{QL} 
= V_{\text{CKM}}^T\ts Y_3^{QL}
= - \frac{\sqrt{6}}{4\ts e^{-i\delta_H}s_H}
  \Big(
  \hat{Y}_D\tts V_{DL}
  - \bar{V}_{QE}^*\ts \hat{Y}_E
  \Big)\,,  
\label{eq:Y3barQL}
\end{align}
where 
$\bar{V}_{QE} = V_{\text{CKM}}^\dagger\ts V_{QE}$. 
The mixings in $\bar{V}_{QE}$ ($V_{DL}$) cause flavor
transitions between different generations of the down-type quarks 
(the charged leptons). 
To suppress dangerous contributions to flavor-changing
processes associated with the first generation~\cite{Mandal:2019gff,Crivellin:2021egp},
such as $K\to\pi\nu\bar{\nu}$ and $\mu^-\to e^-\gamma$, we assume that $\hat{V}_{QE}$ and
$\hat{V}_{DL}$ have only the mixing between the second and the third
generations at the GUT scale: 
\begin{align}
\hat{V}_{QE}
  =
  \begin{pmatrix}
  1&0&0\\
  0&\cos\theta_{QE}&\sin\theta_{QE}\\
  0&-\sin\theta_{QE}&\cos\theta_{QE}
  \end{pmatrix},
  \qquad
\hat{V}_{DL}
  =
  \begin{pmatrix}
  1&0&0\\
  0&\cos\theta_{DL}&\sin\theta_{DL}\\
  0&-\sin\theta_{DL}&\cos\theta_{DL}
 \end{pmatrix}, 
\label{eq:VhatQE-VhatDL}
 \end{align}
where the mixing angles $\theta_{QE}$ and $\theta_{DL}$ are varied from $0$ to $\pi/2$. 
The three Yukawa matrices $Y_{10}$, $Y_5$, and $Y_{45}^D$ are then
determined at the GUT scale by the thirteen input parameters
in addition to $\hat{Y}_U$, $\hat{Y}_D$, $\hat{Y}_E$, and
$V_{\text{CKM}}$, \textit{i.e.}, 
the two mixing angles $\theta_{QE}$ and $\theta_{DL}$, 
the nine phases in $V_{QU}$, $V_{QE}$, and $V_{DL}$, and 
the two parameters $s_H=\sin\theta_H$ and $\delta_H$ 
in the Higgs sector.

The $Y_3^{QL}$ term in Eq.~\eqref{eq:LYukawaEFT} is 
decomposed in terms of the fields in
the EW broken phase as follows~\cite{Dorsner:2016wpm}: 
\begin{align}
\mathcal{L}_Y
&=
-
\bar{\hat{u}}_L^{\ts c}\ts
Y_{3}^{QL}
\hat{e}_{L}\ts
S_3^{\ts 1/3}
-
\sqrt{2}\,
\bar{\hat{d}}_L^{\,c}\ts
\bar{Y}_{3}^{QL}
\hat{e}_{L}\ts
S_3^{\ts 4/3}
+
\sqrt{2}\,
\bar{\hat{u}}_L^{\ts c}\ts
Y_{3}^{QL}
\hat{\nu}_{L}\ts
S_3^{\ts -2/3}
-
\bar{\hat{d}}_L^{\,c}\ts
\bar{Y}_{3}^{QL}
\hat{\nu}_{L}\ts
S_3^{\ts 1/3}
+
\mathrm{h.c.}\,,
\label{eq:LS3}
\end{align}
where 
the hatted quark and lepton fields represent the mass eigenstates as in
Eq.~\eqref{eq:SMfermions}, 
and 
$S_3^{\ts Q}$ denotes a charge eigenstate with charge $Q$ defined in the matrix form 
\begin{align}
\frac{1}{\sqrt{2}}
\big( \sigma^A \big)^\alpha_{\ \beta}\ts(S_3)^A_{\,\chat}
=
\left(
\begin{array}{cc}
\dfrac{1}{\sqrt{2}}\, (S_3^{\ts 1/3})_{\chat}
& (S_3^{\ts 4/3})_{\chat}
\\[2mm]
(S_3^{\ts -2/3})_{\chat} &
- \dfrac{1}{\sqrt{2}}\, (S_3^{\ts 1/3})_{\chat}
\end{array}
\right). 
\end{align}

\section{Phenomenological analysis}
\label{sec:pheno}
\subsection{Input parameters}

We study low-energy phenomenology of the SU(5) GUT model proposed in the last section, where there is an $S_3$ leptoquark with a TeV-scale mass. As explained in Sec.~\ref{sec:Yukawa} the $S_3$ leptoquark has the Yukawa couplings with the left-handed quarks and the left-handed leptons, which lead to rich flavor phenomenology at the low-energy scale. 
In particular, the $S_3$ couplings generate processes with lepton-flavor violation
(LFV) and lepton-flavor-universality violation (LFUV), while such exotic flavor processes are severely constrained by experiments. 
Our aim is to investigate whether current and future flavor experiments have a potential to explore our GUT-inspired scenario. 
The $S_3$ Yukawa matrix $Y_3^{QL}$ in our scenario cannot have an arbitrary structure 
unlike that in phenomenological leptoquark models where $S_3$ is
introduced by hand. 
The coupling $Y_3^{QL}$ originates from $Y_{45}^D$ in the GUT Lagrangian, and 
$Y_{45}^D$ also contribute to the SM Yukawa couplings $Y_D$ and $Y_E$ as in Eq.~\eqref{eq:YUYDYE}, 
which could help to explain the observed masses of the down-type quarks and the charged leptons.
Thus, nontrivial correlations are expected among flavor observables
where the $S_3$ leptoquark contributes.

The parameters in the GUT model, such as the Yukawa couplings $Y_5^U$, $Y_5^D$, and $Y_{45}^D$ and the mixing matrices $V_{QE}$ and $V_{DL}$, are constrained by the low-energy values of the SM fermion masses and the CKM matrix elements. 
We use the fermion masses and the CKM matrix elements listed in Table~\ref{tab:inputs} as inputs, and calculate the running masses at the EW scale by taking into account QCD corrections for quarks with \texttt{RunDec}~\cite{Chetyrkin:2000yt,Herren:2017osy} and one-loop QED corrections for charged leptons~\cite{Arason:1991ic}. 
The masses and the CKM matrix elements as well as the gauge couplings at the EW scale are then evolved up to the GUT scale with the one-loop RGEs in Appendix~\ref{sec:RGE}, where the Yukawa couplings $Y_3^{QL}$, $Y_6^{DU}$, and $Y_8^{DQ}$ are neglected at this stage. At the GUT scale we calculate the couplings $Y_5^U$, $Y_5^D$, $Y_{45}^D$ with Eq.~\eqref{eq:YukawaMatching2} by inputting $V_{QE}$ and $V_{DL}$, $\delta_H$, and $s_H$. The couplings $Y_U$, $Y_D$, $Y_E$, $Y_3^{QL}$, $Y_6^{DU}$, and $Y_8^{DQ}$ are calculated at the GUT scale with Eqs.~\eqref{eq:YUYDYE} and \eqref{eq:Y3QLY6DUY8DQ}, and we then perform the RG evolution from the GUT scale to the low scale. The fermion masses and the CKM elements at the low scale obtained from this procedure are different from the original values due to the effects from $Y_3^{QL}$, $Y_6^{DU}$, and $Y_8^{DQ}$. We iterate the RG running with the obtained values of $Y_3^{QL}$, $Y_6^{DU}$, and $Y_8^{DQ}$ together with the original values of the SM fermion masses and the CKM elements until the difference in the masses and the CKM elements becomes small enough. 
In this way we can determine a set of the GUT parameters that are consistent with the low-energy values of the SM fermion masses, the CKM matrix elements, and the gauge couplings.

We fix the mass of the $S_3$ leptoquark to be $m_{S_3}=2$ TeV to avoid constraints from high-$p_T$ searches at the LHC~\cite{JusteRozas:2853694}. 
In addition, there are the thirteen arbitrary parameters: 
the three mixing angles $\theta_{QE}$, $\theta_{DL}$, and $\theta_H$, 
and the ten phases $\alpha_2^{QU}$, $\alpha_3^{QU}$, $\alpha_2^{QE}$, $\alpha_3^{QE}$, 
$\alpha_2^{DL}$, $\alpha_3^{DL}$, $\beta_1^{DL}$, $\beta_2^{DL}$, $\beta_3^{DL}$, and $\delta_H$. 
In general the Yukawa couplings $(\bar{Y}_{3}^{QL})_{ij}$ in
Eq.~\eqref{eq:Y3barQL} become larger for a smaller Higgs mixing angle $\theta_H$. 
We choose $\cot\theta_H=50$ as a benchmark scenario, while the other parameters are varied arbitrarily in their physical domain. 
The $S_3$ contributions to the flavor observables considered below are
reduced by taking a heavier $m_{S_3}$ and/or a smaller
$\cot\theta_H$.

\subsection{Leptoquark couplings}\label{sec:leptoquarkcouplings}

\begin{figure}[t]
\centering
\includegraphics[scale=0.48]{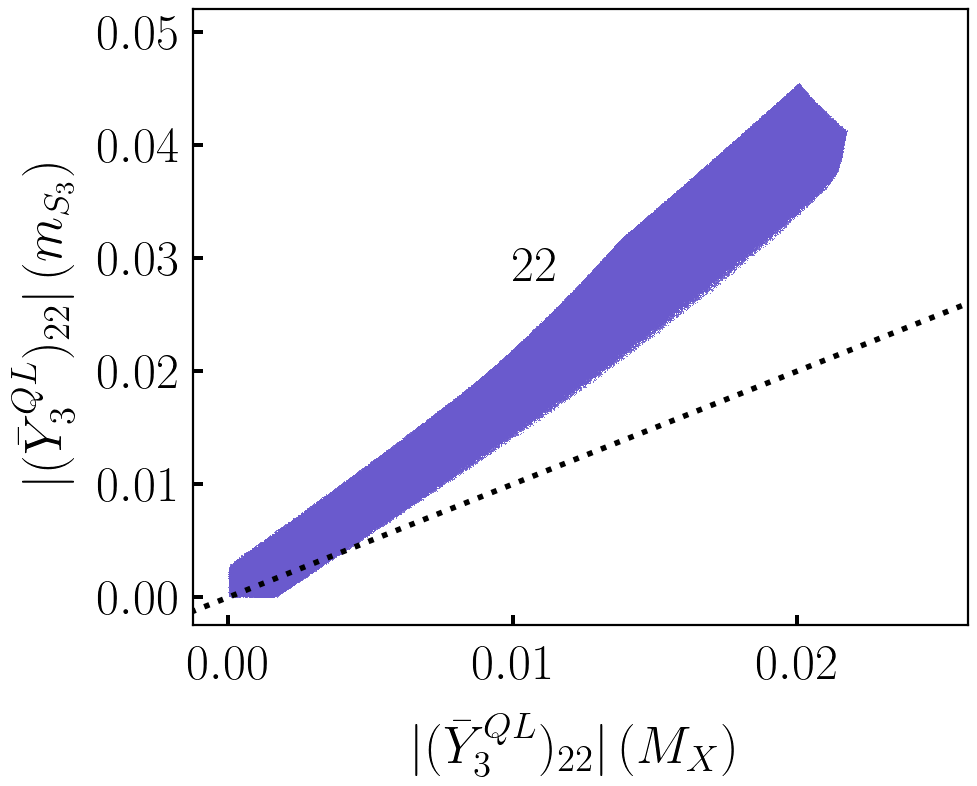}
\hspace{10mm}
\includegraphics[scale=0.48]{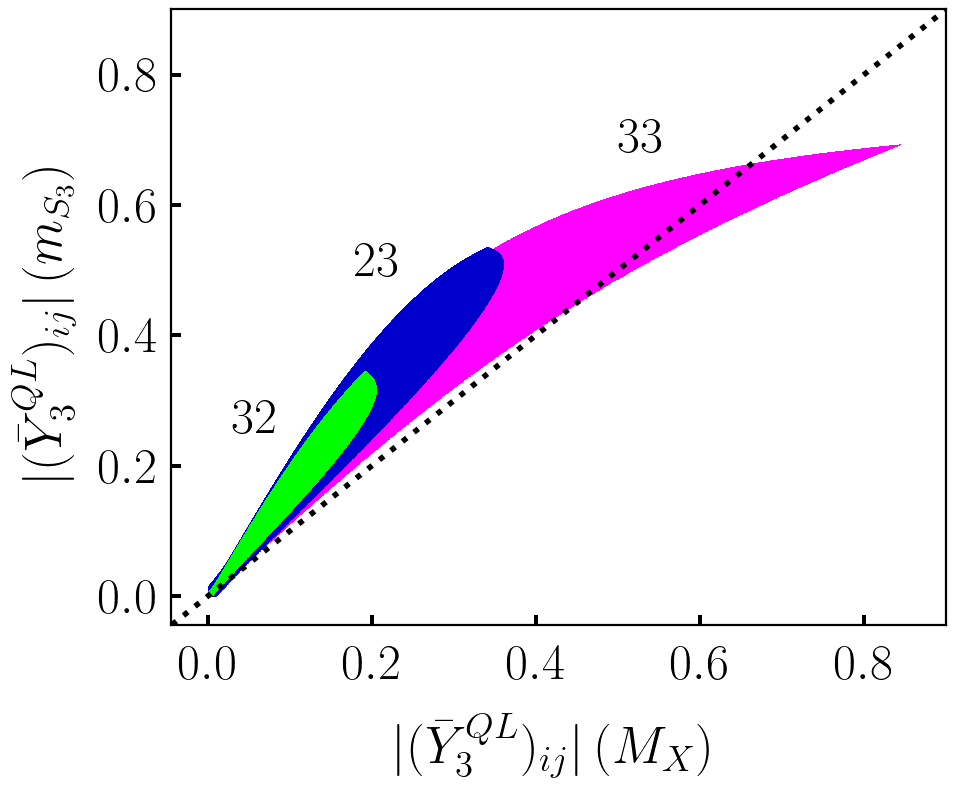}
\caption{Comparisons of the Yukawa couplings of the $S_3$ leptoquark at the GUT scale $\mu=M_X$ and at the $S_3$ mass scale $\mu=m_{S_3}$, where they are identical to each other on the dotted lines.}
\label{fig:Y3QL-2}
\end{figure}

The Yukawa couplings $(\bar{Y}_{3}^{QL})_{ij}$ of the $S_3$ leptoquark are constrained by the GUT relation in Eq.~\eqref{eq:Y3barQL} to accommodate with the measured masses of the down-type quarks and the charged leptons at the low-energy scale. 
The RG effects from the GUT scale $M_X$ to the $S_3$ mass scale $m_{S_3}$ are shown in Fig.~\ref{fig:Y3QL-2}.
It is noted that the magnitudes of the couplings typically enhance at
the lower scale. 
In particular, the 22 coupling is increased by about a factor of 2, receiving one-loop corrections with the other couplings. 
Therefore, the inclusion of the RG evolution is essential to study low-energy phenomenology associated with the 22 coupling,
such as $b\to s\mu^+\mu^-$ processes. 

According to Fig.~\ref{fig:Y3QL-2}, the couplings with the second-generation fermions are typically smaller than those with the third-generation ones: 
\begin{align}
\big|(\bar{Y}_3^{QL})_{22}\big|
\ \ts\ll\ts\ 
\big|(\bar{Y}_3^{QL})_{23}\big|
\sim 
\big|(\bar{Y}_3^{QL})_{32}\big|
\ \ts\lesssim\ts\
\big|(\bar{Y}_3^{QL})_{33}\big|\,. 
\label{eq:hierarchy}
\end{align}
On the other hand, the couplings with the first-generation fermions are negligibly small due to our ignorance of the corresponding mixings in Eq.~\eqref{eq:VhatQE-VhatDL}.

\subsection{Matching onto low-energy theory}

The gauge couplings and the Yukawa couplings in
Eq.~\eqref{eq:LYukawaEFT} at the GUT scale are evolved down to the
mass scale $m_{S_3}$ using the RGEs given in Appendix~\ref{sec:RGE}, 
where $S_6$, $S_8$, and $\Sigma_8$ are decoupled at the intermediate
scale $M_I$. 
The leptoquark $S_3$ is then decoupled at the scale $m_{S_3}$, and the
theory is matched onto the Standard Model Effective Field Theory (SMEFT). 
The corresponding tree-level matching conditions are presented in
Refs.~\cite{deBlas:2014mba,deBlas:2017xtg}, while the one-loop ones
are calculated in Ref.~\cite{Gherardi:2020det}. 
In addition, the one-loop anomalous dimensions in the SMEFT are found in
Refs.~\cite{Jenkins:2013zja,Jenkins:2013wua,Alonso:2013hga}.

\begin{figure}[t]
\centering
\begin{tabular}{c@{\hskip 30mm}c}
\includegraphics[width=37mm]{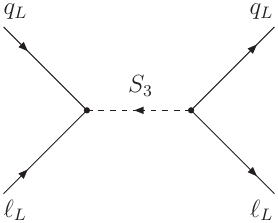}
&
\includegraphics[width=32mm]{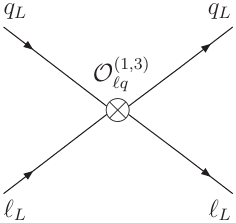}
\\
(a) & (b) 
\end{tabular}
\caption{Diagrams for the tree-level matching at the $S_3$ mass scale. Corresponding diagrams (a) in the model above the $S_3$ mass scale, 
and (b) in the SMEFT below the $S_3$ mass scale. }
\label{fig:matching}
\end{figure}

We adopt the dimension-six SMEFT operators in the so-called Warsaw
basis~\cite{Grzadkowski:2010es}, where the Lagrangian in the SMEFT
is given by the sum of the renormalizable SM 
Lagrangian and terms with higher-dimensional operators $\mathcal{O}_i$: 
$\mathcal{L}_{\text{SMEFT}}
=
\mathcal{L}_{\text{SM}}
+
\sum_i \mathcal{C}_i\ts \mathcal{O}_i$. 
At the tree level, only the semileptonic operators
$[\mathcal{O}_{\ell q}^{(1)}]_{ijkl}
=
( \bar{\ell}_{Li} \gamma^\mu \ell_{Lj} )
( \bar{q}_{Lk} \gamma_\mu q_{Ll} )$
and
$[\mathcal{O}_{\ell q}^{(3)}]_{ijkl}
=
( \bar{\ell}_{Li} \gamma^\mu \sigma_a \ell_{Lj} )
( \bar{q}_{Lk} \gamma_\mu \sigma_a q_{Ll} )$
are generated by integrating out the $S_3$ leptoquark, where the
corresponding Feynman diagram above the $S_3$ mass scale is presented 
in Fig.~\ref{fig:matching}(a), and that below the $S_3$ mass scale, 
\textit{i.e.}, in the SMEFT, is in Fig.~\ref{fig:matching}(b). 
The tree-level matching conditions for the semileptonic operators are given by 
\begin{align}
\big[ \mathcal{C}_{\ell q}^{(1)}(m_{S_3}) \big]_{ijkl}
=
3\ts\big[ \mathcal{C}_{\ell q}^{(3)}(m_{S_3}) \big]_{ijkl}
=
\frac{3}{4\tts m_{S_3}^2}\tts
\big(Y_3^{QL*}\big)_{k\tts i}
\big(Y_3^{QL}\big)_{lj}\,,
\label{eq:ClqSMEFT}
\end{align}
where $Y_3^{QL}$ in the right-hand side is the $S_3$ Yukawa coupling
at the $S_3$ mass scale, 
obtained from the coupling at the GUT scale in Eq.~\eqref{eq:Y3QLY6DUY8DQ}
applying the RG evolution. 
We also define the coefficients $\bar{\mathcal{C}}_{\ell q}^{(1,3)}$ in the mass basis of 
the down-type quarks and the charged leptons, and their matching
conditions read as
\begin{align}
\big[ \bar{\mathcal{C}}_{\ell q}^{(1)}(m_{S_3}) \big]_{ijkl}
=
3\ts\big[ \bar{\mathcal{C}}_{\ell q}^{(3)}(m_{S_3}) \big]_{ijkl}
=
\frac{3}{4\tts m_{S_3}^2}\tts
\big(\bar{Y}_3^{QL*}\big)_{k\tts i}
\big(\bar{Y}_3^{QL}\big)_{lj}\,,
\label{eq:ClqbarSMEFT}
\end{align}
where $\bar{Y}_3^{QL}= V_{\text{CKM}}^T\ts Y_3^{QL}$.

The SMEFT coefficients in Eqs.~\eqref{eq:ClqSMEFT} and \eqref{eq:ClqbarSMEFT} are evolved down to the EW scale, at which the SMEFT is matched onto the low-energy effective field theory (LEFT)~\cite{Jenkins:2017jig} by integrating out the EW gauge bosons, the Higgs boson, and the top quark. 
The LEFT operators used in our phenomenological analysis are listed in Eq.~\eqref{eq:LEFToperators}. 
The tree-level matching conditions for the coefficients $L_i$ in the LEFT Lagrangian of Eq.~\eqref{eq:LEFTLagrangian} can be found in Refs.~\cite{Aebischer:2015fzz,Jenkins:2017jig}, while the one-loop ones are calculated in Ref.~\cite{Dekens:2019ept}. 
Moreover, the RGEs for $L_i$ are calculated at the one-loop level in Refs.~\cite{Aebischer:2017gaw,Jenkins:2017dyc}. 
We decompose $L_i$ into the sum of SM and new physics (NP) contributions as 
$L_i=L_i^{\mathrm{SM}} + L_i^{\mathrm{NP}}$. 
In the current model only the semileptonic operators with the left-handed fermions are generated through the tree-level matching. 
For example, we have the following coefficients at the weak scale $\mu=m_Z$:
\begin{align}
\big[ L_{\tts\nu d}^{V,LL}(m_Z) \big]_{ijkl}^{\mathrm{NP}}
  &=
  \big[ \bar{\mathcal{C}}_{\ell q}^{(1)}(m_Z) \big]_{ijkl}
  -
  \big[ \bar{\mathcal{C}}_{\ell q}^{(3)}(m_Z) \big]_{ijkl}\,,
\label{eq:LnudVLL-tree}
\\
\big[ L_{\tts ed}^{V,LL}(m_Z) \big]_{ijkl}^{\mathrm{NP}}
&=
  \big[ \bar{\mathcal{C}}_{\ell q}^{(1)}(m_Z) \big]_{ijkl}
  +
  \big[ \bar{\mathcal{C}}_{\ell q}^{(3)}(m_Z) \big]_{ijkl}\,,
\label{eq:LedVLL-tree}
\\
\big[ L_{\tts\nu edu}^{V,LL}(m_Z) \big]_{ijkl}^{\mathrm{NP}}
&=
  2\ts V^*_{wk} \big[ \mathcal{C}_{\ell q}^{(3)}(m_Z) \big]_{ijwl}\,,
\label{eq:LnueduVLL-tree}
\end{align}
where $V_{wk}$ denotes a CKM matrix element.

In our numerical analysis, we also include one-loop corrections to the
matching onto the SMEFT, the RG evolution from $\mu=m_{S_3}$ to $\mu=m_Z$,
the matching onto the LEFT, and the RG evolution from $\mu=m_Z$ to the
lower energy scale. 
Let us consider the LEFT coefficient $L_{\tts ed}^{V,LL}$ for $b\to s$
processes as an example. 
Solving the RGEs in the leading-logarithmic approximation, the coefficient
$L_{\tts ed}^{V,LL}$ is given at the bottom scale $\mu=m_b$ by  
\begin{align}
\big[ L_{\tts ed}^{V,LL}(m_b) \big]_{ij23}^{\mathrm{NP}}
  &=
  \frac{
  \big(\bar{Y}_{3}^{QL}\big)^*_{2\tts i}\ts 
  \big(\bar{Y}_{3}^{QL}\big)_{3j}
  }
  {m_{S_3}^2}
  \Bigg\{
  1
  -
  \frac{\alpha}{2\pi}
  \log\bigg(\frac{m_{S_3}^2}{m_b^2}\bigg)
  +
  \frac{g^2( 1 - 4\tts c_W^4)}{32\tts \pi^2 c_W^2}
  \bigg[
  \log\bigg(\frac{m_{S_3}^2}{m_Z^2}\bigg) + \frac{11}{6}
  \bigg]
  \Bigg\}
\nonumber\\
&\hspace{5mm}
  +
  \frac{y_t^2}{64\tts \pi^2}
  \Bigg\{
  2\ts V^*_{ts}V_{tb}
  \frac{
  \big(Y_{3}^{QL}\big)^*_{3\tts i}\ts 
  \big(Y_{3}^{QL}\big)_{3j}
  }
  {m_{S_3}^2}
  +
  \Bigg[
  V^*_{ts}
  \frac{
  \big(Y_{3}^{QL}\big)^*_{3\tts i}\ts 
  \big(\bar{Y}_{3}^{QL}\big)_{3j}
  }
  {m_{S_3}^2}
  +
  V_{tb}
  \frac{
  \big(\bar{Y}_{3}^{QL}\big)^*_{2\tts i}\ts 
  \big(Y_{3}^{QL}\big)_{3j}
  }
  {m_{S_3}^2}
  \Bigg]\ts I_{ed}(x_t)
  \Bigg\}
  \nonumber\\
  &\hspace{5mm}
  -
  \frac{3(N_c+1)}{8}
  \Bigg[
  \frac{
  \big(
    \bar{Y}_{3}^{QL\dagger} \bar{Y}_{3}^{QL} \bar{Y}_{3}^{QL\dagger} 
  \big)_{i\tts 2}\ts
  \big(\bar{Y}_{3}^{QL}\big)_{3j}}
  {(4\pi)^2 m_{S_3}^2}
  +
  \frac{
  \big(\bar{Y}_{3}^{QL}\big)^*_{2\tts i}\ts
  \big(
    \bar{Y}_{3}^{QL} \bar{Y}_{3}^{QL\dagger} \bar{Y}_{3}^{QL} 
  \big)_{3j}}
  {(4\pi)^2 m_{S_3}^2}
  \Bigg]
  \nonumber\\
  &\hspace{5mm}
  -
  \frac{5}{4}
  \frac{
    \big( \bar{Y}_{3}^{QL\dagger} \bar{Y}_{3}^{QL} \big)_{i\tts j}\ts
    \big( \bar{Y}_{3}^{QL} \bar{Y}_{3}^{QL\dagger} \big)_{32}
  }
  {(4\pi)^2m_{S_3}^{2}}
  - 
  \delta_{ij}
  \frac{\alpha}{6\tts \pi} 
  \frac{
    \big( \bar{Y}_{3}^{QL} \bar{Y}_{3}^{QL\dagger}\big)_{32}
  }{m_{S_3}^{2}}
  \bigg[
    \log\bigg( \frac{m_{S_3}^2}{m_b^2} \bigg)
    - \frac{19}{12}
  \bigg]
  \,,
\label{eq:LedVLL}
\end{align}
where $c_W=\cos\theta_W$ is the cosine of the Weinberg angle, 
$y_t$ represents the SM Yukawa coupling of the top quark, 
$x_t=m_t^2/m_W^2$ with $m_t$ and $m_W$ being the masses of the top
quark and the $W$ boson,
$N_c=3$ is the number of colors,  
$\alpha$ is the electromagnetic coupling, 
and $I_{ed}(x)$ is the loop function defined by 
\begin{align}
I_{ed}(x)
&=
  -
  \log\bigg(\frac{m_{S_3}^2}{m_W^2}\bigg)
  -
  \frac{3(x + 1 )}{2(x - 1)}
  +
  \frac{x^2 - 2\tts x + 4}{(x - 1 )^2}
  \log x\,.
\end{align}
In Eq.~\eqref{eq:LedVLL}, the $S_3$ couplings $Y_3^{QL}$ and $\bar{Y}_3^{QL}$ should be
understood as those evaluated at the $S_3$ mass scale. 
The one-loop expressions for the other LEFT
coefficients relevant to our analysis are given in
Appendix~\ref{sec:LEFTcoefficients}.

It is convenient to convert the LEFT coefficients of the 
$b\to s$ semileptonic operators into the
coefficients in the weak Hamiltonian~\cite{Buchalla:1995vs}: 
\begin{align}
\mathcal{H}_{W}
&=
-
\frac{4\tts G_F}{\sqrt{2}}
\frac{\alpha}{4\pi}
V_{ts}^* V_{tb}
\bigg[
\big[ C_{9V} \big]_{ij}\ts
\big(\bar{\hat{s}}_{L} \gamma^\mu \hat{b}_{L}\big)
\big( \bar{\hat{e}}_{i}\gamma_\mu\ts \hat{e}_{j} \big)
+ 
\big[ C_{10A} \big]_{ij}\ts
\big(\bar{\hat{s}}_{L} \gamma^\mu \hat{b}_{L}\big)
\big( \bar{\hat{e}}_{i}\gamma_\mu\gamma_5\ts \hat{e}_{j} \big)
\nonumber\\
&\hspace{30mm}
+
\big[ C_{L} \big]_{ij}\ts
\big(\bar{\hat{s}}_{L} \gamma^\mu \hat{b}_{L}\big)
\big(\bar{\hat{\nu}}_{i} \gamma_\mu(1-\gamma_5) \hat{\nu}_{j}\big)
\bigg]
+ \mathrm{h.c.}, 
\end{align}
where $G_F$ is the Fermi constant, and the NP contributions to the
coefficients at the scale $\mu$ are related to the LEFT ones as  
\begin{align}
\big[ C_{9V}^{\mathrm{NP}}(\mu) \big]_{ij}
&=
\frac{\pi}{\sqrt{2}\ts G_F\alpha\tts V_{ts}^* V_{tb}}\ts
\Big(
\big[ L_{\tts ed}^{V,LL}(\mu) \big]_{ij23}^{\mathrm{NP}}
+
\big[ L_{\tts de}^{V,LR}(\mu) \big]_{23ij}^{\mathrm{NP}}
\Big)\,,
\label{defC9V}
\\
\big[ C_{10A}^{\mathrm{NP}}(\mu) \big]_{ij}
&=
\frac{\pi}{\sqrt{2}\ts G_F\alpha\tts V_{ts}^* V_{tb}}\ts
\Big(
- \big[ L_{\tts ed}^{V,LL}(\mu) \big]_{ij23}^{\mathrm{NP}}
+
\big[ L_{\tts de}^{V,LR}(\mu) \big]_{23ij}^{\mathrm{NP}}
\Big)\,,
\label{defC10A}
\\
\big[ C_{L}^{\mathrm{NP}} \big]_{ij}\ts
&=
\frac{\pi}{\sqrt{2}\ts G_F\alpha\tts V_{ts}^* V_{tb}}\ts
\big[ L_{\tts \nu d}^{V,LL} \big]_{ij23}^{\mathrm{NP}}\,. 
\label{eq:CL}
\end{align}
The argument $\mu$ is omitted in Eq.~\eqref{eq:CL}, since 
the coefficients $[ C_{L}^{\mathrm{NP}} ]_{ij}$ and 
$[ L_{\tts \nu d}^{V,LL} ]_{ij23}^{\mathrm{NP}}$
have no scale dependence. 
Let us consider the coefficients for the $b\to s\mu^+\mu^-$ transition. 
The coefficients
$[C_{9V}^{\mathrm{NP}}(\mu)]_{22}$ and $[C_{10A}^{\mathrm{NP}}(\mu)]_{22}$
in Eqs.~\eqref{defC9V} and \eqref{defC10A} are dominated by 
the LEFT coefficient $[L_{ed}^{V,LL}(\mu)]_{2223}$ generated at the
tree level, while the contributions from $[L_{de}^{V,LR}(\mu)]_{3222}$
induced at the one-loop level are subdominant.
Hence, the approximate relation
$[C_{9V}^{\mathrm{NP}}(\mu)]_{22} \approx - [C_{10A}^{\mathrm{NP}}(\mu)]_{22}$ 
holds in the current model~\cite{Hiller:2014yaa}.

\subsection{Constraints}
\label{sec:constraints}

In the current model, the $S_3$ leptoquark has sizable couplings to quarks and leptons in the second and third generations. 
Strong constraints on the parameter space of the model come from the mass difference of $B_s$ and $\bar{B}_s$ mesons denoted by $\Delta M_s$, the branching ratios for the $B\to K^{(*)}\nu\bar\nu$ decays, the LFUV tests in the $B\to K^{(*)}\ell^+\ell^-$ ($\ell=e,\mu$) decays, and the branching ratio for the $B_s\to\mu^+\mu^-$ decay. 
NP contributions to $\Delta M_s$ are generated at the one-loop level, while those to the others are at the tree level. 
The current experimental data for these observables are summarized in Table~\ref{tab:observables} together with other relevant observables. 
\begin{table}[t]
\centering
\caption{Current measurements and future experimental sensitivities of flavor observables. 
The first column represents the corresponding transition, and the second column 
shows the dominant coupling that induces the transition, where Loop denotes a loop-level transition. }
\label{tab:observables}
\begin{tabular}{ccccc}
\hline\hline
Transition 
& Couplings 
& Observable
& Current measurement 
& Future sensitivity
\\
\hline\hline
$b\to s\mu^+\mu^-$
& $(\bar{Y}_3^{QL*})_{22}(\bar{Y}_3^{QL})_{32}$
& $R_{K^+}[0.1,\,1.1]$
& $0.994^{+0.090+0.029}_{-0.082-0.027}$~\cite{LHCb:2022qnv,LHCb:2022vje}
& 
\\
&
& $R_{K^{*0}}[0.1,\,1.1]$
& $0.927^{+0.093+0.036}_{-0.087-0.035}$~\cite{LHCb:2022qnv,LHCb:2022vje}
& 
\\
&
& $R_{K^+}[1.1,\,6.0]$
& $0.949^{+0.042+0.022}_{-0.041-0.022}$~\cite{LHCb:2022qnv,LHCb:2022vje}
& $\pm 0.007$~\cite{LHCb:2018roe}
\\
&
& $R_{K^{*0}}[1.1,\,6.0]$
& $1.027^{+0.072+0.027}_{-0.068-0.026}$~\cite{LHCb:2022qnv,LHCb:2022vje}
& $\pm 0.008$~\cite{LHCb:2018roe}
\\
&
& $\mathcal{B}(B_s\to\mu^+\mu^-)$
& $(3.01\pm 0.35)\times 10^{-9}$~\cite{ParticleDataGroup:2022pth}
& $\pm 0.16\times 10^{-9}$~\cite{LHCb:2018roe}
\\
\hline
Loop
& $(\bar{Y}_3^{QL*})_{23}(\bar{Y}_3^{QL})_{33}$ 
& $\Delta M_{s}$
& $(17.765\pm 0.006)\ \mathrm{ps}^{-1}$~\cite{ParticleDataGroup:2022pth}
\\
\hline
$b\to s \nu\bar{\nu}$
& $(\bar{Y}_3^{QL*})_{23}(\bar{Y}_3^{QL})_{33}$ 
& $\mathcal{B}(B^+\to K^+\nu\bar\nu)$
& $<1.6\times 10^{-5}\ (90\%)$~\cite{Lees:2013kla} 
& $\pm 11\%\ \mathrm{of\ SM}$~\cite{Kou:2018nap}
\\
&
& $\mathcal{B}(B^0\to K_S\nu\bar\nu)$
& $<1.3\times 10^{-5}\ (90\%)$~\cite{Grygier:2017tzo} 
& 
\\
&
& $\mathcal{B}(B^+\to K^{*+}\nu\bar\nu)$
& $<4.0\times 10^{-5}\ (90\%)$~\cite{Belle:2013tnz} 
& $\pm 9.3\%\ \mathrm{of\ SM}$~\cite{Kou:2018nap}
\\
&
& $\mathcal{B}(B^0\to K^{*0}\nu\bar\nu)$
& $<1.8\times 10^{-5}\ (90\%)$~\cite{Grygier:2017tzo} 
& $\pm 9.6\%\ \mathrm{of\ SM}$~\cite{Kou:2018nap}
\\
\hline
$b\to c\tau^-\bar{\nu}$
& $(\bar{Y}_3^{QL*})_{23}(\bar{Y}_3^{QL})_{33}$
& $R(D)$ 
& $0.357\pm 0.029$~\cite{HeavyFlavorAveragingGroup:2022wzx}
& $(\pm 2.0\pm 2.5)\%$~\cite{Kou:2018nap}
\\
&
& $R(D^*)$ 
& $0.284\pm 0.012$~\cite{HeavyFlavorAveragingGroup:2022wzx}
& $(\pm 1.0\pm 2.0)\%$~\cite{Kou:2018nap}
\\
\hline
$b\to s\tau^+\tau^-$
& $(\bar{Y}_3^{QL*})_{23}(\bar{Y}_3^{QL})_{33}$ 
& $\mathcal{B}(B_s\to\tau^+\tau^-)$
& $< 5.2\times 10^{-3}\ (90\%)$~\cite{Aaij:2017xqt}
& $5\times 10^{-4}$~\cite{LHCb:2018roe}
\\
&
& $\mathcal{B}(B^+\to K^+\tau^+\tau^-)$
& $< 2.25\times 10^{-3}\ (90\%)$~\cite{TheBaBar:2016xwe} 
& $2.0\times 10^{-5}$~\cite{Kou:2018nap}
\\
&
& $\mathcal{B}(B^0\to K^{*0}\tau^+\tau^-)$
& $< 3.1\times 10^{-3}\ (90\%)$~\cite{Belle:2021ecr}
& $5.3\times 10^{-4}$~\cite{Belle-II:2022cgf}
\\
\hline
$b\to s\mu^+\tau^-$
& $(\bar{Y}_3^{QL})_{23}^*(\bar{Y}_3^{QL})_{32}$ 
& $\mathcal{B}(B_s\to\mu^\mp\tau^\pm)$
& $< 3.4\times 10^{-5}\ (90\%)$~\cite{Aaij:2019okb}
& $3\times 10^{-6}$~\cite{LHCb:2018roe}
\\
&
& $\mathcal{B}( B^+ \to K^+ \mu^-\tau^+)$
& $< 5.9\times 10^{-6}\ (90\%)$~\cite{Belle:2022pcr}
& $3.3\times 10^{-6}\ $~\cite{Kou:2018nap}
\\
&
& $\mathcal{B}( B^0 \to K^{*0} \mu^-\tau^+)$
& $< 1.0\times 10^{-5}\ (90\%)$~\cite{LHCb:2022wrs}
& 
\\
\hline
$b\to s\mu^-\tau^+$
& $(\bar{Y}_3^{QL})_{22}^*(\bar{Y}_3^{QL})_{33}$ 
& $\mathcal{B}(B^+\to K^+\mu^+\tau^-)$
& $< 2.45\times 10^{-5}\ (90\%)$~\cite{Belle:2022pcr}
& $3.3\times 10^{-6}\ $~\cite{Kou:2018nap}
\\
&
& $\mathcal{B}(B^0 \to K^{*0} \mu^+\tau^-)$
& $< 8.2\times 10^{-6}\ (90\%)$~\cite{LHCb:2022wrs}
& 
\\
\hline
$\tau^-\to\mu^- \bar{s} s$
& $(\bar{Y}_3^{QL})_{22}^*(\bar{Y}_3^{QL})_{23}$ 
& $\mathcal{B}(\tau^-\to \mu^-\phi)$ 
& $< 2.3\times 10^{-8}\ (90 \%)$~\cite{Belle:2023ziz}
& $8.4\times 10^{-10}$~\cite{Banerjee:2022xuw}
\\
\hline
$b\bar{b}\to\mu^\pm\tau^\mp$
& $(\bar{Y}_3^{QL*})_{32}(\bar{Y}_3^{QL})_{33}$ 
& $\mathcal{B}(\Upsilon(1S)\to \mu^\pm\tau^\mp)$ 
& $<2.7\times 10^{-6}\ (90 \%)$~\cite{Belle:2022cce} 
&
\\
&& $\mathcal{B}(\Upsilon(2S)\to \mu^\pm\tau^\mp)$ 
& $<3.3\times 10^{-6}\ (90 \%)$~\cite{BaBar:2010vxb} 
&
\\
&& $\mathcal{B}(\Upsilon(3S)\to \mu^\pm\tau^\mp)$ 
& $<3.1\times 10^{-6}\ (90 \%)$~\cite{BaBar:2010vxb} 
&
\\
\hline
Loop
& $(\bar{Y}_3^{QL*})_{32}(\bar{Y}_3^{QL})_{33}$ 
& $\mathcal{B}(\tau^-\to \mu^-\gamma)$ 
& $<4.2\times 10^{-8}\ (90 \%)$~\cite{Belle:2021ysv} 
& $6.9\times 10^{-9}$~\cite{Banerjee:2022xuw}
\\
&
& $\mathcal{B}(\tau^-\to \mu^-\mu^+\mu^-)$ 
& $<2.1\times 10^{-8}\ (90 \%)$~\cite{Hayasaka:2010np} 
& $3.6\times 10^{-10}$~\cite{Banerjee:2022xuw}
\\
&
& $\mathcal{B}(Z\to\mu^\mp\tau^\pm)$
& $<6.5\times 10^{-6}\ (95 \%)$~\cite{ATLAS:2021bdj} 
& $\mathcal{O}(10^{-9})$~\cite{Dam:2018rfz}
\\
\hline\hline
\end{tabular}
\end{table}
For the $B\to K^{(*)}\ell^+\ell^-$ decays, we do not consider their branching ratios and the angular observables that exhibit some tensions with the SM~\cite{Gubernari:2022hxn}, since they suffer from hadronic uncertainties~\cite{Jager:2012uw,Lyon:2014hpa,Descotes-Genon:2014uoa,Jager:2014rwa,Ciuchini:2015qxb,Ciuchini:2022wbq}.

For the mass difference $\Delta M_s$, we utilize the following formula that is normalized to the SM value: 
\begin{align}
\frac{\Delta M_s}{\Delta M_s^{\mathrm{SM}}}
&=
\left|
1 + \frac{C_{bs}^{LL,\mathrm{NP}}(m_b)}
{R^{\mathrm{loop}}_{\mathrm{SM}}}
\right|,
\qquad\qquad
C_{bs}^{LL,\mathrm{NP}}(m_b)
=
-\frac{\sqrt{2}}{4\ts G_F (V_{tb}V_{ts}^*)^2}
\big[ L_{\tts dd}^{V,LL}(m_b) \big]_{2323}^{\mathrm{NP}}\,,
\end{align}
where the SM loop contribution
$R^{\mathrm{loop}}_{\mathrm{SM}}=(1.310\pm 0.010)\times 10^{-3}$ and 
the SM prediction 
$\Delta M_s^{\mathrm{SM}}=(18.4^{+0.7}_{-1.2})$ ps$^{-1}$
are evaluated in Ref.~\cite{DiLuzio:2019jyq}. 
Our analysis includes the theoretical uncertainty in $\Delta M_s^{\mathrm{SM}}$, which is much larger than the experimental one. 
In the current model, the LEFT coefficient $[ L_{\tts dd}^{V,LL}(m_b) ]_{2323}^{\mathrm{NP}}$,
given in Eq.~\eqref{eq:LddVLL2323},  
is generated at the one-loop level. 
Contributions from other coefficients with the right-handed quarks are suppressed by the small quark masses and neglected here.
We use the PDG average of the measurements for $\Delta M_s$~\cite{ParticleDataGroup:2022pth}, which gives a constraint on the product of the $S_3$ Yukawa couplings $(\bar{Y}_{3}^{QL} \bar{Y}_{3}^{QL\dagger})_{32}$. 
Because of the hierarchy in the magnitudes of the couplings, 
the product is dominated by 
$(\bar{Y}_{3}^{QL})^*_{23}(\bar{Y}_{3}^{QL})_{33}$ 
compared with 
$(\bar{Y}_{3}^{QL})^*_{21}(\bar{Y}_{3}^{QL})_{31}$ 
and 
$(\bar{Y}_{3}^{QL})^*_{22}(\bar{Y}_{3}^{QL})_{32}$. 

The product $(\bar{Y}_{3}^{QL} \bar{Y}_{3}^{QL\dagger})_{32}$ is also constrained from 
the branching ratios for $B\to K^{(*)}\nu\bar\nu$, which are calculated as 
\begin{align}
\frac{\mathcal{B}(B\to K^{(*)}\nu\bar\nu)}
{\mathcal{B}(B\to K^{(*)}\nu\bar\nu)_{\mathrm{SM}}}
=
\frac{1}{3}
\sum_{ij}
\frac{\big| C_L^{\mathrm{SM}}\ts\delta_{ij} 
+
\big[C_L^{\mathrm{NP}}\big]_{ij}
\big|^2}
{\big|C_L^{\mathrm{SM}}\big|^2}
\,,
\end{align}
where the SM coefficient is given by
$C_L^{\mathrm{SM}}=-X_t/s_W^2$ with 
$X_t=1.469$ and $s_W^2=1-c_W^2$, and the SM predictions are 
$\mathcal{B}(B^+\to K^{+}\nu\bar\nu)_{\mathrm{SM}}
=
(3.98 \pm 0.43 \pm 0.19)\times 10^{-6}$, 
$\mathcal{B}(B^0\to K^{0}\nu\bar\nu)_{\mathrm{SM}}
=
(\tau_{B^0}/\tau_{B^+})
\mathcal{B}(B^+\to K^{+}\nu\bar\nu)_{\mathrm{SM}}$, 
$\mathcal{B}(B^0\to K^{*0}\nu\bar\nu)_{\mathrm{SM}}
=
(9.19 \pm 0.86 \pm 0.50)\times 10^{-6}$, and 
$\mathcal{B}(B^+\to K^{*+}\nu\bar\nu)_{\mathrm{SM}}
=
(\tau_{B^+}/\tau_{B^0})
\mathcal{B}(B^0\to
K^{*0}\nu\bar\nu)_{\mathrm{SM}}$ with 
$\tau_{B^+}$ and $\tau_{B^0}$ being the lifetimes of $B$ mesons~\cite{Buras:2014fpa}. 
The NP contribution
$C_L^{\mathrm{NP}}$ is defined by Eq.~\eqref{eq:CL}, where 
the one-loop expression of the LEFT coefficient 
$[ L_{\tts \nu d}^{V,LL} ]_{ij23}^{\mathrm{NP}}$ is given
in Eq.~\eqref{eq:LnudVLLij23}. 
We select $B^0\to K^{*0}\nu\bar\nu$ as a representative of the 
$B\to K^{(*)}\nu\bar\nu$ processes in our numerical analysis,
where the use of the other processes gives similar results. 
\footnote{Very recently the Belle II collaboration has reported the
  first evidence of the $B^+\to K^+\nu\bar{\nu}$ decay as 
$\mathcal{B}(B^+\to K^+\nu\bar{\nu})=(2.4\pm 0.5^{+0.5}_{-0.4})\times
10^{-5}$~\cite{Belle2:BtoKnunu}. We do not take into account it 
in our analysis.}
The upper limit on $\mathcal{B}(B^0\to K^{*0}\nu\bar\nu)$ is reported
from the Belle experiment~\cite{Grygier:2017tzo}, and provides a
constraint on $(\bar{Y}_{3}^{QL})^*_{23}(\bar{Y}_{3}^{QL})_{33}$. 

\begin{figure}[t]
\centering
\includegraphics[scale=0.55]{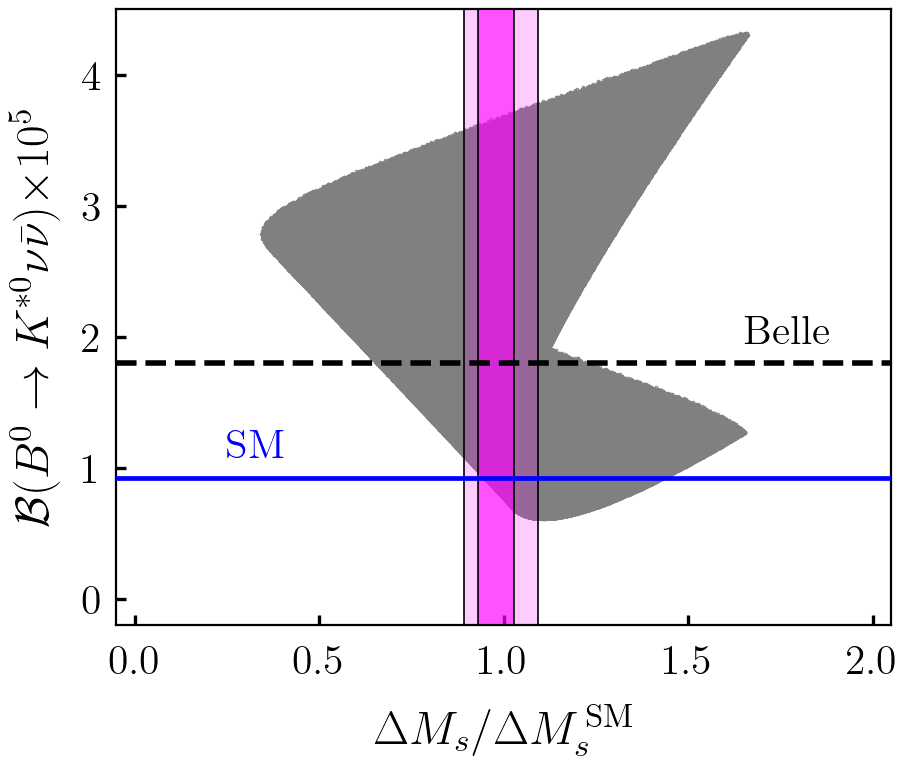}
\hspace{10mm}
\includegraphics[scale=0.55]{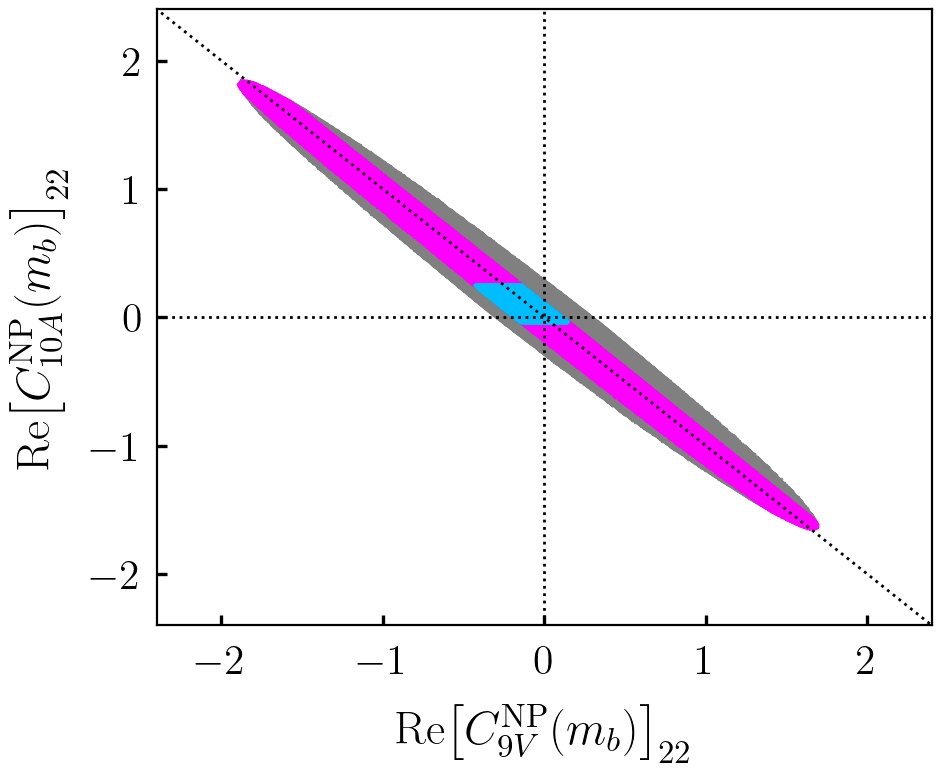}
\caption{Left: constraints from $\Delta M_s/\Delta M_s^{\mathrm{SM}}$
  and $\mathcal{B}(B^0\to K^{*0}\nu\bar\nu)$. The gray region represents the predictions which are consistent with the low-energy values of the gauge couplings and the fermion masses and mixing. 
  The vertical bands in magenta correspond to the experimental measurements at the one and
  two sigma ranges, and the horizontal lines are the
  $90\%$ upper limit at Belle (black dashed line) and the SM prediction (blue solid line). 
  Right: constraints on $\mathrm{Re}[(C_{9V}^{\mathrm{NP}})_{22}]$ and $\mathrm{Re}[(C_{10A}^{\mathrm{NP}})_{22}]$ at the $m_b$ scale, where the oblique dotted line represents $\mathrm{Re}[(C_{9V}^{\mathrm{NP}})_{22}]=-\mathrm{Re}[(C_{10A}^{\mathrm{NP}})_{22}]$. The magenta region can satisfy the experimental bounds from $\Delta M_s$ and $\mathcal{B}(B^0\to K^{*0}\nu\bar{\nu})$, while the cyan region can satisfy further with $R_{K^{+}}[1.1,\,6.0]$, $R_{K^{*0}}[1.1,\,6.0]$, and $\mathcal{B}(B_s\to\mu^+\mu^-)$. These regions are overlaid on top of the gray one, which corresponds to that in the left plot. }
\label{fig:Constraints}
\end{figure}
In the left plot of Fig.~\ref{fig:Constraints}, we present 
constraints in the plane of $\Delta M_s/\Delta M_s^{\mathrm{SM}}$ and
$\mathcal{B}(B^0\to K^{*0}\nu\bar\nu)$, where the gray region is
obtained with the model parameters that are consistent with the
low-energy values of the gauge couplings, the fermion masses, and the
CKM matrix elements. 
Here and hereafter, we take $m_{S_3}=2$ TeV and $\cot\theta_H=50$ as well as the input
parameters in Table~\ref{tab:inputs}. 
A large portion of the parameter space is excluded by the measurement
of $\Delta M_s$ (magenta vertical bands)~\cite{ParticleDataGroup:2022pth} and by the upper limit for
$\mathcal{B}(B^0\to K^{*0}\nu\bar\nu)$ (black horizontal dashed line)~\cite{Grygier:2017tzo},
where the two bands for $\Delta M_s$ correspond to the one-sigma and
two-sigma regions.

Moreover, the measurements for the $b\to s\mu^+\mu^-$ processes listed in Table~\ref{tab:observables} provide constraints on the product of the Yukawa couplings $(\bar{Y}_3^{QL*})_{22}(\bar{Y}_3^{QL})_{32}$. 
In particular, experimental searches for the violation of the lepton-flavor-universality (LFU) in 
$b\to s$ semileptonic decays provide severe constraints on our scenario. 
The LFU ratios $R_H$ ($H=K^+,K^{*0}$) are defined by 
\begin{align}
R_H [q_{\mathrm{min}}^2,\, q_{\mathrm{max}}^2]
  &=
  \frac{\displaystyle
  \int_{q_{\mathrm{min}}^2}^{q_{\mathrm{max}}^2}
  \!d\tts q^2\ts
  \frac{d\tts\mathcal{B}(B\to H\mu^+\mu^-)}{d\tts q^2}
  }{\displaystyle
  \int_{q_{\mathrm{min}}^2}^{q_{\mathrm{max}}^2}
  \!d\tts q^2\ts
  \frac{d\tts\mathcal{B}(B\to He^+e^-)}{d\tts q^2}
  }\,,
\end{align}  
where $q_{\mathrm{min}}^2$ and $q_{\mathrm{max}}^2$ are given in units of GeV$^2$. 
For example, approximate formulas for the region of $1.1\,\mathrm{GeV}^2<q^2<6.0\,\mathrm{GeV}^2$ are 
given in Ref.~\cite{Celis:2017doq}:
\begin{align}
R_{K}[1.1,\, 6.0]
&\approx
1.00
+ 0.23\,\mathrm{Re}\big(
\Delta C_{9V}^{\ts\mathrm{NP}}
\big)
- 0.25\,\mathrm{Re}\big(
\Delta C_{10A}^{\ts\mathrm{NP}}
\big)\,,
\\
R_{K^*}[1.1,\,6.0]
&\approx
1.00
+ 0.20\,\mathrm{Re}\big(
\Delta C_{9V}^{\ts\mathrm{NP}}
\big)
- 0.27\,\mathrm{Re}\big(
\Delta C_{10A}^{\ts\mathrm{NP}}
\big)\,,
\end{align}
where
$\Delta C_{9V}^{\ts\mathrm{NP}} \equiv
[ C_{9V}^{\ts\mathrm{NP}}(m_b) ]_{22}
-
[ C_{9V}^{\ts\mathrm{NP}}(m_b) ]_{11}$
and 
$\Delta C_{10A}^{\ts\mathrm{NP}} \equiv 
[ C_{10A}^{\ts\mathrm{NP}}(m_b) ]_{22}
-
[ C_{10A}^{\ts\mathrm{NP}}(m_b) ]_{11}$. 
These LFU ratios are calculated very accurately in the SM, where 
the hadronic uncertainty is highly canceled by considering the ratios~\cite{Capdevila:2017ert}, 
and the QED correction provides a positive contribution to the ratios 
about less than 3\% for 
$1\,\mathrm{GeV}^2< q^2 < 6\,\mathrm{GeV}^2$~\cite{Bordone:2016gaq,Isidori:2020acz}.
The above approximate formulas are derived by neglecting the QED corrections. The theoretical uncertainties are negligible in our study. 
The recent measurements at LHCb~\cite{LHCb:2022vje} listed in Table~\ref{tab:observables} 
are compatible with the SM predictions.
We adopt only $R_{K}[1.1,\,6.0]$ and $R_{K^*}[1.1,\,6.0]$ as constraints, since 
the ratios in the low $q^2$ regions 
$R_{K}[0.1,\,1.1]$ and $R_{K^*}[0.1,\,1.1]$
have larger experimental uncertainties. 
In addition, we also consider the branching ratio for the leptonic decay
$B_s \to \mu^+\mu^-$, which is written simply with the NP contribution to $C_{10A}$: 
\begin{align}
\mathcal{B}( B_s \to \mu^+\mu^- )
&=
\mathcal{B}( B_s \to \mu^+\mu^- )_{\mathrm{SM}}
\left|\,
1 + 
\frac{\big[C_{10A}^{\mathrm{NP}}(m_b)\big]_{22}}
{C_{10A}^{\mathrm{SM}}(m_b)}
\,\right|^2,
\label{eq:Bsmumu}
\end{align}
where the SM values are 
$\mathcal{B}( B_s \to \mu^+\mu^- )_{\mathrm{SM}}
=
(3.65 \pm 0.23)\times 10^{-9}$~\cite{Bobeth:2013uxa} and 
$C_{10A}^{\mathrm{SM}}(m_b)=-4.2$~\cite{Blake:2016olu}.
It is noted that a nonvanishing decay width difference $\Delta\Gamma_s$ of the $B_s$ system 
has to be taken into account when comparing the theoretical value calculated using Eq.~\eqref{eq:Bsmumu} 
with the experimental data in Table~\ref{tab:observables}, 
since the time dependence of the decay rate is integrated in the experiment~\cite{DeBruyn:2012wj,DeBruyn:2012wk}. 
This gives only a minor effect on our numerical analysis. 
In the current model, $[C_{9V}^{\mathrm{NP}}(m_b)]_{22}$ and $[C_{10A}^{\mathrm{NP}}(m_b)]_{22}$ appearing in 
$R_{K}[1.1,\,6.0]$, $R_{K^*}[1.1,\,6.0]$, and 
$\mathcal{B}(B_s \to \mu^+\mu^-)$
are dominated by the LEFT coefficient $[L_{ed}^{V,LL}(m_b)]_{2223}$, which is given in terms of the product of the $S_3$ Yukawa couplings 
$(\bar{Y}_3^{QL*})_{22}(\bar{Y}_3^{QL})_{32}$ at the tree level.

The right plot of Fig.~\ref{fig:Constraints} shows constraints on 
$\mathrm{Re}[C_{9V}^{\mathrm{NP}}(m_b)]_{22}$ and 
$\mathrm{Re}[C_{10A}^{\mathrm{NP}}(m_b)]_{22}$. 
The magenta region can satisfy the experimental bounds from 
$\Delta M_s$ within two sigma and 
$\mathcal{B}(B^0\to K^{*0}\nu\bar{\nu})$ at 90\% C.L., while the cyan
region can satisfy further $R_{K^{+}}[1.1,\,6.0]$,
$R_{K^{*0}}[1.1,\,6.0]$, and $\mathcal{B}(B_s\to\mu^+\mu^-)$ within
two sigma. 
These regions are overlaid on top of the gray one, which corresponds  
to that in the left plot.

\begin{figure}[t]
\centering
\includegraphics[scale=0.47]{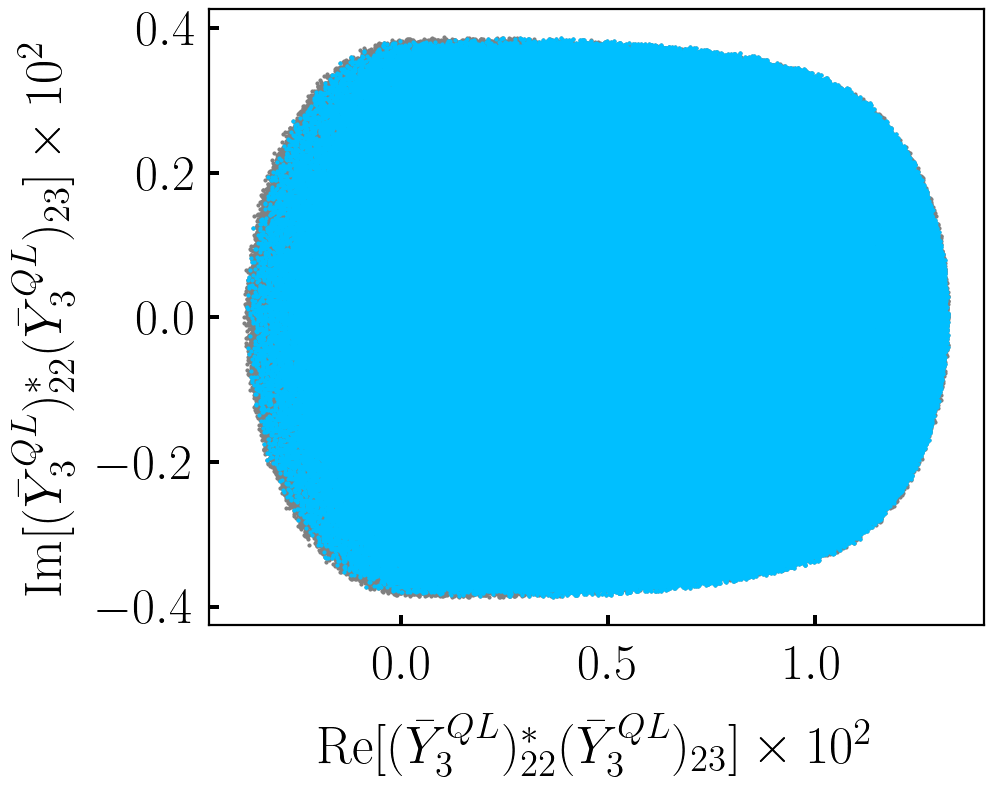}
\hfill
\includegraphics[scale=0.47]{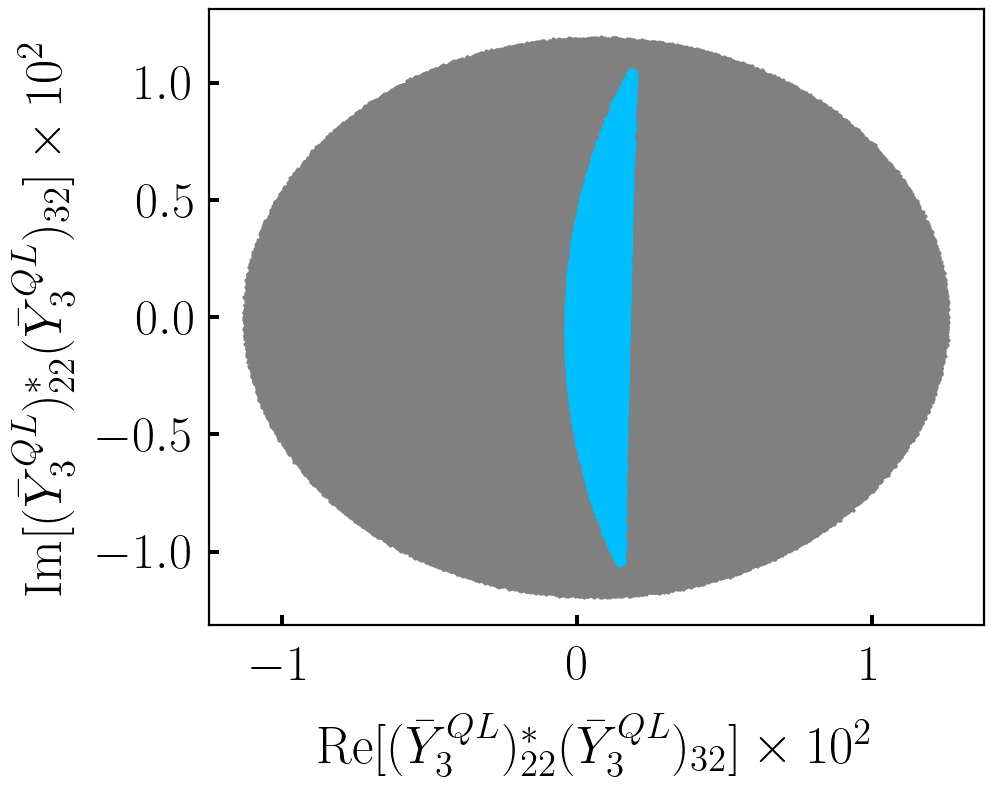}
\hfill
\includegraphics[scale=0.47]{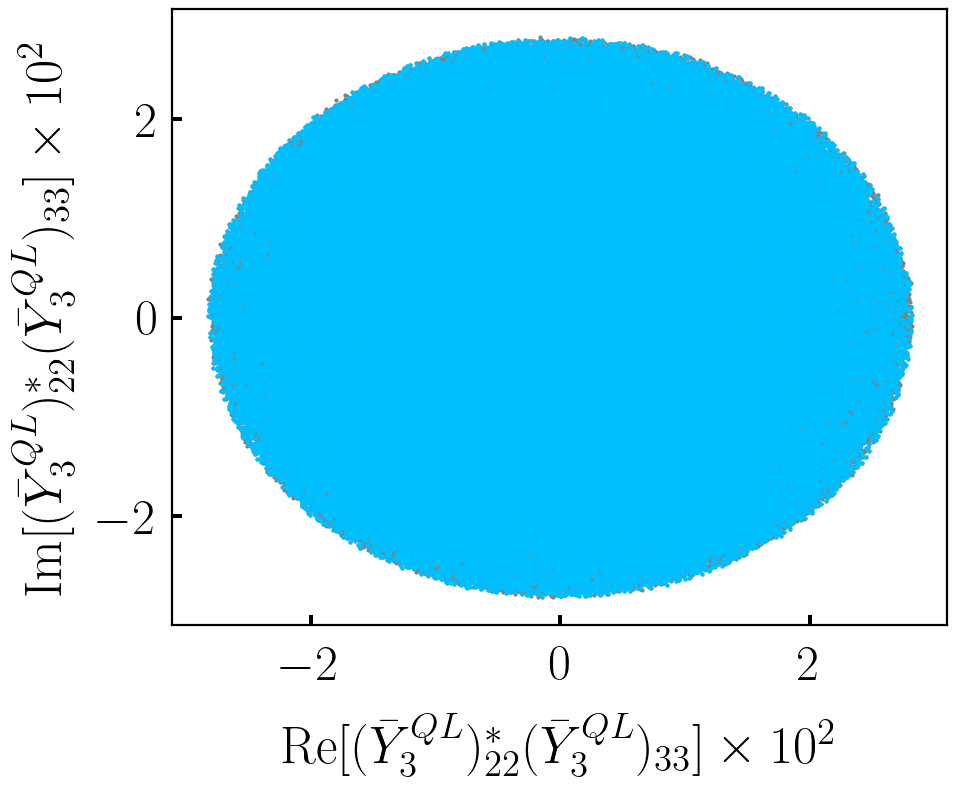}
\\[2mm]
\includegraphics[scale=0.47]{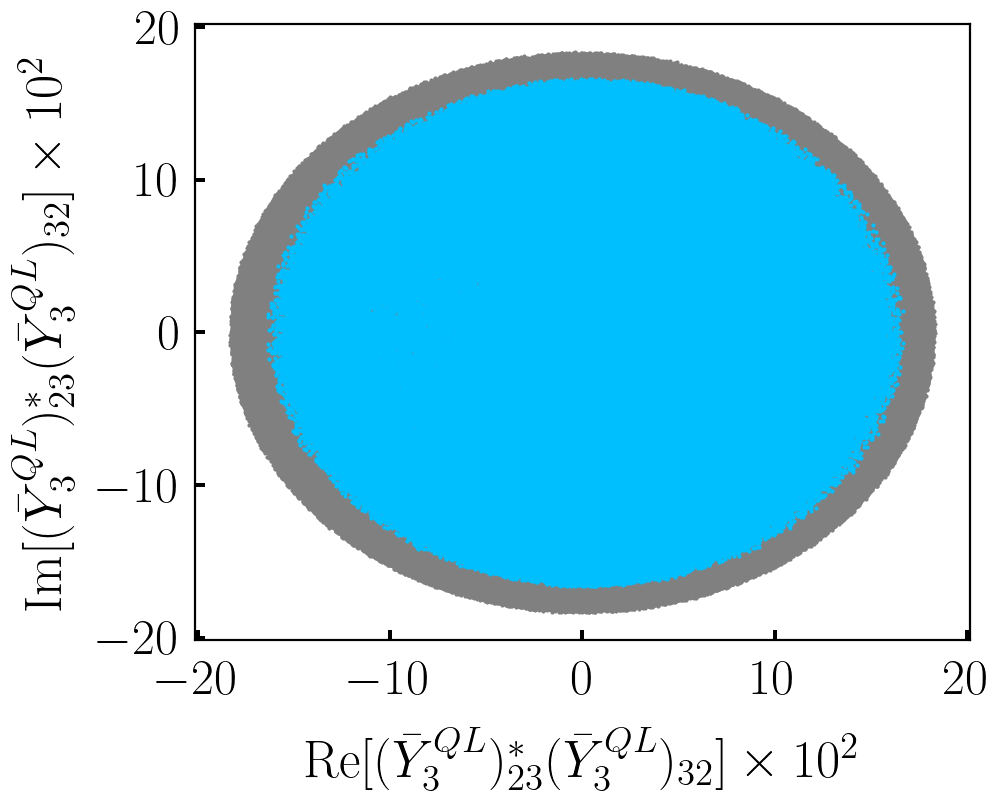}
\hfill
\includegraphics[scale=0.47]{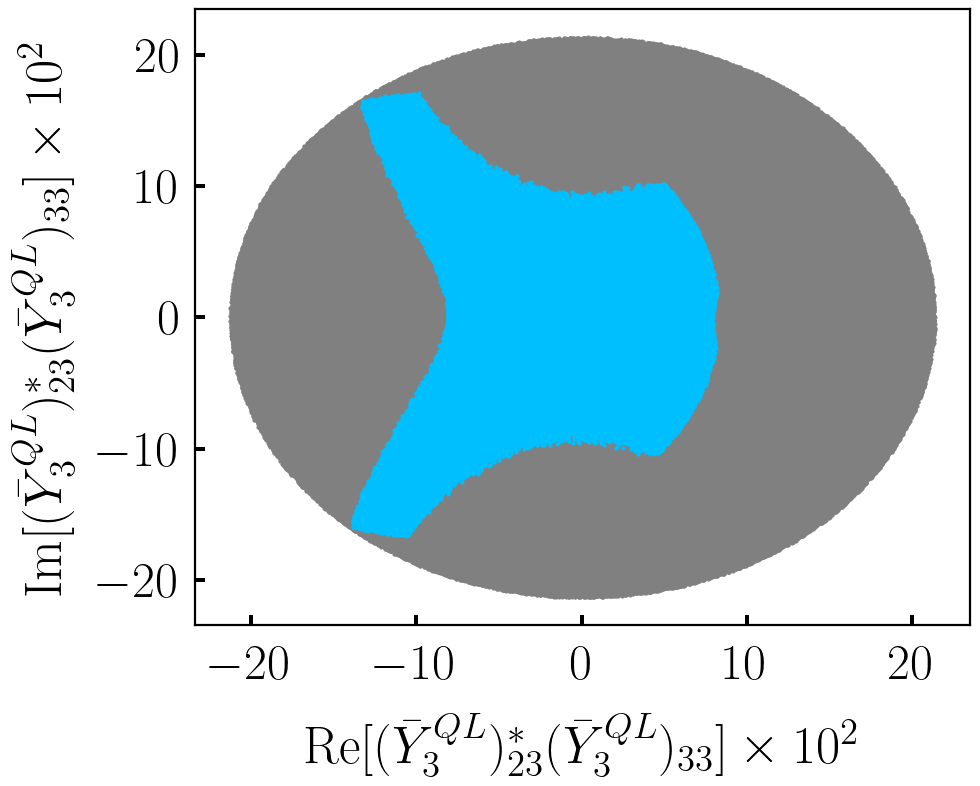}
\hfill
\includegraphics[scale=0.47]{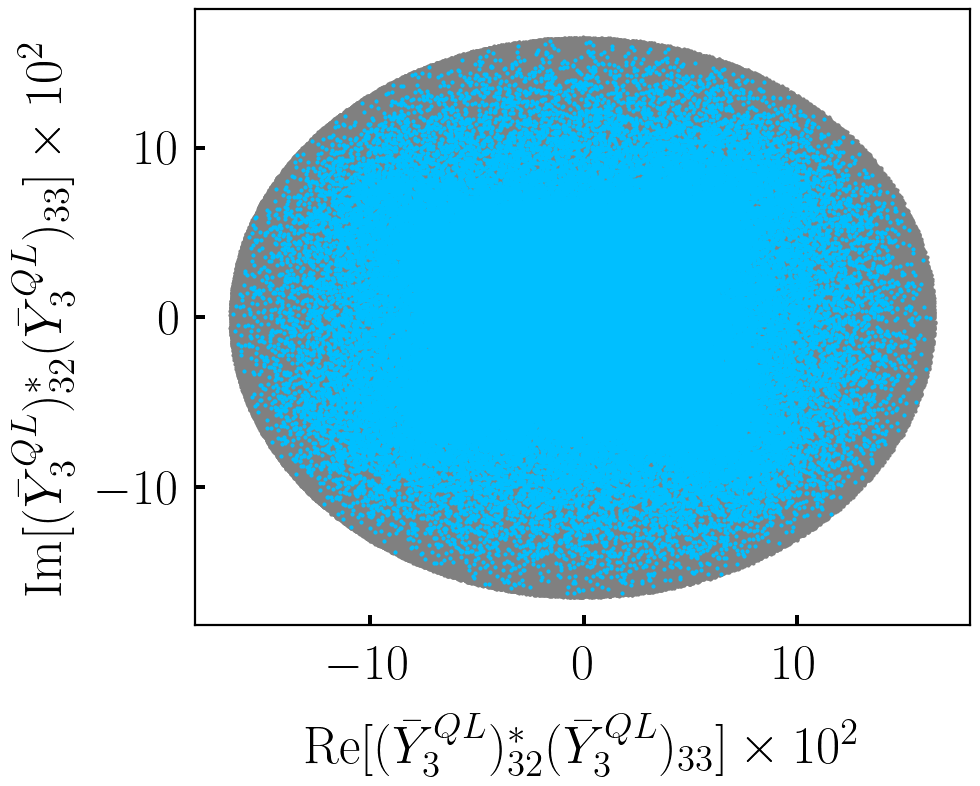}
\caption{Allowed region for the products of the Yukawa couplings of the $S_3$
  leptoquark at the $S_3$ mass scale, where the cyan region
  shows the parameter points that are consistent with  
  $\Delta M_s$, 
  $R_{K^{+}}[1.1,\,6.0]$, $R_{K^{*0}}[1.1,\,6.0]$, and $\mathcal{B}(B_s\to\mu^+\mu^-)$
  within two sigma and $\mathcal{B}(B^0\to K^{*0}\nu\bar\nu)$ at 90\% C.L..
  The cyan regions are overlaid on top of the gray ones, which correspond 
  to those in Fig.~\ref{fig:Constraints}.}
\label{fig:Y3QL-1}
\end{figure}
We also present allowed regions for
the products of the $S_3$ Yukawa couplings at the $S_3$ mass scale in
Fig.~\ref{fig:Y3QL-1}. 
Here the cyan regions show the parameter points that
are consistent with $\Delta M_s$,
$R_{K^{+}}[1.1,\,6.0]$, $R_{K^{*0}}[1.1,\,6.0]$, and
$\mathcal{B}(B_s\to\mu^+\mu^-)$ within two sigma and
$\mathcal{B}(B^0\to K^{*0}\nu\bar\nu)$ at 90\% confidence level (C.L.). 
It is noted that the cyan regions are overlaid on top of 
the gray regions that correspond to those in Fig.~\ref{fig:Constraints}. 
The magnitudes of the products in the upper row of Fig.~\ref{fig:Y3QL-1} are smaller
than those in the lower row because of the hierarchy given in Eq.~\eqref{eq:hierarchy}. 
The product 
$(\bar{Y}_3^{QL*})_{22}(\bar{Y}_3^{QL})_{32}$ is highly constrained 
by $R_{K^{+}}[1.1,\,6.0]$, $R_{K^{*0}}[1.1,\,6.0]$, and
$\mathcal{B}(B_s\to\mu^+\mu^-)$, while 
$(\bar{Y}_3^{QL*})_{23}(\bar{Y}_3^{QL})_{33}$ is by 
$\Delta M_s$ and $\mathcal{B}(B^0\to K^{*0}\nu\bar\nu)$. 
The other products are less constrained by these observables.

\subsection{Predictions}

\begin{figure}[t]
\centering
\includegraphics[scale=0.55]{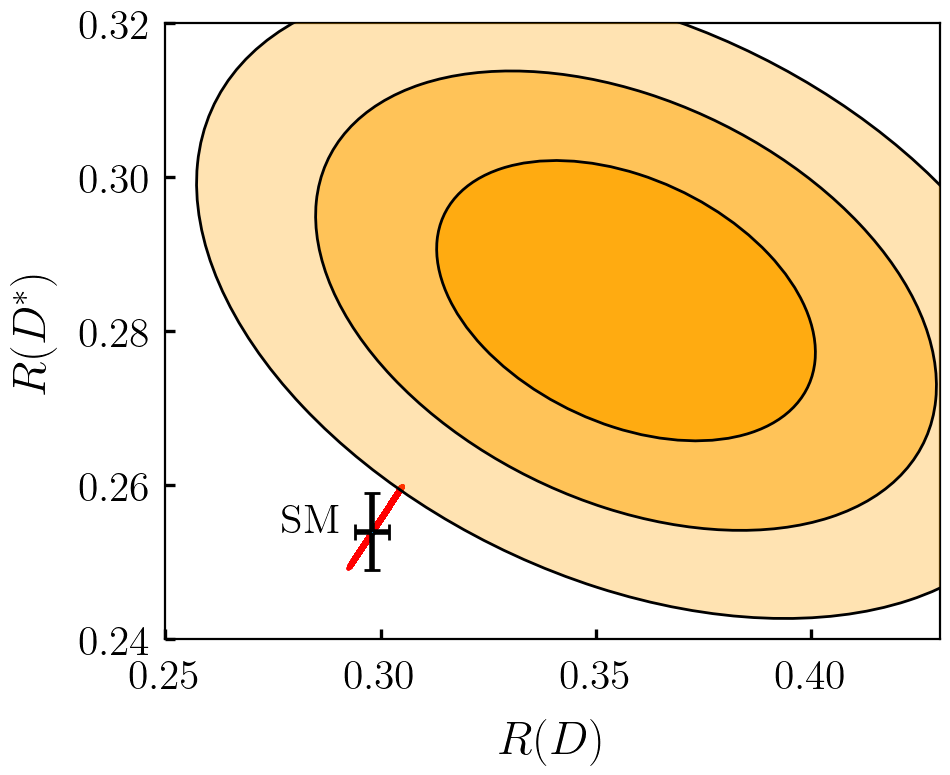}
\caption{Predictions for $R(D)$ and $R(D^*)$ (denoted by the red points) with the HFLAV average of their experimental measurements at the levels of one sigma, two sigma, and three sigma (denoted by the orange ellipses) and the SM values (denoted by the black cross). Theoretical uncertainties associated with the SM errors are not included in the predictions.}
\label{fig:BtoCanomaly}
\end{figure}

\begin{figure}[t]
\centering
\begin{tabular}{c@{\hskip 10mm}c}
\includegraphics[scale=0.55]{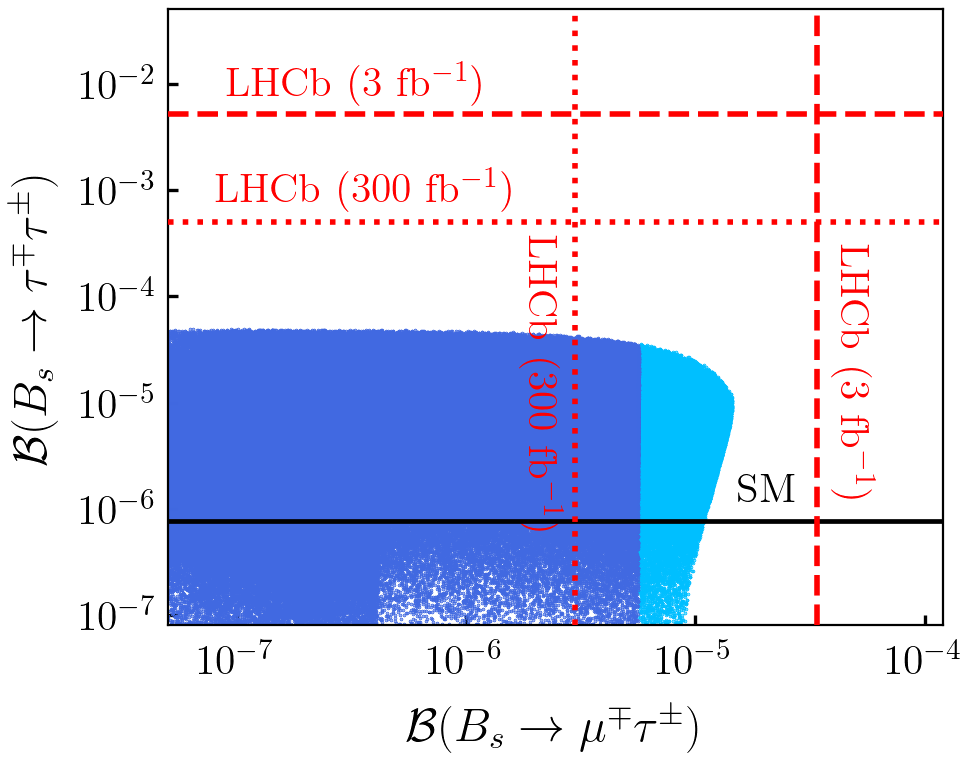}
&
\includegraphics[scale=0.55]{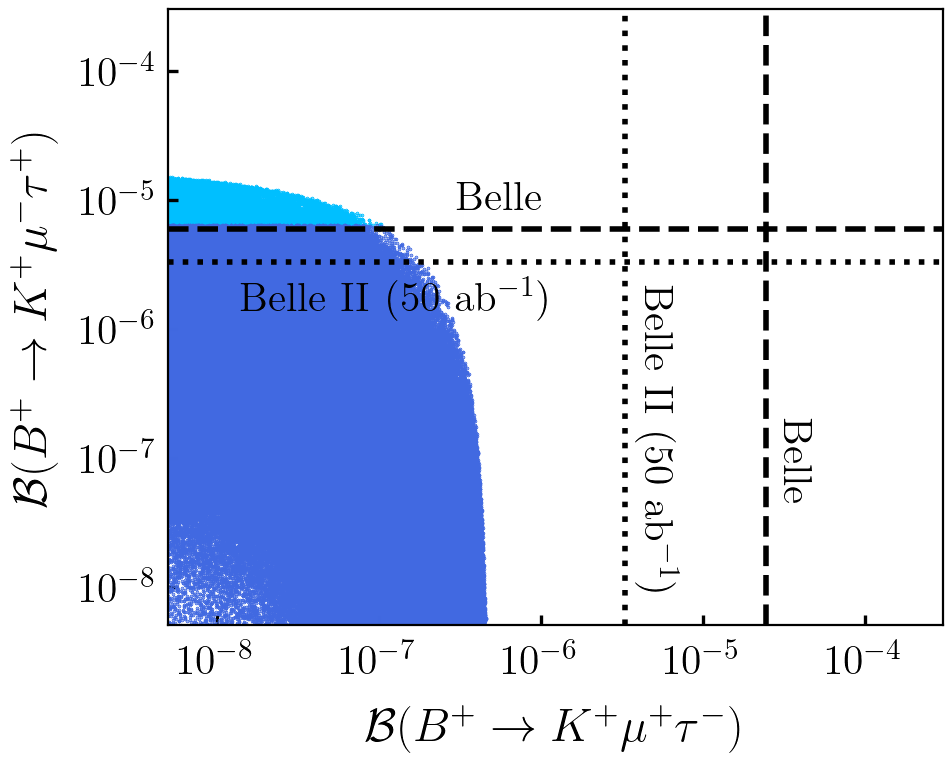}
\\
(a) & (b)
\\[2mm]
\includegraphics[scale=0.55]{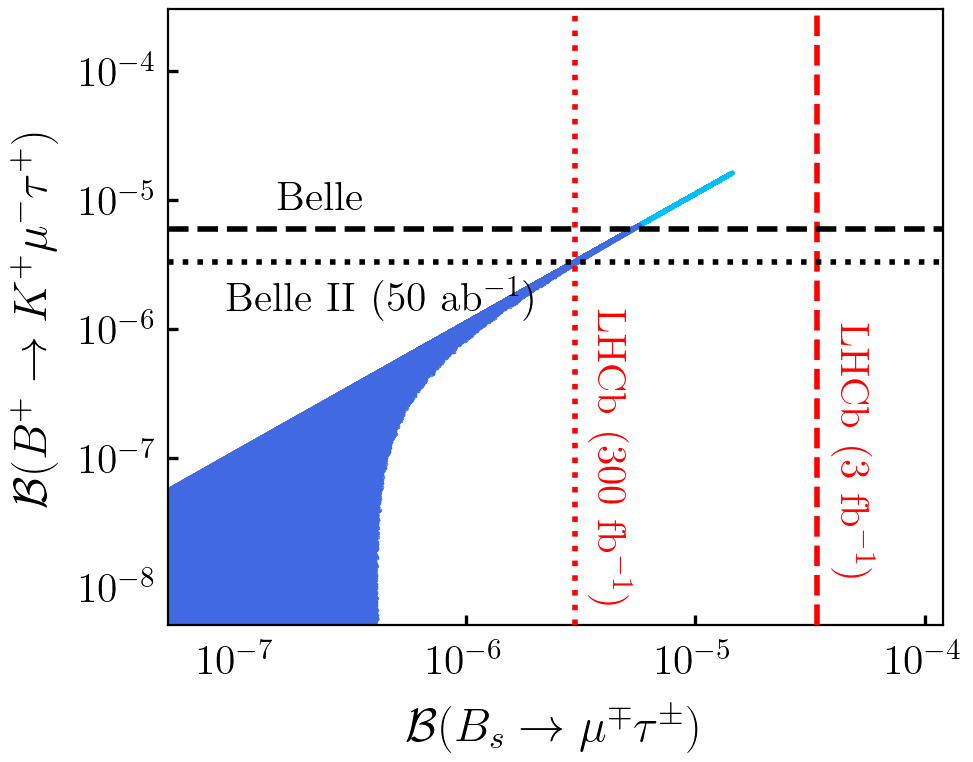}
&
\includegraphics[scale=0.55]{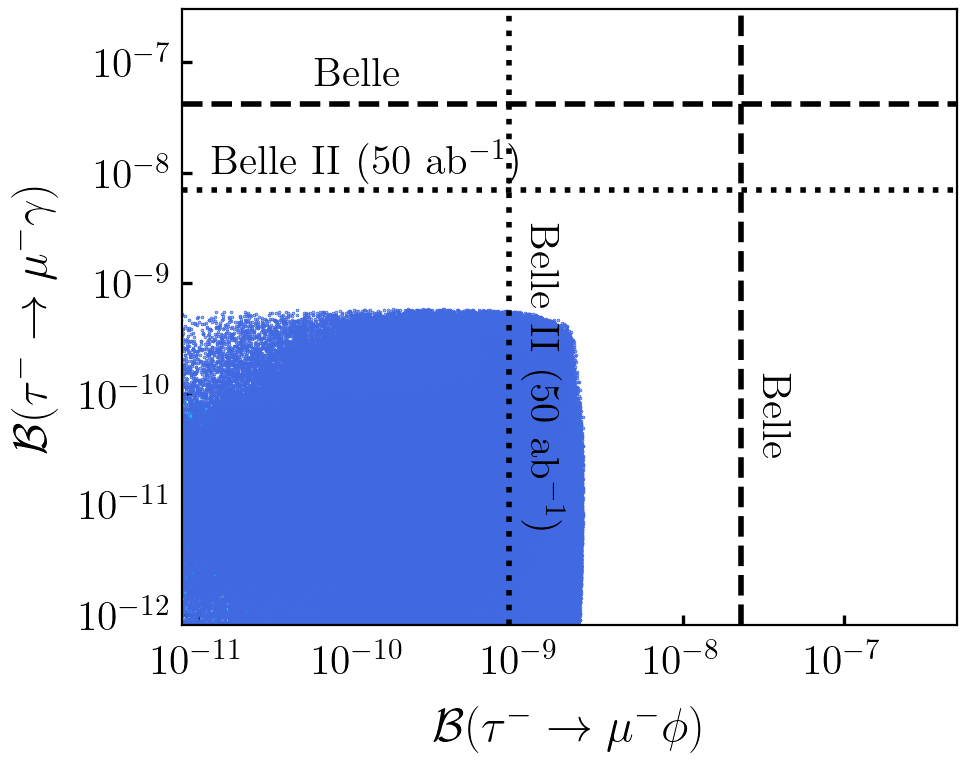}
\\
(c) & (d)
\\[2mm]
\includegraphics[scale=0.55]{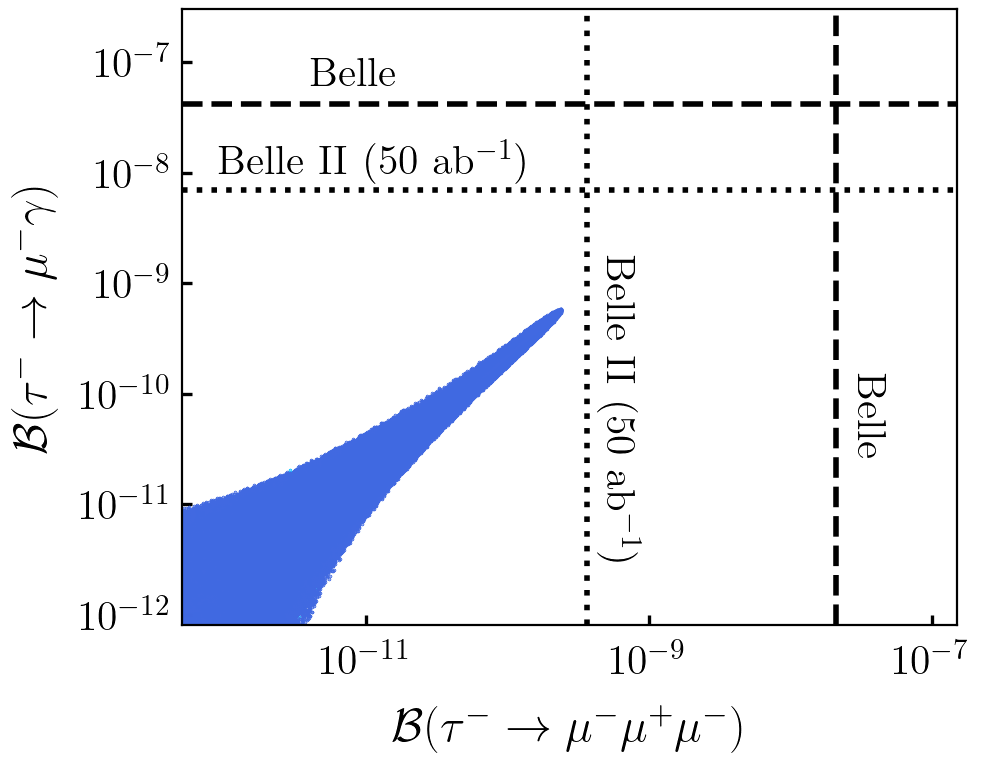}
&
\includegraphics[scale=0.55]{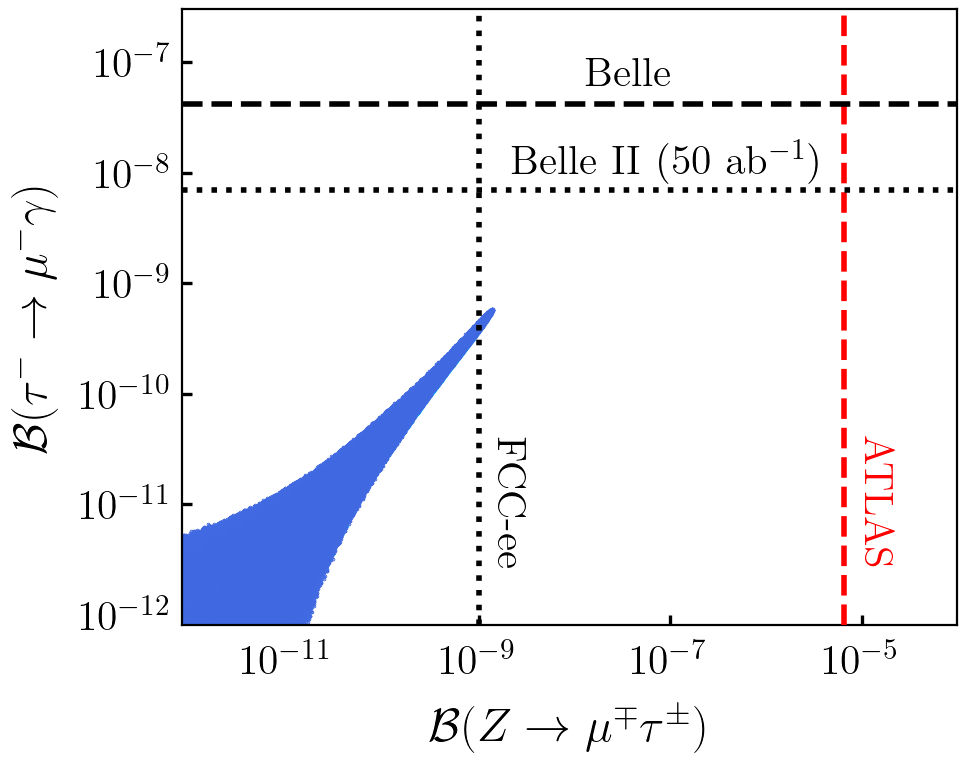}
\\
(e) & (f)
\end{tabular}
\caption{Predictions on relevant flavor processes, where the colored regions 
  satisfy the experimental bounds from $R_{K^{+}}[1.1,\,6.0]$,
  $R_{K^{*0}}[1.1,\,6.0]$, $\mathcal{B}(B_s\to\mu^+\mu^-)$,
  and $\Delta M_s$ within two sigma and $\mathcal{B}(B^0\to K^{*0}\nu\bar\nu)$ at 90\% C.L.. 
  The red and black dashed lines show the present upper bound on each processes by LHC experiments and $B$ factories, respectively, and the red and black dotted lines show the sensitivities expected at the LHCb with 300~fb$^{-1}$ and the $e^+e^-$ experiments (such as the Belle II with 50~ab$^{-1}$ and FCC-ee), respectively.  
  The cyan regions in (a), (b), and (c) are excluded by the upper limit
  on $\mathcal{B}(B^+\to K^+\mu^-\tau^+)$ at Belle. 
}
\label{fig:Predictions}
\vspace{-2mm}
\end{figure}

The $S_3$ leptoquark can generate various LFV and LFUV with
the second- and third-generation fermions. 
Under the constraints studied in Sec.~\ref{sec:constraints}, 
we here consider the following observables: 
$R(D^{(*)})$, 
$\mathcal{B}(B_s\to \tau^{+}\tau^{-})$, 
$\mathcal{B}(B_s\to \mu^{\mp}\tau^{\pm})$, 
$\mathcal{B}(B\to K^{(*)}\mu^\mp\tau^\pm)$,
$\mathcal{B}(\Upsilon(nS)\to \mu^\pm\tau^\mp)$, 
$\mathcal{B}(\tau^-\to \mu^-\phi)$, 
$\mathcal{B}(\tau^-\to \mu^-\gamma)$, 
$\mathcal{B}(\tau^-\to \mu^-\mu^+\mu^-)$,
and
$\mathcal{B}(Z\to \mu^\mp\tau^\pm)$.
The first six observables receive tree-level contributions, 
while the rest are induced at the one-loop level. 
Figures~\ref{fig:BtoCanomaly} and \ref{fig:Predictions} show 
predictions for these observables in the current model. 
Here we only consider flavor-changing-neutral-current processes except
for $R(D^{(*)})$, since the $S_3$ effects on 
charged-current processes, such as $B^0\to D^{(*)-}\mu^+\nu$ and
$D_s^+\to\mu^+\nu$, are not significant.

In Fig.~\ref{fig:BtoCanomaly}, we present the predictions for the ratios 
$R(D^{(*)})=
\mathcal{B}(B^0\to D^{(*)}\tau^+\nu) /
\mathcal{B}(B^0\to D^{(*)}\ell^+\nu)$ for $\ell=e,\mu$ calculated under the constraints
from $\Delta M_s/\Delta M_s^{\mathrm{SM}}$, 
$\mathcal{B}(B^0\to K^{*0}\nu\bar\nu)$, 
$R_{K^{+}}[1.1,\,6.0]$, $R_{K^{*0}}[1.1,\,6.0]$, and 
$\mathcal{B}(B_s\to\mu^+\mu^-)$.
At the tree level $R(D^{(*)})$ are given by 
\begin{align}
R(D^{(*)})
\approx
R(D^{(*)})_{\mathrm{SM}}
\Big( 1 + 2\, \mathrm{Re}\big[C_{V_1}^{\mathrm{NP}}(m_b)\big]_{33} \Big)\,,
\label{eq:RDandRDst}
\end{align}
where we adopt the SM predictions $R(D)_{\mathrm{SM}}=0.298\pm 0.004$ and
$R(D^{*})_{\mathrm{SM}}=0.254\pm 0.005$~\cite{HeavyFlavorAveragingGroup:2022wzx}. 
The coefficient $C_{V_1}^{\mathrm{NP}}$ is defined through the effective Lagrangian,  
\begin{align}
\mathcal{L}_{\mathrm{eff}}
&=
-
\frac{4\tts G_F}{\sqrt{2}}
V_{cb}^*
\Big( \delta_{ij} + \big[C_{V_1}^{\mathrm{NP}}(m_b)\big]_{ij} \Big)
\big( \bar{\hat{b}}_L\ts\gamma^\mu \hat{c}_L \big)
\big( \bar{\hat{\nu}}_{Li}\ts\gamma_\mu \hat{e}_{Lj} \big)\,,
&
\big[C_{V_1}^{\mathrm{NP}}(m_b)\big]_{33}
&=
-
\frac{1}{2\sqrt{2}\ts G_F V_{cb}^*}
\big[
L_{\nu e d u}^{V,LL}(m_b)
\big]^{\mathrm{NP}}_{3332}\,,
\end{align}
where we use the tree-level result for the LEFT coefficient
$[L_{\nu e d u}^{V,LL}(m_b)]^{\mathrm{NP}}_{3332}$ given in Eq.~\eqref{eq:LnueduVLL-tree}. 
We keep only the $33$ component of $C_{V_1}^{\mathrm{NP}}$ in
Eq.~\eqref{eq:RDandRDst}, since the dominant NP contributions arise in
the $23$, $32$, and $33$ ones in the current model and only the $33$ one has an
interference with the SM contribution.
We use the average of the experimental data by the Heavy Flavor Averaging Group (HFLAV)~\cite{HeavyFlavorAveragingGroup:2022wzx}. 
Here the $b\to c\tau^-\bar{\nu}$ transition is dominated by the contribution
from the product $(\bar{Y}_3^{QL})_{23}^*(\bar{Y}_3^{QL})_{33}$, which also contributes to 
$b\to s \nu\bar\nu$ and $\Delta M_s$. 
It is known that the $S_3$ contribution that explains the $b\to c$ anomaly
is severely constrained by the $b\to s \nu\bar\nu$ processes and $\Delta M_s$~\cite{Dorsner:2017ufx}. 
Consequently, the $S_3$ contribution does not alter $R(D^{(*)})$ significantly, and thus 
the resolution of the $R(D^{(*)})$ anomaly requires an extension of the model~\cite{Crivellin:2017zlb,Buttazzo:2017ixm}. We do not consider such a possibility in the current paper.

Next, let us consider decay processes involving $b\to s\tau^+\tau^-$ transition.
The studies of NP contributions to this transition are found,
for example, in Refs.~\cite{Bobeth:2011st,Capdevila:2017iqn}. 
In the current model, the contributions to the $b\to s\tau^+\tau^-$ leptonic and
semileptonic decays arise at the tree level through the product
$(\bar{Y}_3^{QL})_{23}^*(\bar{Y}_3^{QL})_{33}$. 
As in the case of $B_s\to\mu^+\mu^-$ in Eq.~\eqref{eq:Bsmumu}, 
the leptonic mode receives NP contribution to $C_{10A}$: 
\begin{align}
\mathcal{B}( B_s \to \tau^+\tau^- )
&=
\mathcal{B}( B_s \to \tau^+\tau^- )_{\mathrm{SM}}
\left|\,
1 + 
\frac{\big[C_{10A}^{\mathrm{NP}}(m_b)\big]_{33}}
{C_{10A}^{\mathrm{SM}}(m_b)}
\,\right|^2,
\end{align}
where the SM prediction is 
$\mathcal{B}( B_s \to \tau^+\tau^- )_{\mathrm{SM}}
=
(7.73 \pm 0.49)\times 10^{-7}$~\cite{Bobeth:2013uxa}. 
Moreover, the branching ratios of the semileptonic modes in the large
$q^2$ region are calculated in Ref.~\cite{Capdevila:2017iqn}: 
\begin{align}
\mathcal{B}( B \to K \tau^+\tau^- )^{[15,22]}
&=
10^{-7}
\Big(
1.20
+ 
0.15\, \mathrm{Re}\big[C_{9V}^{\mathrm{NP}}(m_b)\big]_{33}
-
0.42\, \mathrm{Re}\big[C_{10A}^{\mathrm{NP}}(m_b)\big]_{33}
\nonumber\\
&\hspace{17mm}
+
0.02\, \big| \big[C_{9V}^{\mathrm{NP}}(m_b)\big]_{33} \big|^2
+
0.05\, \big| \big[C_{10A}^{\mathrm{NP}}(m_b)\big]_{33} \big|^2
\Big)\,, 
\\
\mathcal{B}( B \to K^* \tau^+\tau^- )^{[15,19]}
&=
10^{-7}
\Big(
0.98
+ 
0.38\, \mathrm{Re}\big[C_{9V}^{\mathrm{NP}}(m_b)\big]_{33}
-
0.14\, \mathrm{Re}\big[C_{10A}^{\mathrm{NP}}(m_b)\big]_{33}
\nonumber\\
&\hspace{17mm}
+
0.05\, \big| \big[C_{9V}^{\mathrm{NP}}(m_b)\big]_{33} \big|^2
+
0.02\, \big| \big[C_{10A}^{\mathrm{NP}}(m_b)\big]_{33} \big|^2
\Big)\,, 
\end{align}
which are the averages of the charged and the neutral modes. 
The predicted branching ratios in the SM are of 
$\mathcal{O}(10^{-7})$~\cite{Capdevila:2017iqn}.  
The branching ratios for these leptonic and semileptonic modes can
largely deviate from their SM values. 
Figure~\ref{fig:Predictions}(a) shows that 
$\mathcal{B}( B_s \to \tau^+\tau^- )$
can be as large as $\mathcal{O}(10^{-5})$, which is an order of magnitude
smaller than the future sensitivity at LHCb with 300 fb$^{-1}$~\cite{LHCb:2018roe}. 
Similarly, the predictions for $\mathcal{B}( B \to K^{(*)}
\tau^+\tau^- )$ in the large $q^2$ region can be enhanced by an order
of magnitude, but it is still much smaller than the future sensitivity at Belle II with 50
fb$^{-1}$~\cite{Kou:2018nap}.

We also study the LFV processes $b\to s\mu^+\tau^-$ and $b\to s\mu^-\tau^+$, which are generated through the products of the $S_3$ Yukawa couplings 
$(\bar{Y}_3^{QL})_{23}^*(\bar{Y}_3^{QL})_{32}$ and 
$(\bar{Y}_3^{QL})_{22}^*(\bar{Y}_3^{QL})_{33}$, respectively. 
Because of the hierarchy in the magnitudes of the $S_3$ Yukawa couplings presented in Eq.~\eqref{eq:hierarchy} and Fig.~\ref{fig:Y3QL-1}, 
the relation $|(\bar{Y}_3^{QL})_{23}^*(\bar{Y}_3^{QL})_{32}| \gg |(\bar{Y}_3^{QL})_{22}^*(\bar{Y}_3^{QL})_{33}|$ holds typically. 
At the LHC experiments, the branching ratio for the leptonic decay is measured as a sum of the two channels
$B_s \to \mu^-\tau^+$
and
$B_s \to \mu^+\tau^-$. 
The corresponding theoretical formula is given by~\cite{Dedes:2008iw}
\begin{align}
\mathcal{B}( B_s \to \mu^\mp\tau^\pm )
&=
\mathcal{B}( B_s \to \mu^-\tau^+ )
+
\mathcal{B}( B_s \to \mu^+\tau^- )
\,,
\nonumber\\
&=
\frac{\tau_{B_s}\, f_{B_s}^2 m_{B_s}\ts m_{\tau}^2\,
  \alpha^2\tts G_F^2\ts \big|V^*_{ts}V_{tb}\big|^2}
{64\ts\pi^3}
\left(
1 - \frac{m_\tau^2}{m_{B_s}^2}
\right)^2
\nonumber\\
&\hspace{10mm}\times
\Big(\,
\big|\big[C_{9V}^{\mathrm{NP}}(m_b)\big]_{23}\big|^2
+
\big|\big[C_{10A}^{\mathrm{NP}}(m_b)\big]_{23}\big|^2
+
\big|\big[C_{9V}^{\mathrm{NP}}(m_b)\big]_{32}\big|^2
+
\big|\big[C_{10A}^{\mathrm{NP}}(m_b)\big]_{32}\big|^2
\Big)
\,,
\end{align}
where
$m_\tau$ and $\tau_\tau$ are the mass and the lifetime of $\tau$ lepton, 
$m_{B_s}$, $\tau_{B_s}$, and $f_{B_s}$ are the mass, the lifetime, and the decay constant of $B_s$ meson, 
and the muon mass is neglected. 
As shown in Fig.~\ref{fig:Predictions}(a), the prediction on 
$\mathcal{B}(B_s\to \mu^{\mp}\tau^{\pm})$ can be as large as 
$\mathcal{O}(10^{-5})$, which may be probed by the 
LHCb measurement with 300 fb$^{-1}$~\cite{LHCb:2018roe}.
For the semileptonic channels, approximate numerical formulas are
given by~\cite{Becirevic:2016zri} 
\begin{align}
\mathcal{B}( B^+ \to K^{+} \mu^-\tau^+ )
&=
10^{-9}
\Big(
12.5\,
\big|\big[C_{9V}^{\mathrm{NP}}(m_b)\big]_{32}\big|^2
+
12.9\,
\big|\big[C_{10A}^{\mathrm{NP}}(m_b)\big]_{32}\big|^2
\Big)
\frac{\tau_{B^+}}{\tau_{B^0}}\,,
\\
\mathcal{B}( B^+ \to K^{+} \mu^+ \tau^- )
&=
10^{-9}
\Big(
12.5\,
\big|\big[C_{9V}^{\mathrm{NP}}(m_b)\big]_{23}\big|^2
+
12.9\,
\big|\big[C_{10A}^{\mathrm{NP}}(m_b)\big]_{23}\big|^2
\Big)
\frac{\tau_{B^+}}{\tau_{B^0}}\,,
\\
\mathcal{B}( B^0 \to K^{*0} \mu^-\tau^+ )
&=
10^{-9}
\Big(
22.1\,
\big|\big[C_{9V}^{\mathrm{NP}}(m_b)\big]_{32}\big|^2
+
20.6\,
\big|\big[C_{10A}^{\mathrm{NP}}(m_b)\big]_{32}\big|^2
\Big)\,,
\\
\mathcal{B}( B^0 \to K^{*0} \mu^+ \tau^- )
&=
10^{-9}
\Big(
22.1\,
\big|\big[C_{9V}^{\mathrm{NP}}(m_b)\big]_{23}\big|^2
+
20.6\,
\big|\big[C_{10A}^{\mathrm{NP}}(m_b)\big]_{23}\big|^2
\Big)\,. 
\label{eq:bsmutauSemi}
\end{align}
It is noted that
$B^+ \to K^{+} \mu^-\tau^+$ ($B^0 \to K^{*0} \mu^-\tau^+$) 
and
$B^+ \to K^{+} \mu^+ \tau^-$ ($B^0 \to K^{*0} \mu^+ \tau^-$) 
receive contributions from 
$|(\bar{Y}_3^{QL})_{23}^*(\bar{Y}_3^{QL})_{32}|$ and 
$|(\bar{Y}_3^{QL})_{22}^*(\bar{Y}_3^{QL})_{33}|$, respectively. 
We here present results on $B^+ \to K^{+} \mu^\mp\tau^\pm$, since future sensitivities
at Belle II can be found for these processes in Ref.~\cite{Kou:2018nap}. 
As shown in Fig.~\ref{fig:Predictions}(b), $\mathcal{B}(B^+\to K^+\mu^-\tau^+)$
can be large enough to be observed at Belle II with 50 ab$^{-1}$, while 
$\mathcal{B}(B^+\to K^+\mu^+\tau^-)$ is out of the reach of Belle II. 
A part of the parameter space is already excluded by the current measurement of $\mathcal{B}(B^+\to K^+\mu^-\tau^+)$ at Belle, but it does not alter the other predictions in Fig.~\ref{fig:Predictions} except for that of $\mathcal{B}(B_s\to \mu^{\mp}\tau^{\pm})$. 
Figure~\ref{fig:Predictions}(c) shows a strong correlation between 
$\mathcal{B}(B_s\to \mu^{\mp}\tau^{\pm})$ and 
$\mathcal{B}(B^+\to K^+\mu^-\tau^+)$, since both of them are induced mainly by $|(\bar{Y}_3^{QL})_{23}^*(\bar{Y}_3^{QL})_{32}|$. 
The current upper limit on $\mathcal{B}(B^+\to K^+\mu^-\tau^+)$ directly leads to the limit on $\mathcal{B}(B_s\to \mu^{\mp}\tau^{\pm})$. 
These correlations among the $b\to s\mu^+\tau^-$ and $b\to s\mu^-\tau^+$
observables can be explored by the combination of the Belle II and the LHCb measurements.

Besides, we consider the LFV decays of heavy quarkonia, 
$\Upsilon(nS)\to \mu^\mp\tau^\pm$ ($n=1,2,3$). 
The branching ratios for these processes are given 
by~\cite{Abada:2015zea,Hazard:2016fnc,Calibbi:2022ddo} 
\begin{align}
\mathcal{B}(\Upsilon(nS)\to\mu^\pm\tau^\mp)
&=
\mathcal{B}(\Upsilon(nS)\to e^+e^-)_{\mathrm{SM}}\,
\frac{1}{2}
\left|
\frac{
3\ts m_{\Upsilon(nS)}^2
\big[L_{\tts ed}^{V,LL}(m_{\Upsilon(nS)})\big]_{2333}
}{
8\ts\pi \alpha
}
\right|^2, 
\end{align}
where $m_{\Upsilon(nS)}$ is the mass of $\Upsilon(nS)$, and 
the charged-lepton masses are neglected. 
From the bottom-right plot in Fig.~\ref{fig:Y3QL-1}, 
we estimate the magnitude of the LEFT coupling as 
$|[L_{\tts ed}^{V,LL}(m_{\Upsilon(nS)})]_{2333}|
\sim 
|(\bar{Y}_3^{QL*})_{32}(\bar{Y}_3^{QL})_{33}|/m_{S_3}^2
\lesssim \mathcal{O}(10^{-8}\,\mathrm{GeV}^2)$. 
Therefore, the branching ratios are as large as
$\mathcal{O}(10^{-11})$, which are too small to be measured at
current and near-future experiments.

Furthermore, the $S_3$ leptoquark contributions also induce LFV decays of tau lepton. 
At the tree level, 
the $\tau^-\to \mu^-\phi$ decay with $\tau^-\to\mu^- \bar{s}s$ transition is generated through the $S_3$ exchange. 
The branching ratio for $\tau^-\to \mu^-\phi$ is given by~\cite{Goto:2010sn}
\begin{align}
\mathcal{B}(\tau^- \to \mu^-\phi)
&=
\frac{f_{\phi}^2 \ts m_{\tau}^3 \tau_{\tau}}{128\tts\pi}\ts
\bigg( 1 - \frac{m_\phi^2}{m_\tau^2} \bigg)^{\!\! 2}
\Bigg\{
  \bigg( 1 + \frac{2\ts m_\phi^2}{m_\tau^2} \bigg)
  \Big| \big[L_{\tts ed}^{V,LL}(m_\tau)\big]_{3222}
  + \big[L_{\tts ed}^{V,LR}(m_\tau)\big]_{3222} \Big|^2
  \nonumber\\
  &\hspace{37mm}
  +
  \frac{8\tts e}{m_\tau}\ts
  \mathrm{Re}
  \Big[
   \big[ L_{\tts e\gamma}(m_\tau) \big]_{23}
   \Big( \big[L_{\tts ed}^{V,LL}(m_\tau)\big]_{3222}
   + \big[L_{\tts ed}^{V,LR}(m_\tau)\big]_{3222} \Big)
  \Big]
  \nonumber\\
  &\hspace{37mm}
  +
  \frac{16\tts e^2}{9\tts m_\phi^2}\ts
  \bigg( 2 + \frac{m_\phi^2}{m_\tau^2} \bigg)
  \Big| \big[ L_{\tts e\gamma}(m_\tau) \big]_{23} \Big|^2
\Bigg\}\,,
\end{align}
where $m_\phi$ and $f_\phi$ are the mass and the decay constant of $\phi$ meson,
$e$ is the electric charge, 
and the LEFT coefficients
$[L_{\tts ed}^{V,LL}(m_\tau)]_{3222}$, 
$[L_{\tts ed}^{V,LR}(m_\tau)]_{3222}$, and 
$[L_{\tts e\gamma}(m_\tau) ]_{23}$
are given in Eqs.~\eqref{eq:LedVLL3222}, \eqref{eq:LedVLR3222}, and \eqref{eq:Legammaij}, respectively. 
In the current model, the branching ratio for $\tau^-\to\mu^-\phi$ is not significantly enhanced due to the smallness of the
$(\bar{Y}_3^{QL})_{22}$ coupling in the tree-level contribution. As shown in Fig.~\ref{fig:Predictions}(d), 
$\mathcal{B}(\tau^- \to \mu^-\phi)$ might be observed at the Belle II experiment~\cite{Kou:2018nap}. 
We also consider the loop-induced LFV processes of tau lepton, $\tau^-\to \mu^-\gamma$ and $\tau^-\to \mu^-\mu^+\mu^-$. 
The branching ratio for $\tau^- \to \mu^-\gamma$ is given by 
\begin{align}
\mathcal{B}(\tau^- \to \mu^-\gamma)
&=
\frac{m_{\tau}^3 \tau_{\tau}}{4\tts\pi}\,
\Big| \big[L_{\tts e\gamma}(m_\tau)\big]_{23}^{\mathrm{NP}} \Big|^2,
\end{align}
and that for $\tau^- \to \mu^-\mu^+\mu^-$ can be found, 
\textit{e.g.}, in Refs.~\cite{Okada:1999zk,Kuno:1999jp}: 
\begin{align}
\mathcal{B}(\tau^- \to \mu^-\mu^+\mu^-)
&=
\frac{m_\tau^5\tau_\tau}{1536\tts\pi^3}
\bigg\{
2\, \Big| 
  \big[L_{\tts ee}^{V,LL}(m_\tau)\big]_{3222} 
  + \big[L_{\tts ee}^{V,LL}(m_\tau)\big]_{2232} 
\Big|^2  
+
\Big| \big[L_{\tts ee}^{V,LR}(m_\tau)\big]_{3222} \Big|^2 
\nonumber\\
&\hspace{18mm}
+
\frac{8\ts e}{m_\tau}\,
\mathrm{Re}
\Big[
\big[ L_{\tts e\gamma}(m_\tau) \big]_{23}
\Big(
2\ts 
\big[L_{\tts ee}^{V,LL}(m_\tau)\big]_{3222} 
+
2\ts 
\big[L_{\tts ee}^{V,LL}(m_\tau)\big]_{2232} 
+
\big[L_{\tts ee}^{V,LR}(m_\tau)\big]_{3222} 
\Big)
\Big]
\nonumber\\
&\hspace{18mm}
+
\frac{32\ts e^2}{m_{\tau}^2}
\bigg( \log\frac{m_\tau^2}{m_\mu^2} - \frac{11}{4} \bigg)
\Big| \big[ L_{\tts e\gamma}(m_\tau) \big]_{23} \Big|^2
\bigg\}\,. 
\end{align}
The LEFT coefficients
$\big[ L_{\tts ee}^{V,LL}(m_\tau) \big]_{2232}=
\big[ L_{\tts ee}^{V,LL}(m_\tau) \big]_{3222}$, 
$\big[ L_{\tts ee}^{V,LR}(m_\tau) \big]_{3222}$ 
and 
$[L_{\tts e\gamma}(m_\tau)]_{23}^{\mathrm{NP}}$, evaluated at the $\tau$ mass scale,  are given in
Eqs.~\eqref{eq:LeeVLL3222}, \eqref{eq:LeeVLR3222}, and \eqref{eq:Legammaij}, respectively.  
In the expression of $\mathcal{B}(\tau^- \to \mu^-\mu^+\mu^-)$, 
contributions from the $RL$ and $RR$ operators are neglected,
since LFV occurs dominantly in the left-handed leptons in the current model. 
The predictions for $\mathcal{B}(\tau^- \to \mu^-\gamma)$ and $\mathcal{B}(\tau^- \to \mu^-\mu^+\mu^-)$ are shown in Figs.~\ref{fig:Predictions}(d) and (e). 
They exhibit a strong correlation with each other, but are 
slightly smaller than the planned sensitivities of Belle II 
with 50 fb$^{-1}$~\cite{Kou:2018nap}.

In the current model, the muon anomalous magnetic moment (known as the
muon $g-2$) is generated through the product 
$(Y_{3}^{QL})^*_{32}(Y_{3}^{QL})_{32}$ via 
the dipole coupling 
$[ L_{\tts e\gamma}(m_\tau) ]_{22}^{\mathrm{NP}}$. 
We find that this contribution is too small to explain the
long-standing tension between the measured value and the SM 
prediction of the muon $g-2$~\cite{Aoyama:2020ynm,Muong-2:2023cdq}.

The $S_3$ leptoquark also affects $W$-boson and $Z$-boson couplings with the SM
fermions. We evaluate them with the one-loop expressions in
Ref.~\cite{Arnan:2019olv}, which include radiative corrections beyond the
leading-logarithmic approximation. 
The effects on the $W$-boson couplings are not significant to be
measured at the current and planned future experiments. 
We here present only the result for $\mathcal{B}(Z\to \mu^\mp\tau^\pm)$, which
is calculated with the formulas given in Appendix~\ref{sec:ZtoMuTau}. 
Figure~\ref{fig:Predictions}(f) shows a strong correlation 
between $\mathcal{B}(Z\to \mu^\mp\tau^\pm)$ and $\mathcal{B}(\tau^-\to \mu^-\gamma)$. 
In our scenario, the $\mathcal{B}(Z\to \mu^\mp\tau^\pm)$ can be 
as large as $\mathcal{O}(10^{-9})$.
The present experimental bounds are given by the LEP experiment as 
$\mathcal{B}(Z\to\mu^\mp\tau^\pm)<1.2\times 10^{-5}$~\cite{DELPHI:1996iox}
and the LHC experiment as
$\mathcal{B}(Z\to\mu^\mp\tau^\pm)<6.5\times 10^{-6}$~\cite{ATLAS:2021bdj}. 
On the other hand, the FCC-ee experiment has a sensitivity to
$\mathcal{O}(10^{-9})$~\cite{Dam:2018rfz}. 
In the case that $\mathcal{B}(Z\to \mu^\mp\tau^\pm)$ is enhanced enough, 
$\mathcal{B}(\tau^-\to \mu^-\gamma)$ is also significantly enhanced.

\section{Summary}
\label{sec:summary}

We have constructed a realistic GUT model which addresses two serious
issues in the minimal SU(5) GUT: the realization of the gauge coupling
unification and that of the flavor structures in the down-type-quark
and the charged-lepton sectors. By introducing a 45-dimensional scalar
representation $\Phi_{45}$ to the minimal SU(5) GUT, the Yukawa
matrices of the down-type quarks and the charged leptons are
reproduced correctly by the Georgi-Jarlskog mechanism.  
In addition, we have shown that the three gauge couplings can be
unified through the RG running under the constraint from proton decay, 
if $S_3$, $S_6$, and $S_8$ in $\Phi_{45}$ and $\Sigma_8$ in the
24-dimensional scalar representation $\Sigma$ lie much below the GUT scale. 
In particular, the mass of $S_3$, which is a scalar leptoquark, can be 
of the order of TeV.

The Yukawa couplings of the $S_3$ leptoquark at the low-energy scale
is constrained by the matching condition at the GUT scale in  
Eq.~(\ref{eq:YukawaMatching2}). 
In our scenario, the $S_3$ leptoquark couples strongly to the SM
fermions in the second and third generations, where the magnitudes of
the couplings obey the hierarchy shown in Eq.~(\ref{eq:hierarchy}) and 
Fig.~\ref{fig:Y3QL-2}. 
In particular, the coupling $(\bar{Y}_3^{QL})_{22}$ is suppressed
compared with $(\bar{Y}_3^{QL})_{23}$, $(\bar{Y}_3^{QL})_{32}$, and 
$(\bar{Y}_3^{QL})_{33}$. 
The smallness of $(\bar{Y}_3^{QL})_{22}$ leads to 
the characteristic patterns of correlations among flavor observables.

We have investigated flavor phenomenology in this realistic GUT
scenario with the $S_3$ leptoquark at the TeV scale. 
We have derived constraints on the $S_3$ Yukawa couplings from 
$\Delta M_s$, $\mathcal{B}(B\to K^{(*)}\nu\bar{\nu})$, 
$R_{K^{(*)}}$, and $\mathcal{B}(B_s\to\mu^+\mu^-)$, where 
the results are shown in Fig.~\ref{fig:Y3QL-1}. 
We have then calculated various decays of $B$ mesons, $\Upsilon(nS)$, 
tau lepton, and $Z$ boson. 
In the current model, the $R(D^{(*)})$ anomaly cannot be 
explained by the $S_3$ contribution due to the strong constraints from
$\Delta M_s$ and $\mathcal{B}(B\to K^{(*)}\nu\bar{\nu})$. 
The LFV processes 
$B_s\to \mu^{\mp}\tau^{\pm}$, $B^+\to K^+\mu^-\tau^+$, and 
$\tau^-\to \mu^-\phi$ may be observed at Belle II with 50~ab$^{-1}$ 
and LHCb with 300~fb$^{-1}$. 
It is noted that $\mathcal{B}(B^+\to K^+\mu^+\tau^-)$ cannot reach the
future sensitivity at Belle II 
unlike $\mathcal{B}(B^+\to K^+\mu^-\tau^+)$. 
Therefore, the observation of $B^+\to K^+\mu^-\tau^+$ together with 
the nonobservation of $B^+\to K^+\mu^+\tau^-$ is a clear signal of the 
current model. 
On the other hand, it is rather hard to observe the other processes 
$\tau^-\to \mu^-\gamma$, $\tau^-\to \mu^-\mu^+\mu^-$, and
$Z\to \mu^\mp\tau^\pm$, and much more data are needed for their
observations.

In general, it is challenging to probe a GUT model, since the
unification occurs at a very high-energy scale.  
The proton decay is a direct probe for GUT, but it has not
been observed yet. 
We have provided a well-motivated benchmark
scenario which may be able to be probed by the precise measurements of
the flavor observables at the Belle II and LHCb experiments.  
Besides, the $S_3$ leptoquark at the TeV scale can be directly searched
for at the current and future hadron-collider experiments. 
We thus conclude that the precise flavor measurements as well as the
direct searches for the $S_3$ leptoquark play complementary roles to
the searches for proton decay in probing our GUT scenario.

\section*{Acknowledgments}
This work was supported in part by the Japan Society for the Promotion of Science (JSPS) KAKENHI Grants 
No.~17K05429 (S.M.) and No. 20H00160 (T.S.).
Computations were partially carried out on the computer system (sushiki)
at YITP in Kyoto University.

\appendix

\section{Scalar potential and masses}
\label{sec:scalarMasses}

The scalar potential $V(\Sigma,\Phi_5,\Phi_{45})$ in the
SU(5)-symmetric renormalizable Lagrangian in 
Eq.~\eqref{eq:SU5Lag} is given by 
\begin{align}
V(\Sigma,\Phi_{5},\Phi_{45})
&=
V_{24}
+
V_{5}
+
V_{45}
+
V_{24\cdot 5}
+
V_{24\cdot 45}
+
V_{5\cdot 45}
+
V_{24\cdot 5\cdot 45}
\,,
\label{eq:SU5ScalarPotential}
\end{align}
where each term is defined as 
\begin{align}
V_{24} &=
  m_{24}^2\,\tr\,\Sigma^2
  + \chi_{24}\ts\tr\,\Sigma^3
  + \lambda_{24}^{(1)} \big(\tr\,\Sigma^2\big)^2
  + \lambda_{24}^{(2)}\, \tr\,\Sigma^4,
  \\
V_{5} &=
  m_{5}^2\ts \Phi_5^\dagger \Phi_5
  + \lambda_5 \big(\Phi_5^\dagger \Phi_5\big)^2,
  \\
V_{45} &=
  m_{45}^2 (\Phi_{45}^\dagger)^{A}_{BC}(\Phi_{45})^{BC}_{A}
  + \lambda_{45}^{(1)}
     \big[(\Phi_{45}^\dagger)^{A}_{BC}(\Phi_{45})^{BC}_{A}\big]^2
  + \lambda_{45}^{(2)}
    (\Phi_{45}^\dagger)^{A}_{BC} (\Phi_{45})^{BC}_{D}
    (\Phi_{45}^\dagger)^{D}_{EF} (\Phi_{45})^{EF}_{A}
  \nonumber\\
  &\hspace{4mm}
  + \lambda_{45}^{(3)}
    (\Phi_{45}^\dagger)^{A}_{BC} (\Phi_{45})^{BF}_{A}
    (\Phi_{45}^\dagger)^{D}_{EF} (\Phi_{45})^{EC}_{D}
  + \lambda_{45}^{(4)}
    (\Phi_{45}^\dagger)^{A}_{BC} (\Phi_{45})^{BC}_{F}
    (\Phi_{45}^\dagger)^{D}_{EA} (\Phi_{45})^{EF}_{D}
  \nonumber\\
  &\hspace{4mm}
  + \lambda_{45}^{(5)}
    (\Phi_{45}^\dagger)^{A}_{BC} (\Phi_{45}^\dagger)^{B}_{AD} 
    (\Phi_{45})^{EC}_{F} (\Phi_{45})^{FD}_{E}
  + \lambda_{45}^{(6)}
    (\Phi_{45}^\dagger)^{A}_{BC} (\Phi_{45}^\dagger)^{B}_{DE} 
    (\Phi_{45})^{CD}_{F} (\Phi_{45})^{EF}_{A},
  \\
V_{24\cdot 5} &=
  \chi_{5}\ts \Phi_5^\dagger\ts\Sigma\, \Phi_5
  + a^{(1)} \big( \tr\,\Sigma^2\big)\ts \Phi_5^\dagger \Phi_5
  + a^{(2)} \Phi_5^\dagger\ts\Sigma^2\ts\Phi_5\,,
  \\
V_{24\cdot 45} &=
  \chi_{45}^{(1)}
    (\Phi_{45}^\dagger)^{A}_{BC}\ts \Sigma^D_{\ \,A}\tts 
    (\Phi_{45})^{BC}_{D} 
  + \chi_{45}^{(2)}
    (\Phi_{45}^\dagger)^{A}_{BC}\ts \Sigma^C_{\ \,D}\tts 
    (\Phi_{45})^{BD}_{A}
  + b^{(1)} \big( \tr\,\Sigma^2 \big)
    (\Phi_{45}^\dagger)^{A}_{BC} (\Phi_{45})^{BC}_{A} 
  \nonumber\\
  &\hspace{4mm}
  + b^{(2)}
    (\Phi_{45}^\dagger)^{A}_{BC}\ts(\Sigma^2)^{D}_{\ \,A}\tts 
    (\Phi_{45})^{BC}_{D} 
  + b^{(3)}
    (\Phi_{45}^\dagger)^{A}_{BC}\ts(\Sigma^2)^{B}_{\ \,D}\tts 
    (\Phi_{45})^{DC}_{A} 
  + b^{(4)}
    (\Phi_{45}^\dagger)^{A}_{BC}\ts \Sigma^{E}_{\ \,A} 
    \Sigma^{B}_{\ \,D}\tts (\Phi_{45})^{DC}_{E}
  \nonumber\\
  &\hspace{4mm}
  + b^{(5)}
    (\Phi_{45}^\dagger)^{A}_{BC}\ts \Sigma^{C}_{\ \,E} 
    \Sigma^{B}_{\ \,D}\tts (\Phi_{45})^{DE}_{A}
  + b^{(6)}
    (\Phi_{45}^\dagger)^{A}_{BC}\ts \Sigma^{B}_{\ \,A} 
    \Sigma^{E}_{\ \,D}\tts (\Phi_{45})^{DC}_{E}, 
  \\
V_{5\cdot 45} &=
  c^{(1)}
    (\Phi_{45}^\dagger)^{A}_{BC} (\Phi_{45})^{BC}_{A} 
    ( \Phi_{5}^\dagger \Phi_{5} )
  + c^{(2)}
    (\Phi_{5}^\dagger)_{A} (\Phi_{45}^\dagger)^{A}_{BC} 
    (\Phi_{45})^{BC}_{D} (\Phi_{5})^{D}
  + c^{(3)}
    (\Phi_{5}^\dagger)_{C} (\Phi_{45})^{BC}_{A} 
    (\Phi_{45}^\dagger)^{A}_{BD} (\Phi_{5})^{D}
  \nonumber\\
  &\hspace{4mm}
  + \Big[
    c^{(4)} (\Phi_{45})^{BC}_{A} (\Phi_{45})^{AD}_{B}
    (\Phi_{5}^\dagger)_{C} (\Phi_{5}^\dagger)_{D}
    + c^{(5)} (\Phi_{45}^\dagger)^{A}_{BC} (\Phi_{45})^{BC}_{D} 
      (\Phi_{45})^{DE}_{A} (\Phi_{5}^\dagger)_{E}
    \nonumber\\
    &\hspace{10mm}
    + c^{(6)} (\Phi_{45}^\dagger)^{A}_{BC} (\Phi_{45})^{BD}_{A} 
      (\Phi_{45})^{CE}_{D} (\Phi_{5}^\dagger)_{E}
    + \hc
  \Big]\,,
\\
V_{24\cdot 5\cdot 45} &=
  \tilde{\chi}\, (\Phi_{5}^\dagger)_{C}\ts 
  \Sigma^{A}_{\ \,B}\tts (\Phi_{45})^{BC}_{A}
  + d^{\ts(1)} (\Phi_{5}^\dagger)_{C}\ts
    (\Sigma^2)^{A}_{\ \,B}\tts (\Phi_{45})^{BC}_{A}
  + d^{\ts(2)} (\Phi_{5}^\dagger)_{D}\ts \Sigma^{D}_{\ \,C}
    \Sigma^{A}_{\ \,B}\tts (\Phi_{45})^{BC}_{A}
  + \hc
\label{eq:SU5ScalarPotentialEach}
\end{align}
The $\mathbf{24}$-representation scalar $\Sigma$ gets the VEV as 
\begin{align}
  \langle \Sigma\rangle = \begin{pmatrix}
    2\ts v_{24}&0&0&0&0\\
    0&2\ts v_{24}&0&0&0\\
    0&0&2\ts v_{24}&0&0\\
    0&0&0&-3\ts v_{24}&0\\
    0&0&0&0&-3\ts v_{24}
  \end{pmatrix}\,,
\end{align}
when the condition $2\ts m_{24}^2 + 4\tts (30\tts\lambda_{24}^{(1)} +
7\tts\lambda_{24}^{(2)})v_{24}^2-3\ts\chi_{24}\ts v_{24}=0$ is 
satisfied~\cite{Buras:1977yy}.\footnote{The minimization of the scalar
  potential for the $\mathbf{45}$ representation is studied in
  Refs.~\cite{Frampton:1979wf,Kalyniak:1982pt,Eckert:1983bn}.} 
From the potential, the squared masses of the component fields in the
scalars $\Sigma$, $\Phi_{5}$, and $\Phi_{45}$ can be read at the tree
level as 
\begin{align}
&m^2_{\Sigma_1}
  =
  -2\ts m_{24}^2 + \frac{3}{2}\ts\chi_{24}\ts v_{24}\,,
  \qquad
m^2_{\Sigma_3}
  =
  40\ts\lambda_{24}^{(2)} v_{24}^2 - \frac{15}{2}\ts\chi_{24}\ts v_{24}\,,
  \qquad
m^2_{\Sigma_8}
  =
  10\ts\lambda_{24}^{(2)} v_{24}^2 + \frac{15}{2}\ts\chi_{24}\ts v_{24}\,,
  \nonumber\\
&R_H^\dagger\begin{pmatrix} 
  m^2_{H} & 0 \\ 0 & m^2_{H'} \end{pmatrix}\! R_H
  =
  \begin{pmatrix}
  \tilde{m}_{5}^2 + 3\ts\tilde{a}^{(2)} v_{24}^2
  &
  -\dfrac{5\sqrt{3}}{2\sqrt{2}}
  \big( \tilde{d}^{(1)} + 3\ts d^{(2)} \big) v_{24}^2
  \\[4mm]
  -\dfrac{5\sqrt{3}}{2\sqrt{2}}
  \big( \tilde{d}^{(1)*} + 3\ts d^{(2)*} \big) v_{24}^2
  &
  \tilde{m}_{45}^2
  + \bigg( \dfrac{7}{4}\ts\tilde{b}^{(2)} + \dfrac{19}{8}\ts\tilde{b}^{(3)}
    + 8\ts b^{(4)} + \dfrac{17}{4}\ts b^{(5)}
    + \dfrac{75}{8}\ts b^{(6)} \!\bigg)\ts v_{24}^2
  \end{pmatrix},
  \nonumber\\
&R_S^\dagger\begin{pmatrix} 
  m^2_{H_C} & 0 \\ 0 & m^2_{S_1} \end{pmatrix}\! R_S
  =
  \begin{pmatrix}
  \tilde{m}_{5}^2 - 2\ts\tilde{a}^{(2)} v_{24}^2
  &
  -\dfrac{5}{\sqrt{2}}
  \big( \tilde{d}^{(1)} - 2\ts d^{(2)} \big) v_{24}^2
  \\[4mm]
  -\dfrac{5}{\sqrt{2}}
  \big( \tilde{d}^{(1)*} - 2\ts d^{(2)*} \big) v_{24}^2
  &
  \tilde{m}_{45}^2
  + \bigg( \dfrac{1}{2}\ts\tilde{b}^{(2)} - \dfrac{3}{4}\ts\tilde{b}^{(3)}
    + \dfrac{17}{4}\ts b^{(4)} - 2\ts b^{(5)}
    + \dfrac{25}{2}\ts b^{(6)} \!\bigg)\ts v_{24}^2
  \end{pmatrix},
  \nonumber\\
&m^2_{\tilde{S}_1}
  =
  \tilde{m}_{45}^2
  + \bigg(\! -2\ts\tilde{b}^{(2)} + 3\ts\tilde{b}^{(3)}
    - \frac{9}{2}\ts b^{(4)} + 8\ts b^{(5)} \!\bigg)\ts v_{24}^2\,,
  \qquad
m^2_{R_2}
  =
  \tilde{m}_{45}^2
  + \bigg( 3\ts\tilde{b}^{(2)} - 2\ts\tilde{b}^{(3)}
    - \frac{9}{2}\ts b^{(4)} + 3\ts b^{(5)} \!\bigg)\ts v_{24}^2\,,
  \nonumber\\
&m^2_{S_3}
  =
  \tilde{m}_{45}^2
  + \bigg( 3\ts\tilde{b}^{(2)} + \frac{1}{2}\ts\tilde{b}^{(3)}
    + 3\ts b^{(4)} - 7\ts b^{(5)} \!\bigg)\ts v_{24}^2\,,
  \qquad\quad
m^2_{S_6}
  =
  \tilde{m}_{45}^2
  + \bigg(\! - 2\ts\tilde{b}^{(2)} - 2\ts\tilde{b}^{(3)}
    + \frac{11}{2}\ts b^{(4)} + 3\ts b^{(5)} \!\bigg)\ts v_{24}^2\,,
  \nonumber\\
&m^2_{S_8}
  =  
  \tilde{m}_{45}^2
  + \bigg(\! - 2\ts\tilde{b}^{(2)} + \frac{1}{2}\ts\tilde{b}^{(3)}
    + \frac{1}{2}\ts b^{(4)} - 7\ts b^{(5)} \!\bigg)\ts v_{24}^2\,,
\label{eq:scalarMasses}
\end{align}
where the following combinations of the parameters are introduced: 
\begin{align}
\tilde{m}_{5}^2
  &= m_{5}^2 + \big( 30\ts a^{(1)} + 6\ts a^{(2)} \big) v_{24}^2\,,
  \qquad
\tilde{m}_{45}^2
  = m_{45}^2 + \bigg( 30\ts b^{(1)} + 6\ts b^{(2)} + 6\ts b^{(3)}
  -\frac{3}{2}\ts b^{(4)} + b^{(5)} \bigg) v_{24}^2\,,
  \nonumber\\
\tilde{a}^{(2)}
  &= a^{(2)} - \frac{\chi_5}{v_{24}}\,,
  \qquad
\tilde{b}^{(2)}
  = b^{(2)} - \frac{\chi_{45}^{(1)}}{v_{24}}\,,
  \qquad
\tilde{b}^{(3)} 
  = b^{(3)} - \frac{\chi_{45}^{(2)}}{v_{24}}\,,
  \qquad
\tilde{d}^{(1)} 
  = d^{(1)} - \frac{\tilde{\chi}}{v_{24}}\,,
\end{align}
and the rotation matrices $R_H$ and $R_S$ are given as in Eq.~\eqref{eq:mixing}: 
\begin{align}
R_H
 &=
  \begin{pmatrix}
  c_H & e^{-i\delta_H} s_H\\
  -e^{i\delta_H}s_H & c_H 
  \end{pmatrix},
  \qquad
R_S
 =
  \begin{pmatrix}
  c_S & e^{-i\delta_S} s_S\\
  -e^{i\delta_S}s_S & c_S
  \end{pmatrix}.
\end{align}
The masses of $\Sigma_1$, $\Sigma_3$, and $\Sigma_8$ can be
freely chosen, since $V_{24}$ in Eq.~\eqref{eq:SU5ScalarPotentialEach}
contains a sufficient number of parameters. 
On the other hand, the masses of the other scalars are constrained by
the following relation:  
\begin{align}
- 8\big( s_H^2\ts m^2_{H} + c_H^2\ts m^2_{H'} \big)
+ 6\big( s_S^2\ts m^2_{H_C} + c_S^2\ts m^2_{S_1} \big)
+ 6\ts m^2_{\tilde{S}_1}
- 6\ts m^2_{R_2}
+ 9\ts m^2_{S_3}
+ 3\ts m^2_{S_6}
- 10\ts m^2_{S_8}
=
0\,.
\label{eq:massRelation}
\end{align}

\section{Matching conditions at the GUT scale}
\label{sec:YukawaMatching}

Below the GUT scale, the Yukawa interactions
are given in terms of the component
fields in Eqs.~\eqref{eq:10and5bar}, \eqref{eq:Phi5}, and \eqref{eq:Phi45} 
as follows:
\begin{align}
- \mathcal{L}_Y 
  =&\
  (Y_U)_{ij}\epsilon_{\alpha\beta}\ts
    \bar{u}_{R\ahat i}\ts H^{\alpha}q^{\ahat\beta}_{Lj}
  + (Y_D)_{ij}\ts
    \bar{d}_{R\ahat i}\ts H_{\alpha}^*\ts q^{\ahat\alpha}_{Lj}
  + (Y_E)_{ij}\ts
    \bar{e}_{Ri}\ts H_{\alpha}^*\, \ell^{\ts\alpha}_{Lj}
  \nonumber\\
  &\! 
  + (Y_U^{\prime})_{ij}\epsilon_{\alpha\beta}\ts
    \bar{u}_{R\ahat i}\ts H^{\prime\alpha} q^{\ahat\beta}_{Lj}
  + (Y_D^{\prime})_{ij}\ts
    \bar{d}_{R\ahat i}\ts H_{\alpha}^{\prime *}\ts q^{\ahat\alpha}_{Lj}
  + (Y_E^{\prime})_{ij}\ts
    \bar{e}_{Ri}\ts H_{\alpha}^{\prime *}\, \ell^{\ts\alpha}_{Lj}
  \nonumber\\
  &\! 
  + (Y_C^{QL})_{ij}\epsilon_{\alpha\beta}\ts
    \bar{q}^{\ts c\ts\ahat\alpha}_{Li} H_{C\ahat}\ts \ell_{Lj}^{\ts\beta}
  + (Y_C^{UE})_{ij}\ts
    \bar{u}^{}_{R\ahat i}\ts H_C^{*\ahat} e_{Rj}^c
  + (Y_C^{DU})_{ij}\epsilon^{\ahat\hat{b}\hat{c}}\ts
    \bar{d}_{R\ahat i}\ts H_{C\hat{b}}\ts u_{R\hat{c} j}^{c}
  + \frac{(Y_C^{QQ})_{ij}}{2}\ts\epsilon_{\ahat\hat{b}\hat{c}}\ts\epsilon_{\alpha\beta}\ts
    \bar{q}^{\ts c\ts\ahat\alpha}_{Li}H_C^{*\hat{b}} q_{Lj}^{\hat{c}\beta}
  \nonumber\\
  &\!
  + (Y_1^{QL})_{ij}\epsilon_{\alpha\beta}\ts
    \bar{q}^{\ts c\ts\ahat\alpha}_{Li} S_{1\ahat}\ts \ell_{Lj}^{\ts\beta}
  + (Y_1^{UE})_{ij}\ts
    \bar{u}^{}_{R\ahat i}\ts S_1^{*\ahat} e_{Rj}^c
  + (Y_1^{DU})_{ij}\epsilon^{\ahat\hat{b}\hat{c}}\ts
    \bar{d}_{R\ahat i}\ts S_{1\hat{b}}\ts u_{R\hat{c} j}^{c}
  + \frac{(Y_1^{QQ})_{ij}}{2}\ts\epsilon_{\ahat\hat{b}\hat{c}}\ts\epsilon_{\alpha\beta}\ts
    \bar{q}^{\ts c\ts\ahat\alpha}_{Li}S_1^{*\hat{b}} q_{Lj}^{\hat{c}\beta}
  \nonumber\\
  &\! 
  + (\tilde{Y}_1^{ED})_{ij}\ts
    \bar{e}_{Ri}^{}\ts \tilde{S}_{1}^{*\ahat}d_{R\ahat j}^{\ts c}
  + \frac{(\tilde{Y}_1^{UU})_{ij}}{2}\ts\epsilon^{\ahat\hat{b}\hat{c}}\ts
    \bar{u}_{R\ahat i}\ts\tilde{S}_{1\hat{b}}\ts u_{R\hat{c} j}^c
  + (Y_2^{UL})_{ij}\epsilon_{\alpha\beta}\ts
    \bar{u}_{R\ahat i}\tts R_{2}^{\tts\ahat\alpha}\ell_{Lj}^{\ts\beta}
  + (Y_2^{EQ})_{ij}\ts
    \bar{e}_{Ri}\tts R_{2\ahat\alpha}^{\tts *}\ts q_{Lj}^{\ahat\alpha}
  \nonumber\\
  &\! 
  + (Y_3^{QL})_{ij}\epsilon_{\alpha\beta}\ts
    \bar{q}_{Li}^{\ts c\ts\ahat\gamma}(\sigma_a)^{\alpha}{}_{\gamma}\ts
    S_{3\ahat}^a\ts\ell_{Lj}^{\ts\beta}
  + \frac{(Y_3^{QQ})_{ij}}{2}\ts
    \epsilon_{\ahat\hat{b}\hat{c}}\ts\epsilon_{\alpha\beta}\ts
    \bar{q}_{Li}^{c\ts\ahat\alpha}(\sigma^a)^{\beta}{}_{\gamma}\ts
    S_3^{* a\bhat}q_{Lj}^{\hat{c}\ts\gamma}
  \nonumber\\
  &\! 
  + (Y_6^{DU})_{ij}\ts
    \bar{d}_{R\ahat i}\ts (\eta^A)^{\ahat\hat{b}}
    S_{6}^{A}\ts u_{R\hat{b} j}^c
  +\frac{(Y_6^{QQ})_{ij}}{2}\ts\epsilon_{\alpha\beta}\ts
    \bar{q}_{Li}^{\ts c\ts\ahat\alpha} (\eta^A)_{\ahat\hat{b}}\ts 
    S_{6}^{A*}\tts q_{Lj}^{\hat{b}\beta}
  \nonumber\\
  &\! 
  + (Y_8^{UQ})_{ij}\epsilon_{\alpha\beta}\ts
    \bar{u}_{R\ahat i}^{}\ts (\lambda^A)^{\ahat}{}_{\hat{b}}\ts
    S_{8}^{A \alpha }\ts q_{Lj}^{\hat{b}\beta}
  + (Y_8^{DQ})_{ij}\ts
    \bar{d}_{R\ahat i}\ts (\lambda^A)^{\ahat}{}_{\hat{b}}\ts
    S_{8\alpha}^{A*}\ts q_{Lj}^{\hat{b}\alpha}
  + \hc\,,    
\label{eq:LYbelowGUT}
\end{align}
where $Y_C^{QQ}$ and $Y_1^{QQ}$ are symmetric matrices in the flavor
space, while $\tilde{Y}_1^{UU}$, $Y_3^{QQ}$, and $Y_6^{QQ}$ are
antisymmetric matrices: 
\begin{align}
(Y_C^{QQ})^T &= Y_C^{QQ},
&
(Y_1^{QQ})^T &= Y_1^{QQ},
&
(\tilde{Y}_1^{UU})^T &= -\tilde{Y}_1^{UU},
&
(Y_3^{QQ})^T &= -Y_3^{QQ},
&
(Y_6^{QQ})^T &= -Y_6^{QQ}. 
\end{align}
The Yukawa couplings in Eq.~\eqref{eq:LYbelowGUT} are matched onto 
those in Eq.~\eqref{eq:SU5Yukawa} at the tree level as 
\begin{align}
  Y_U =&\ 
    -\frac{1}{2}\ts V_{QU}^{\ts T}
    \bigg( c_H\ts Y_{5}^U 
      + \sqrt{\frac{2}{3}}\,e^{i\delta_H}s_H\ts Y_{45}^{U}\bigg)^{\! T},
  &
  Y_U^{\prime} =&\  
    \frac{1}{2}\ts V_{QU}^{\ts T}
    \bigg(e^{-i\delta_H}s_H\ts Y_{5}^U
    - \sqrt{\frac{2}{3}}\,c_H\ts Y_{45}^{U}\bigg)^{\! T},
  \nonumber\\ 
  Y_D =&\ 
    - \frac{1}{\sqrt{2}}
    \bigg( c_H\ts Y_{5}^{D}
    - \frac{1}{2\sqrt{6}}\,e^{-i\delta_H}s_H\ts Y_{45}^{D} \bigg)^{\! T},
  &
  Y_D^{\prime} =&\ 
    \frac{1}{\sqrt{2}}
    \bigg( e^{i\delta_H}s_H\ts Y_{5}^{D}
    + \frac{1}{2\sqrt{6}}\,c_H\ts Y_{45}^{D} \bigg)^{\! T},
  \nonumber\\ 
  Y_E =&\ 
    - \frac{1}{\sqrt{2}}\ts V_{QE}^T
    \bigg( c_H\ts Y_{5}^D
    + \frac{\sqrt{3}}{2\sqrt{2}}\,e^{-i\delta_H}s_H\ts Y_{45}^{D}\bigg)
    V_{DL}\,,
  &
  Y_E^{\prime} =&\ 
    \frac{1}{\sqrt{2}}\ts V_{QE}^T
    \bigg(e^{i\delta_H}s_H\ts Y_{5}^{D}
    - \frac{\sqrt{3}}{2\sqrt{2}}\, c_H\ts Y_{45}^{D}\bigg)
    V_{DL}\,,
  \nonumber\\
  Y_C^{QL} =&\ 
    \frac{1}{\sqrt{2}}
    \bigg(
    c_S\ts Y_{5}^{D}
    + \frac{1}{2\sqrt{2}}\,e^{i\delta_S}s_S\ts Y_{45}^{D}\bigg)
    V_{DL}\,,
  &
  Y_1^{QL} =&\ 
    \frac{1}{\sqrt{2}}
    \bigg(\!
    - e^{-i\delta_S}s_S\ts Y_{5}^{D}
    + \frac{1}{2\sqrt{2}}\,c_S\ts Y_{45}^{D} \bigg) 
    V_{DL}\,,
  \nonumber\\
  Y_C^{UE} =&\ 
    \frac{1}{2}\ts V_{QU}^{\ts T}
    \Big( c_S\ts Y_{5}^{U}
    - \sqrt{2}\,e^{-i\delta_S}s_S\ts Y_{45}^{U} \Big)
    V_{QE}\,,
  &
  Y_1^{UE} =&\ 
    - \frac{1}{2}\ts V_{QU}^{\ts T}
    \Big( e^{i\delta_S}s_S\ts Y_{5}^{U}
    + \sqrt{2}\, c_S\ts Y_{45}^{U} \Big)
    V_{QE}\,,
  \nonumber\\
  Y_C^{DU} =&\ 
    \frac{1}{\sqrt{2}} \bigg(\!
    - c_S\ts Y_{5}^{D}
    + \frac{1}{2\sqrt{2}}\, e^{i\delta_S}s_S\ts Y_{45}^{D}
    \bigg)^{\! T}
    V_{QU}\,,
  &
  Y_1^{DU} =&\ 
    \frac{1}{\sqrt{2}} \bigg(
    e^{-i\delta_S}s_S\ts Y_{5}^{D}
    + \frac{1}{2\sqrt{2}}\, c_S\ts Y_{45}^{D}
    \bigg)^{\! T}
    V_{QU}\,,
  \nonumber\\
  Y_C^{QQ} =&\ 
    \frac{1}{2}\ts c_S\ts Y_{5}^{U},
  \hspace{13mm}
  Y_1^{QQ} =
    - \frac{1}{2}\ts e^{i\delta_S}s_S\ts Y_{5}^{U},
  &
  \tilde{Y}_1^{UU} =&\ 
    \frac{1}{\sqrt{2}}\ts V_{QU}^{\ts T}\ts Y_{45}^{U}\ts
    V_{QU}\,,
  \qquad
  \tilde{Y}_1^{ED} =
    \frac{1}{2}\ts V_{QE}^{\ts T}\ts Y_{45}^{D}\,,
  \nonumber\\
  Y_2^{EQ} =&\ 
    \frac{1}{\sqrt{2}}\ts V_{QE}^{\ts T}\ts Y_{45}^{U}\,,
  \qquad
  Y_2^{UL} =
    \frac{1}{2}\ts V_{QU}^{\ts T}\ts Y_{45}^{D}\ts 
    V_{DL}\,,
  &
  Y_3^{QQ} =&\ 
    \frac{1}{2}\ts Y_{45}^{U}\,,
  \hspace{21mm}
  Y_3^{QL} =
    - \frac{1}{2\sqrt{2}}\ts Y_{45}^{D}\ts V_{DL}\,,
  \nonumber\\
  Y_6^{QQ} =&\ 
    - \frac{1}{\sqrt{2}}\ts Y_{45}^{U}\,,
  \hspace{8.7mm}
  Y_6^{DU} =
    \frac{1}{2}\ts (Y_{45}^{D})^{T}\tts  V_{QU}\,,
  &
  Y_8^{UQ} =&\ 
    - \frac{1}{2}\ts 
    V_{QU}^{\ts T}\ts
    Y_{45}^{U}\,,
  \hspace{8mm}
  Y_8^{DQ} = 
    \frac{1}{2\sqrt{2}}\ts (Y_{45}^{D})^{T}.
\label{eq:YukawaMatching1}
\end{align}

\section{Renormalization group equations}
\label{sec:RGE}

The scale dependence of the gauge couplings is governed by
the RGEs, 
\begin{align}
  \frac{d\tts g_i}{d\log\mu}=\frac{\beta_{g_i}}{(4\pi)^2}\,,
\end{align}
where $g_i=g_s$, $g$ and $g'$, and $\beta_{g_i}$ denotes the
corresponding beta function. The one-loop contributions to the beta
functions are given by 
\begin{align}
\beta_{g_i}
=\bigg[B_{g_i}^{\mathrm{SM}}
+ \sum_{\phi}B_{g_i}^{\phi}\, \theta(m_{\phi}-\mu)
\bigg]\,g_i^3\,,
\end{align}
where $\phi=H'$, $H_C$, $S_1$, $\tilde{S}_1$, $R_2$, $S_3$, $S_6$, $S_8$, 
$\Sigma_1$, $\Sigma_3$, 
and $\Sigma_8$,
and the coefficients $B_{g_i}^{\mathrm{SM}}$ and $B_{g_i}^{\phi}$ are listed in Table~\ref{Table:beta_gi}.
\begin{table}
  \caption{$B_{g_i}^{\mathrm{SM}}$ and $B_{g_i}^{\phi}$ for the RGEs of the gauge couplings. 
  \label{Table:beta_gi}}
  \setlength{\tabcolsep}{5pt}
  \renewcommand{\arraystretch}{1.5}
  \begin{tabular}{c||c|c|c|c|c|c|c|c|c|c|c|c}
  \hline\hline
  $g_i$ 
  &$B_{g_i}^{\mathrm{SM}}$
  &$B_{g_i}^{H'}$
  &$B_{g_i}^{H_C}$
  &$B_{g_i}^{S_1}$
  &$B_{g_i}^{\tilde{S}_1}$
  &$B_{g_i}^{R_2}$
  &$B_{g_i}^{S_3}$
  &$B_{g_i}^{S_6}$
  &$B_{g_i}^{S_8}$
  &$B_{g_i}^{\Sigma_1}$
  &$B_{g_i}^{\Sigma_3}$
  &$B_{g_i}^{\Sigma_8}$
  \\ 
  \hline\hline
  $g_s$
  &$-7$ 
  &$0$ 
  &$1/6$ 
  &$1/6$ 
  &$1/6$ 
  &$1/3$ 
  &$1/2$ 
  &$5/6$ 
  &$2$ 
  &$0$ 
  &$0$ 
  &$1/2$ 
  \\ 
  \hline
  $g$
  &$-19/6$ 
  &$1/6$ 
  &$0$ 
  &$0$ 
  &$0$ 
  &$1/2$ 
  &$2$ 
  &$0$ 
  &$4/3$ 
  &$0$ 
  &$1/3$ 
  &$0$ 
  \\ 
  \hline
  $g'$
  &$41/6$ 
  &$1/6$ 
  &$1/9$ 
  &$1/9$ 
  &$16/9$ 
  &$49/18$ 
  &$1/3$ 
  &$2/9$ 
  &$4/3$ 
  &$0$ 
  &$0$ 
  &$0$ 
  \\ 
  \hline\hline
\end{tabular}
\end{table}

The RGEs of the Yukawa couplings 
$Y^{\ts\phi}_{\bar\psi\psi'}$ associated
with the interaction of 
the form $[Y^{\ts \phi}_{\bar\psi\psi'}]_{jk}\ts \bar{\psi}_j\ts \phi\ts \psi'_k$ 
are given by
\begin{align}
\frac{d}{d\log\mu}\ts
Y^{\ts\phi}_{\bar\psi\psi'}
=
\frac{1}{(4\pi)^2}\ts
\beta_{Y^{\ts\phi}_{\bar\psi\psi'}}
\,,
\end{align}
where the one-loop beta functions 
can generally be written as~\cite{Machacek:1983tz,Machacek:1983fi,Machacek:1984zw,Luo:2002ti}
\begin{align}
\beta_{Y^{\ts\phi}_{\bar\psi\psi'}}
= - 3\ts \sum_{i} g_i^2
\big[
C_2^{\ts i}(\psi)\ts Y^{\ts\phi}_{\bar\psi\psi'} + Y^{\ts\phi}_{\bar\psi\psi'}\ts C_2^{\ts i}(\psi')
\big]
+
\frac{1}{2}\ts
\big[
Y_2(\psi) \ts Y^{\ts\phi}_{\bar\psi\psi'} + Y^{\ts\phi}_{\bar\psi\psi'}\ts Y_2(\psi')
\big]
+
Y^{\ts\phi}_{\bar\psi\psi'}\ts \Theta(\phi)
+
2\, \Gamma_{Y^{\ts\phi}_{\bar\psi\psi'}}
\,. 
\label{eq:BetaYukawa}
\end{align}
Below we list explicit formulas for the Yukawa couplings defined in
Eq.~\eqref{eq:LYbelowGUT}:\footnote{The RGEs for the SM, $S_1$, and $S_3$ Yukawa couplings were recently studied in Refs.~\cite{Kowalska:2020gie,Fedele:2023rxb}.}
$Y^{\ts\phi}_{\bar\psi\psi'}=Y_U$, $Y_D$, $Y_E$, 
$Y'_U$, $Y'_D$, $Y'_E$, 
$Y_C^{QL}$, $Y_C^{UE}$, $Y_C^{DU}$, $Y_C^{QQ}$, 
$Y_1^{QL}$, $Y_1^{UE}$, $Y_1^{DU}$, $Y_1^{QQ}$, 
$\tilde{Y}_1^{ED}$, $\tilde{Y}_1^{UU}$, 
$Y_2^{UL}$, $Y_2^{EQ}$, 
$Y_3^{QL}$, $Y_3^{QQ}$, 
$Y_6^{DU}$, $Y_6^{QQ}$, 
$Y_8^{UQ}$, and $Y_8^{DQ}$. 
In the beta functions, the coupling $Y^{\ts\phi}_{\bar\psi\psi'}$ should be understood as 
$Y^{\ts\phi}_{\bar\psi\psi'}\ts \theta(m_{\phi}-\mu)$ by considering
the decoupling of heavy particles, and for $\phi=H,H',H_C$ and $S_1$, 
the term $Y^{\ts\phi}_{\bar\psi\psi'}\ts \Theta(\phi)$ is replaced as 
\begin{align}
Y^{\ts\phi}_{\bar\psi\psi'}\ts \Theta(\phi)
\to
\left\{
\begin{array}{ll}
Y^{\ts H}_{\bar\psi\psi'}\ts \Theta(H) + Y^{\ts H'}_{\bar\psi\psi'}\ts \Theta(H^{\prime *} H)
&\mathrm{for}\ \phi=H\ts,
\\[2mm]
Y^{\ts H'}_{\bar\psi\psi'}\ts \Theta(H') + Y^{\ts H}_{\bar\psi\psi'}\ts \Theta(H^{*} H')
&\mathrm{for}\ \phi=H'\ts,
\\[2mm]
Y^{\ts H_C}_{\bar\psi\psi'}\ts \Theta(H_C) + Y^{\ts S_1}_{\bar\psi\psi'}\ts \Theta(S_1^{*} H_C)
&\mathrm{for}\ \phi=H_C\ts,
\\[2mm]
Y^{\ts S_1}_{\bar\psi\psi'}\ts \Theta(S_1) + Y^{\ts H_C}_{\bar\psi\psi'}\ts \Theta(H_C^{*} S_1)
&\mathrm{for}\ \phi=S_1\ts.
\end{array}
\right.
\end{align}

\begin{itemize}
\item Gauge-boson-loop contributions:
\begin{align}
\sum_i g_i^2C_2^{\ts i}(q_L) =&\ 
\frac{4}{3}\ts g_s^2+\frac{3}{4}\ts g^2+\frac{1}{36}\ts g^{\prime 2}\,,\quad
\sum_i g_i^2C_2^{\ts i}(u_R) = 
\frac{4}{3}\ts g_s^2+\frac{4}{9}\ts g^{\prime 2}\,,\quad
\sum_i g_i^2C_2^{\ts i}(d_R) =
\frac{4}{3}\ts g_s^2+\frac{1}{9}\ts g^{\prime 2}\,,\nonumber\\
\sum_i g_i^2C_2^{\ts i}(\ell_L) =&\ 
\frac{3}{4}\ts g^2+\frac{1}{4}\ts g^{\prime 2}\,,\quad
\sum_i g_i^2C_2^{\ts i}(e_R) =
g^{\prime 2}\,,
\end{align}
where $C_2^{\ts i}(\psi^c) = C_2^{\ts i}(\psi)$.

\item Self-energy contributions to the fermions:
\begin{align} 
Y_2(q_L)=&\ 
Y_U^{\dagger}Y_U^{}
+Y_D^{\dagger}Y_D^{}
+Y_U^{\prime \dagger}Y_U^{\prime}
+Y_D^{\prime \dagger}Y_D^{\prime}
+Y_C^{QL *}(Y_C^{QL})^T
+2\ts Y_C^{QQ\dagger}Y_C^{QQ}
+Y_1^{QL *}(Y_1^{QL})^T
+2\ts Y_1^{QQ\dagger}Y_1^{QQ}
\nonumber\\
&\ 
+Y_2^{EQ\dagger}Y_2^{EQ}
+6\ts Y_3^{QQ\dagger}Y_3^{QQ}
+3\ts Y_3^{QL*}(Y_3^{QL})^{T}
+2\ts Y_6^{QQ\dagger}Y_6^{QQ}
+\frac{16}{3}\ts Y_8^{UQ\dagger}Y_8^{UQ}
+\frac{16}{3}\ts Y_8^{DQ\dagger}Y_8^{DQ}
\,,\nonumber\\
Y_2(u_R) 
=&\ 
2\ts Y_U^{}Y_U^{\dagger}
+2\ts Y_U^{\prime}Y_U^{\prime \dagger}
+Y_C^{UE}Y_C^{UE\dagger}
+2\ts (Y_C^{DU})^{T}Y_C^{DU *}
+Y_1^{UE}Y_1^{UE\dagger}
+2\ts (Y_1^{DU})^{T}Y_1^{DU *}
+2\ts \tilde{Y}_1^{UU}\tilde{Y}_1^{UU\dagger}
\nonumber\\
&\ 
+2\ts Y_2^{UL}Y_2^{UL\dagger}
+2\ts (Y_6^{DU})^{T}Y_6^{DU *}
+\frac{32}{3}\ts Y_8^{UQ}Y_8^{UQ\dagger}
\,,\nonumber\\
Y_2(d_R)=&\ 
2\ts Y_DY_D^\dagger
+2\ts Y_D^{\prime}Y_D^{\prime \dagger}
+2\ts Y_C^{DU}Y_C^{DU\dagger}
+2\ts Y_1^{DU}Y_1^{DU\dagger}
+ (\tilde{Y}_1^{ED})^T\tilde{Y}_1^{ED *}
+2\ts Y_6^{DU}Y_6^{DU\dagger}
+\frac{32}{3}\ts Y_8^{DQ}Y_8^{DQ\dagger}
\,,\nonumber\\
Y_2(\ell_L)=&\ 
Y_E^{\dagger}Y_E^{}+Y_E^{\prime \dagger}Y_E^{\prime}
+3\ts Y_C^{QL\dagger}Y_C^{QL}
+3\ts Y_1^{QL\dagger}Y_1^{QL}
+3\ts Y_2^{UL\dagger}Y_2^{UL}
+9\ts Y_3^{QL\dagger}Y_3^{QL}
\,,\nonumber\\
Y_2(e_R)=&\ 
2\ts Y_EY_E^{\dagger}
+2\ts Y_E^{\prime}Y_E^{\prime \dagger}
+3\ts (Y_C^{UE})^TY_C^{UE *}
+3\ts (Y_1^{UE})^TY_1^{UE *}
+3\ts \tilde{Y}_1^{ED}\tilde{Y}_1^{ED\dagger}
+6\ts Y_2^{EQ}Y_2^{EQ\dagger}
\,,
\end{align}
where $Y_2(\psi^c) = [Y_2(\psi)]^T$.

\item Self-energy contributions to the scalars:
\begin{align}
\Theta(H)
=&\ 
\mathrm{tr}\Big(3\ts Y_U^{\dagger}Y_U^{}+3\ts
Y_D^{\dagger}Y_D^{}+Y_E^{\dagger}Y_E^{}\Big)\,,
\qquad\quad\ \ 
\Theta(H^{\prime})
=
\mathrm{tr}\Big(3\ts Y_U^{\prime \dagger}Y_U^{\prime}+3\ts
Y_D^{\prime\dagger}Y_D^{\prime}+Y_E^{\prime\dagger}Y_E^{\prime}\Big)
\,,\nonumber\\
\Theta(H_C)
=&\ 
\mathrm{tr}\Big(2\ts Y_C^{QL\dagger}Y_C^{QL}
+Y_C^{UE\dagger}Y_C^{UE}
+2\ts Y_C^{DU\dagger}Y_C^{DU}
+2\ts Y_C^{QQ\dagger}Y_C^{QQ}\Big)
\,,\nonumber\\
\Theta(S_1)
=&\ 
\mathrm{tr}\Big(2\ts Y_1^{QL\dagger}Y_1^{QL}
+Y_1^{UE\dagger}Y_1^{UE}
+2\ts Y_1^{DU\dagger}Y_1^{DU}
+2\ts Y_1^{QQ\dagger}Y_1^{QQ}\Big)
\,,\nonumber\\
\Theta(\tilde{S}_1)
=&\ 
\mathrm{tr}\Big(\tilde{Y}_1^{ED\dagger}\tilde{Y}_1^{ED}
+\tilde{Y}_1^{UU\dagger}\tilde{Y}_1^{UU}\Big)
\,,
\qquad\qquad\ \,
\Theta(R_2)
=
\mathrm{tr}\Big(Y_2^{UL\dagger}Y_2^{UL}
+Y_2^{EQ\dagger}Y_2^{EQ}
\Big)
\,,\nonumber\\
\Theta(S_3)
=&\ 
\mathrm{tr}\Big(
2\ts Y_3^{QL\dagger}Y_3^{QL}
+2\ts Y_3^{QQ\dagger}Y_3^{QQ}
\Big)
\,,
\qquad\quad\ \ 
\Theta(S_6)
=
\mathrm{tr}\Big(Y_6^{DU\dagger}Y_6^{DU}
+Y_6^{QQ\dagger}Y_6^{QQ}\Big)
\,,\nonumber\\
\Theta(S_8)
=&\ 
\mathrm{tr}\Big(
  2\ts Y_8^{DQ\dagger}Y_8^{DQ}
  +2\ts Y_8^{UQ\dagger}Y_8^{UQ}\Big)
\,,
\qquad
\Theta(H^*H')
=
\mathrm{tr}\Big(3\ts Y_U^{\dagger}Y_U^{\prime}+3\ts Y_D^{}Y_D^{\prime\dagger}+Y_E^{}Y_E^{\prime\dagger}\Big)\,,
\nonumber\\
\Theta(S_1^*H_C)
=&\ 
\mathrm{tr}\Big(2\ts Y_1^{QL\dagger}Y_C^{QL}
+Y_1^{UE}Y_C^{UE\dagger}
+2\ts Y_1^{DU\dagger}Y_C^{DU}
+2\ts Y_1^{QQ}Y_C^{QQ\dagger}\Big)
\,,
\end{align}
where
$\Theta(\phi^*)=\Theta(\phi)$, 
$\Theta(H^{\prime *} H)=[\Theta(H^* H')]^*$,
and
$\Theta(H_C^{*} S_1)=[\Theta(S_1^{*} H_C)]^*$.

\item Vertex corrections:
\begin{align}
\Gamma_{Y_U^{(\prime)}}=&\ 
- Y_U^{} Y_D^{(\prime) \dagger} Y_D^{}
- Y_U^{\prime} Y_D^{(\prime) \dagger} Y_D^{\prime}
- Y_C^{UE} Y_E^{(\prime) *} (Y_C^{QL})^{T}
+ 2\ts (Y_C^{DU})^{T}Y_D^{(\prime) *} Y_C^{QQ}
- Y_1^{UE} Y_E^{(\prime) *} (Y_1^{QL})^{T}
\nonumber\\
&\ 
+ 2\ts (Y_1^{DU})^{T} Y_D^{(\prime) *} Y_1^{QQ}
- Y_2^{UL} Y_E^{(\prime) \dagger} Y_2^{EQ}
+ 2\ts (Y_6^{DU})^{T}Y_D^{(\prime) *} Y_6^{QQ}
- \frac{16}{3}\ts Y_8^{UQ} Y_D^{(\prime) \dagger} Y_8^{DQ}
\,,\nonumber\\
\Gamma_{Y_D^{(\prime)}}=&\ 
- Y_D^{} Y_U^{(\prime) \dagger} Y_U
- Y_D^{\prime} Y_U^{(\prime) \dagger} Y_U^{\prime}
+2\ts  Y_C^{DU} Y_U^{(\prime) *} Y_C^{QQ}
+2\ts  Y_1^{DU} Y_U^{(\prime) *} Y_1^{QQ}
-2\ts  Y_6^{DU} Y_U^{(\prime) *} Y_6^{QQ}
\nonumber\\
&\ 
-\frac{16}{3}\ts  Y_8^{DQ} Y_U^{(\prime) \dagger} Y_8^{UQ}
\,,\nonumber\\
\Gamma_{Y_E^{(\prime)}}=&\ 
-3\ts (Y_C^{UE})^T Y_U^{(\prime) *} Y_C^{QL}
-3\ts (Y_1^{UE})^T Y_U^{(\prime) *} Y_1^{QL}
-3\ts Y_2^{EQ} Y_U^{(\prime) \dagger} Y_2^{UL}
\,,\nonumber\\
\Gamma_{ Y_i^{QL}}=&\ 
-Y_U^T Y_i^{UE *} Y_E
-Y_U^{\prime T} Y_i^{UE *} Y_E^{\prime}
-2\ts Y_C^{QQ} Y_i^{QQ\dagger} Y_C^{QL}
-2\ts Y_1^{QQ} Y_i^{QQ\dagger} Y_1^{QL}
+ (Y_2^{EQ})^T Y_i^{UE\dagger} Y_2^{UL}
\nonumber\\
&\ 
+6\ts Y_3^{QQ} Y_i^{QQ\dagger} Y_3^{QL}
\qquad (i=C,1)
\,,\nonumber\\
\Gamma_{ Y_i^{UE}}=&\ 
- 2\ts Y_U^{} Y_i^{QL *} Y_E^T
- 2\ts Y_U^{\prime} Y_i^{QL *} Y_E^{\prime T}
- 2 (Y_C^{DU})^T Y_i^{DU *} Y_C^{UE}
- 2 (Y_1^{DU})^T Y_i^{DU *} Y_1^{UE}
- 2\ts \tilde{Y}_1^{UU} Y_i^{DU\dagger}( \tilde{Y}_1^{ED})^T
\nonumber\\
&\ 
+ 2\ts Y_2^{UL} Y_i^{QL\dagger} (Y_2^{EQ})^T
\qquad (i=C,1)
\,,\nonumber\\
\Gamma_{ Y_i^{DU}}=&\ 
2\ts Y_D^{} Y_i^{QQ\dagger} Y_U^{T}
+2\ts Y_D^{\prime} Y_i^{QQ\dagger} Y_U^{\prime T}
- Y_C^{DU} Y_i^{UE*} (Y_C^{UE})^{T}
- Y_1^{DU} Y_i^{UE*} (Y_1^{UE})^{T}
-( \tilde{Y}_1^{ED})^T Y_i^{UE\dagger} \tilde{Y}_1^{UU}
\nonumber\\
&\ 
-\frac{16}{3}\ts Y_8^{DQ} Y_i^{QQ\dagger} (Y_8^{UQ})^T
\qquad (i=C,1)
\,,\nonumber\\
\Gamma_{ Y_i^{QQ}}=&\ 
 Y_D^T Y_i^{DU*} Y_U^{}
+ Y_U^{T} Y_i^{DU\dagger} Y_D^{}
+ Y_D^{\prime T} Y_i^{DU*} Y_U^{\prime }
+ Y_U^{\prime T} Y_i^{DU\dagger} Y_D^{\prime}
- Y_C^{QL} Y_i^{QL\dagger} Y_C^{QQ}
- Y_C^{QQ} Y_i^{QL*} (Y_C^{QL})^T
\nonumber\\
&\ 
- Y_1^{QL} Y_i^{QL\dagger} Y_1^{QQ}
- Y_1^{QQ} Y_i^{QL*} (Y_1^{QL})^T
-3\ts Y_3^{QQ} Y_i^{QL*} (Y_3^{QL})^T
+3\ts Y_3^{QL} Y_i^{QL\dagger} Y_3^{QQ}
\nonumber\\
&\ 
-\frac{8}{3}\ts (Y_8^{UQ})^T Y_i^{DU\dagger} Y_8^{DQ}
-\frac{8}{3}\ts (Y_8^{DQ})^T Y_i^{DU *} Y_8^{UQ}
\qquad (i=C,1)
\,,\nonumber\\
\Gamma_{ \tilde{Y}_1^{ED}}=&\ 
2\ts (Y_C^{UE})^T \tilde{Y}_1^{UU\dagger} (Y_C^{DU})^T
+2\ts (Y_1^{UE})^T \tilde{Y}_1^{UU\dagger} (Y_1^{DU})^T
\,,\nonumber\\
\Gamma_{ \tilde{Y}_1^{UU}}=&\ 
- Y_C^{UE} \tilde{Y}_1^{ED*} Y_C^{DU}
+ (Y_C^{DU})^T \tilde{Y}_1^{ED\dagger} (Y_C^{UE})^T
- Y_1^{UE} \tilde{Y}_1^{ED*} Y_1^{DU}
+ (Y_1^{DU})^T \tilde{Y}_1^{ED\dagger} (Y_1^{UE})^T
\,,\nonumber\\
\Gamma_{ Y_2^{UL}}=&\ 
- Y_U^{} Y_2^{EQ\dagger} Y_E^{}
- Y_U^{\prime } Y_2^{EQ\dagger} Y_E^{\prime }
+ Y_C^{UE} Y_2^{EQ*} Y_C^{QL}
+ Y_1^{UE} Y_2^{EQ*} Y_1^{QL}
\,,\nonumber\\ 
\Gamma_{ Y_2^{EQ}}=&\ 
- Y_E^{} Y_2^{UL\dagger} Y_U^{}
- Y_E^{\prime} Y_2^{UL\dagger} Y_U^{\prime}
+ (Y_C^{UE})^T Y_2^{UL*} (Y_C^{QL})^T
+ (Y_1^{UE})^T Y_2^{UL*} (Y_1^{QL})^T
\,,\nonumber\\
\Gamma_{ Y_3^{QL}}=&\ 
2\ts Y_C^{QQ} Y_3^{QQ\dagger} Y_C^{QL}
+2\ts Y_1^{QQ} Y_3^{QQ\dagger} Y_1^{QL}
+2\ts Y_3^{QQ} Y_3^{QQ\dagger} Y_3^{QL}
\,,\nonumber\\
\Gamma_{ Y_3^{QQ}}=&\ 
- Y_C^{QQ} Y_3^{QL*} (Y_C^{QL})^T
+ Y_C^{QL} Y_3^{QL\dagger} Y_C^{QQ}
- Y_1^{QQ} Y_3^{QL*} (Y_1^{QL})^T
+ Y_1^{QL} Y_3^{QL\dagger} Y_1^{QQ}
+ Y_3^{QQ} Y_3^{QL*} (Y_3^{QL})^T
\nonumber\\
&\ 
+ Y_3^{QL} Y_3^{QL\dagger} Y_3^{QQ}
\,,\nonumber\\ 
\Gamma_{ Y_6^{DU}}=&\ 
-2\ts Y_D^{} Y_6^{QQ\dagger} Y_U^{T}
-2\ts Y_D^{\prime} Y_6^{QQ\dagger} Y_U^{\prime T}
-\frac{8}{3}\ts Y_8^{DQ} Y_6^{QQ\dagger} (Y_8^{UQ})^T
\,,\nonumber\\
\Gamma_{ Y_6^{QQ}}=&\ 
- Y_U^{T} Y_6^{DU\dagger} Y_D^{}
+ Y_D^{T} Y_6^{DU*} Y_U^{}
- Y_U^{\prime T} Y_6^{DU\dagger} Y_D^{\prime }
+ Y_D^{\prime T} Y_6^{DU*} Y_U^{\prime }
-\frac{4}{3}\ts (Y_8^{UQ})^T Y_6^{DU\dagger} Y_8^{DQ}
\nonumber\\
&\ 
+\frac{4}{3}\ts (Y_8^{DQ})^T Y_6^{DU*} Y_8^{UQ}
\,,\nonumber\\
\Gamma_{ Y_8^{UQ}}=&\ 
- Y_U^{} Y_8^{DQ\dagger} Y_D^{}
- Y_U^{\prime} Y_8^{DQ\dagger} Y_D^{\prime}
- (Y_C^{DU})^T Y_8^{DQ *} Y_C^{QQ}
- (Y_1^{DU})^T Y_8^{DQ *} Y_1^{QQ}
+\frac{1}{2}\ts (Y_6^{DU})^T Y_8^{DQ*} Y_6^{QQ}
\nonumber\\
&\ 
+\frac{2}{3}\ts Y_8^{UQ} Y_8^{DQ\dagger} Y_8^{DQ}
\,,\nonumber\\
\Gamma_{ Y_8^{DQ}}=&\ 
- Y_D^{} Y_8^{UQ\dagger} Y_U^{}
- Y_D^{\prime} Y_8^{UQ\dagger} Y_U^{\prime}
- Y_C^{DU} Y_8^{UQ*} Y_C^{QQ}
- Y_1^{DU} Y_8^{UQ*} Y_1^{QQ}
-\frac{1}{2}\ts Y_6^{DU} Y_8^{UQ *} Y_6^{QQ}
\nonumber\\
&\ 
+\frac{2}{3}\ts Y_8^{DQ} Y_8^{UQ\dagger} Y_8^{UQ}
\,.
\end{align}
\end{itemize}

Notice that the coupling $Y_3^{QQ}$ is shown to be vanishing in the whole range of the renormalization scale below the GUT scale if $Y_3^{QQ}=0$ and $H_C$, $S_1$, and $\tilde{S}_1$ decouple at the GUT scale. In fact, the Yukawa interaction Lagrangian in Eq.~\eqref{eq:LYbelowGUT} becomes invariant under U(1)$_B$ and U(1)$_L$ separately if the terms with $H_C$, $S_1$, and $\tilde{S}_1$ are removed and if $Y_3^{QQ}$ is set to vanish. The latter condition is satisfied if $Y_{45}^U=0$ at the GUT scale in the tree-level approximation. The appropriate assignment of the baryon and lepton numbers in the above case is listed in Table~\ref{tab:BL}.
\begin{table}[t]
\centering
\caption{Assignment of the baryon and lepton numbers to the scalars, where the Yukawa interactions in Eq.~\eqref{eq:LYbelowGUT} are invariant under U(1)$_B$ and U(1)$_L$ separately if $H_C$, $S_1$, and $\tilde{S}_1$ decouple and $Y_3^{QQ}$ is set to vanish at the GUT scale. }
\label{tab:BL}
\setlength{\tabcolsep}{4pt}
\begin{tabular}{c|rrrr}
\hline\hline
& $R_2^*$ & $S_3^*$ & $S_6^*$ & $S_8$ 
\\
\hline
$3B$ & $-1$ & $1$ & $-2$ & $0$
\\
$L$ & $1$ & $1$ & $0$ & $0$
\\
\hline\hline
\end{tabular}
\end{table}

\section{LEFT Lagrangian}
\label{sec:LEFTcoefficients}

The LEFT Lagrangian is given by 
\begin{align}
\mathcal{L}_{\text{LEFT}}
&=
\mathcal{L}_{\text{QCD}+\text{QED}}
+
\sum_i L_i\tts \mathcal{Q}_i\,,
\label{eq:LEFTLagrangian}
\end{align}
where $\mathcal{L}_{\text{QCD}+\text{QED}}$ is the renormalizable QCD
and QED Lagrangian with the SM fermions except for top quark. 
The LEFT operators $\mathcal{Q}_i$ up to dimension six are
classified in Ref.~\cite{Jenkins:2017jig}. The operators relevant to
the current study are 
\begin{align}
\big[ \mathcal{Q}_{\tts\nu d}^{V,LL} \big]_{ijkl}
  &=
  \big( \bar{\hat{\nu}}_{Li}\gamma^\mu\ts \hat{\nu}_{Lj} \big)
  \big( \bar{\hat{d}}_{Lk}\gamma_\mu\ts \hat{d}_{Ll} \big)\,,
&\big[ \mathcal{Q}_{\tts ed}^{V,LL} \big]_{ijkl}
  &=
  \big( \bar{\hat{e}}_{Li}\gamma^\mu\ts \hat{e}_{Lj} \big)
  \big( \bar{\hat{d}}_{Lk}\gamma_\mu\ts \hat{d}_{Ll} \big)\,,
  \nonumber\\
\big[ \mathcal{Q}_{\tts ee}^{V,LL} \big]_{ijkl}
  &=
  \big( \bar{\hat{e}}_{Li}\gamma^\mu\ts \hat{e}_{Lj} \big)
  \big( \bar{\hat{e}}_{Lk}\gamma_\mu\ts \hat{e}_{Ll} \big)\,,
&\big[ \mathcal{Q}_{\tts ee}^{V,LR} \big]_{ijkl}
  &=
  \big( \bar{\hat{e}}_{Li}\gamma^\mu\ts \hat{e}_{Lj} \big)
  \big( \bar{\hat{e}}_{Rk}\gamma_\mu\ts \hat{e}_{Rl} \big)\,,
  \nonumber\\
\big[ \mathcal{Q}_{\tts dd}^{V,LL} \big]_{ijkl}
  &=
  \big( \bar{\hat{d}}_{Li}\gamma^\mu\ts \hat{d}_{Lj} \big)
  \big( \bar{\hat{d}}_{Lk}\gamma_\mu\ts \hat{d}_{Ll} \big)\,,
&\big[ \mathcal{Q}_{\tts e\gamma} \big]_{ij}
  &=
  \big( \bar{\hat{e}}_{Li}\tts\sigma^{\mu\nu}\tts\hat{e}_{Rj} \big)
  F_{\mu\nu}\,, 
  \nonumber\\
\big[ \mathcal{Q}_{\tts ed}^{V,LR} \big]_{ijkl}
  &=
  \big( \bar{\hat{e}}_{Li}\gamma^\mu\ts \hat{e}_{Lj} \big)
  \big( \bar{\hat{d}}_{Rk}\gamma_\mu\ts \hat{d}_{Rl} \big)\,,
&
\big[ \mathcal{Q}_{\tts de}^{V,LR} \big]_{ijkl}
  &=
  \big( \bar{\hat{d}}_{Li}\gamma^\mu\ts \hat{d}_{Lj} \big)
  \big( \bar{\hat{e}}_{Rk}\gamma_\mu\ts \hat{e}_{Rl} \big)\,,
  \nonumber\\
\big[ \mathcal{Q}_{\tts \nu edu}^{V,LL} \big]_{ijkl}
  &=
  \big( \bar{\hat{\nu}}_{Li}\gamma^\mu\ts \hat{e}_{Lj} \big)
  \big( \bar{\hat{d}}_{Lk}\gamma_\mu\ts \hat{u}_{Ll} \big)\,, 
\label{eq:LEFToperators}
\end{align}
where there also exist the Hermitian conjugates
of the non-self-conjugate operators. 

The Wilson coefficients for these operators are 
calculated as follows.
\begin{enumerate}
  \item The $S_3$ field is integrated out at the $S_3$ mass scale, and 
  the model is matched onto the SMEFT at the one-loop level.
  The one-loop matching formulas from a model with the $S_1$ and $S_3$ leptoquarks to 
  the SMEFT are listed in Refs.~\cite{deBlas:2014mba,deBlas:2017xtg,Gherardi:2020det}. 
  \item The RG running effects of the SMEFT operators are 
  taken into account. 
  The anomalous dimensions for the dimension-six operators in the SMEFT are listed in
  Refs.~\cite{Jenkins:2013wua,Jenkins:2013zja,Alonso:2013hga}.
  \item The SMEFT is matched onto the LEFT at the weak scale. The
    one-loop matching formulas are listed in Refs.~\cite{Jenkins:2017jig,Aebischer:2015fzz,Dekens:2019ept}.
  \item The RG effects in the LEFT are taken into account. The
    corresponding anomalous dimensions are given in Refs.~\cite{Jenkins:2017dyc,Aebischer:2017gaw}.
\end{enumerate}
The coefficients in the LEFT Lagrangian at the relevant scale for the
process under consideration are given 
in the leading-logarithmic approximation by 
\begin{align}
\big[ L_{\tts \nu d}^{V,LL} \big]_{ij23}^{\mathrm{NP}}
  &=
  \frac{
  \big(\bar{Y}_{3}^{QL}\big)^*_{2\tts i}\ts 
  \big(\bar{Y}_{3}^{QL}\big)_{3j}
  }
  {2\tts m_{S_3}^2}
  \Bigg\{
  1
  +
  \frac{g^2( 1 + 2\tts c_W^2)}{32\tts \pi^2 c_W^2}
  \bigg[
  \log\bigg(\frac{m_{S_3}^2}{m_Z^2}\bigg) + \frac{11}{6}
  \bigg]
  -
  \frac{3\ts g^2}{16\tts \pi^2}
  \bigg[
    \log\bigg(\frac{m_{S_3}^2}{m_W^2}\bigg) + \frac{11}{6}
  \bigg]
  \Bigg\}
\nonumber\\
&\hspace{5mm}
  +
  \frac{y_t^2}{64\tts \pi^2}
  \Bigg\{
  4\ts V_{ts}^*V_{tb}
  \frac{
  \big(Y_{3}^{QL}\big)^*_{3\tts i}\ts 
  \big(Y_{3}^{QL}\big)_{3j}}
  {m_{S_3}^2}
  +
  \frac{1}{2}
  \Bigg[
  V^*_{ts}
  \frac{
  \big(Y_{3}^{QL}\big)^*_{3\tts i}\ts 
  \big(\bar{Y}_{3}^{QL}\big)_{3j}}
  {m_{S_3}^2}
  +
  V_{tb}
  \frac{
  \big(\bar{Y}_{3}^{QL}\big)^*_{2\tts i}\ts 
  \big(Y_{3}^{QL}\big)_{3j}}
  {m_{S_3}^2}
  \Bigg]\ts I_{\nu d}(x_t)
  \Bigg\}
  \nonumber\\
  &\hspace{5mm}
  -
  \frac{3\tts (N_c+1)}{16}
  \Bigg[
  \frac{
  \big(
    \bar{Y}_{3}^{QL\dagger} \bar{Y}_{3}^{QL} \bar{Y}_{3}^{QL\dagger} 
  \big)_{i\tts 2}\ts
  \big(\bar{Y}_{3}^{QL}\big)_{3j}}
  {(4\pi)^2 m_{S_3}^2}
  +
  \frac{
  \big(\bar{Y}_{3}^{QL}\big)^*_{2\tts i}\ts
  \big(
    \bar{Y}_{3}^{QL} \bar{Y}_{3}^{QL\dagger} \bar{Y}_{3}^{QL} 
  \big)_{3j}}
  {(4\pi)^2 m_{S_3}^2}
  \Bigg]
  \nonumber\\
  &\hspace{5mm}
  -
  \frac{1}{4}
  \frac{
    \big( \bar{Y}_{3}^{QL\dagger} \bar{Y}_{3}^{QL} \big)_{i\tts j}\ts
    \big( \bar{Y}_{3}^{QL} \bar{Y}_{3}^{QL\dagger} \big)_{32}
  }
  {(4\pi)^2 m_{S_3}^{2}}
  \,,
  \label{eq:LnudVLLij23}
  \\
\big[ L_{\tts de}^{V,LR}(m_b) \big]_{23ij}^{\mathrm{NP}}
  &=
  - 
  \delta_{ij}
  \frac{\alpha}{6\tts\pi}
  \frac{
  \big( \bar{Y}_{3}^{QL} \bar{Y}_{3}^{QL\dagger}\big)_{32}
  }{m_{S_3}^{2}}
  \Bigg[
  \log\bigg(\frac{m_{S_3}^2}{m_b^2}\bigg)
  - \frac{19}{12} 
  \Bigg]\,,
  \label{eq:LdeVLR23ij}
  \\
\big[ L_{\tts ed}^{V,LL}(m_\tau) \big]_{3222}^{\mathrm{NP}}
  &=
  \frac{
  \big(\bar{Y}_{3}^{QL}\big)^*_{23}\ts 
  \big(\bar{Y}_{3}^{QL}\big)_{22}}
  {m_{S_3}^2}
  \Bigg\{
  1
  -
  \frac{\alpha}{2\pi}
  \log\bigg(\frac{m_{S_3}^2}{m_\tau^2}\bigg)
  +
  \frac{g^2( 1 - 4\tts c_W^4)}{32\tts \pi^2 c_W^2}
  \bigg[
  \log\bigg(\frac{m_{S_3}^2}{m_Z^2}\bigg) + \frac{11}{6}
  \bigg]
  \Bigg\}
  \nonumber\\
  &\hspace{5mm}
  +
  \frac{y_t^2}{64\tts \pi^2}
  \Bigg\{
  2\ts V^*_{ts}V_{ts}
  \frac{
  \big(Y_{3}^{QL}\big)^*_{33}\ts 
  \big(Y_{3}^{QL}\big)_{32}
  }
  {m_{S_3}^2}
  +
  \Bigg[
  V^*_{ts}
  \frac{
  \big(Y_{3}^{QL}\big)^*_{33}\ts 
  \big(\bar{Y}_{3}^{QL}\big)_{22}
  }
  {m_{S_3}^2}
  +
  V_{ts}
  \frac{
  \big(\bar{Y}_{3}^{QL}\big)^*_{23}\ts 
  \big(Y_{3}^{QL}\big)_{32}
  }
  {m_{S_3}^2}
  \Bigg]\ts I_{ed}(x_t)
  \Bigg\}
  \nonumber\\
  &\hspace{5mm}
  -
  \frac{3\tts (N_c+1)}{8}
  \Bigg[
  \frac{
  \big(
    \bar{Y}_{3}^{QL\dagger} \bar{Y}_{3}^{QL} \bar{Y}_{3}^{QL\dagger} 
  \big)_{32}\ts
  \big(\bar{Y}_{3}^{QL}\big)_{22}}
  {(4\pi)^2 m_{S_3}^2}
  +
  \frac{
  \big(\bar{Y}_{3}^{QL}\big)^*_{23}\ts
  \big(
    \bar{Y}_{3}^{QL} \bar{Y}_{3}^{QL\dagger} \bar{Y}_{3}^{QL} 
  \big)_{22}}
  {(4\pi)^2 m_{S_3}^2}
  \Bigg]
  \nonumber\\
  &\hspace{5mm}
  -
  \frac{5}{4}
  \frac{
    \big( \bar{Y}_{3}^{QL\dagger} \bar{Y}_{3}^{QL} \big)_{32}\ts
    \big( \bar{Y}_{3}^{QL} \bar{Y}_{3}^{QL\dagger} \big)_{22}
  }
  {(4\pi)^2 m_{S_3}^{2}}
  -
  \frac{\alpha}{6\tts\pi} N_c\ts Q_d^2
  \Bigg[
  \frac{
  \big(Y_{3}^{QL}\big)^*_{33}\ts
  \big(Y_{3}^{QL}\big)_{32}
  }{m_{S_3}^{2}}
  \log\bigg( \frac{m_t^2}{m_b^2} \bigg)
  -
  \frac{3}{4}
  \frac{\big(Y_{3}^{QL\dagger}Y_{3}^{QL}\big)_{32}}{m_{S_3}^{2}}
  \Bigg]
  \nonumber\\
  &\hspace{5mm}
  -
  N_c
  \big( I_{d_L}^3 - Q_d s_W^2 \big)\tts y_t^2
  \frac{\big( Y_{3}^{QL} \big)^*_{33}\ts
  \big(Y_{3}^{QL} \big)_{32}}
  {(4\tts\pi)^2\ts m_{S_3}^{2}}
  \Bigg[\log\bigg(\frac{m_{S_3}^2}{m_{t}^2}\bigg) - 1 \Bigg] 
  \label{eq:LedVLL3222}
  \,,
  \\
\big[ L_{\tts ed}^{V,LR}(m_\tau) \big]_{3222}^{\mathrm{NP}}
&=
  -
  \frac{\alpha}{6\tts\pi} N_c\ts Q_d^2
  \Bigg[
  \frac{
  \big(Y_{3}^{QL}\big)^*_{33}\ts
  \big(Y_{3}^{QL}\big)_{32}
  }{m_{S_3}^{2}}
  \log\bigg( \frac{m_t^2}{m_b^2} \bigg)
  -
  \frac{3}{4}
  \frac{\big(Y_{3}^{QL\dagger}Y_{3}^{QL}\big)_{32}}{m_{S_3}^{2}}
  \Bigg]
  \nonumber\\
  &\hspace{5mm}
  - N_c
  \big(\! -Q_d s_W^2\big)\tts y_t^2
  \frac{
  \big(Y_{3}^{QL} \big)^*_{33}\ts
  \big(Y_{3}^{QL} \big)_{32}
  }
  {(4\tts\pi)^2\ts m_{S_3}^{2}}
  \Bigg[
  \log\bigg(\frac{m_{S_3}^2}{m_{t}^2}\bigg)
  - 1 \Bigg]
  \,,
  \label{eq:LedVLR3222}
  \\
\big[ L_{\tts ee}^{V,LL}(m_\tau) \big]_{3222}
&=
  -
  \frac{5\tts N_{c}}{8}\tts
  \frac{
  \big(Y_{3}^{QL\dagger}Y_{3}^{QL}\big)_{32}\ts
  \big(Y_{3}^{QL\dagger}Y_{3}^{QL}\big)_{22}}
  {(4\tts\pi)^2\ts m_{S_3}^{2}}
  \nonumber\\
  &\hspace{5mm}
  -
  \frac{\alpha}{12\tts\pi} N_c\ts Q_d\tts Q_e
  \Bigg[
  \frac{
  \big(Y_{3}^{QL}\big)^*_{33}\ts
  \big(Y_{3}^{QL}\big)_{32}
  }{m_{S_3}^{2}}
  \log\bigg( \frac{m_t^2}{m_b^2} \bigg)
  - 
  \frac{3}{4}
  \frac{\big(Y_{3}^{QL\dagger}Y_{3}^{QL}\big)_{32}}{m_{S_3}^{2}}
  \Bigg]
  \nonumber\\
  &\hspace{5mm}
  - 
  \frac{N_c}{2}
  \big( I_{e_L}^3 - Q_e s_W^2 \big)\tts y_t^2
  \frac{\big( Y_{3}^{QL} \big)^*_{33}\ts
  \big(Y_{3}^{QL} \big)_{32}}
  {(4\tts\pi)^2\ts m_{S_3}^{2}}
  \Bigg[ 
  \log\bigg(\frac{m_{S_3}^2}{m_t^2}\bigg) - 1 \Bigg]
  \,,
  \label{eq:LeeVLL3222}
  \\
\big[ L_{\tts ee}^{V,LR}(m_\tau) \big]_{3222}
  &=
  -
  \frac{\alpha}{6\tts\pi} N_c\ts Q_d\tts Q_e
  \Bigg[
  \frac{
  \big(Y_{3}^{QL}\big)^*_{33}\ts
  \big(Y_{3}^{QL}\big)_{32}
  }{m_{S_3}^{2}}
  \log\bigg( \frac{m_t^2}{m_b^2} \bigg)
  -
  \frac{3}{4}
  \frac{\big(Y_{3}^{QL\dagger}Y_{3}^{QL}\big)_{32}}{m_{S_3}^{2}}
  \Bigg]
  \nonumber\\
  &\hspace{5mm}
  -
  N_c
  \big(\! -Q_e s_W^2\big)\tts y_t^2
  \frac{\big( Y_{3}^{QL} \big)^*_{33}\ts
  \big(Y_{3}^{QL} \big)_{32}}
  {(4\tts\pi)^2\ts m_{S_3}^{2}}
  \Bigg[ 
  \log\bigg(\frac{m_{S_3}^2}{m_t^2}\bigg) - 1 \Bigg]
  \,,
  \label{eq:LeeVLR3222}
  \\
\big[ L_{\tts e\gamma}(m_\tau) \big]_{ij}^{\mathrm{NP}}
  &=
  \frac{e N_c m_{e_j}}{8}
  \frac{
  \big(Y_{3}^{QL\dagger}Y_{3}^{QL} \big)_{ij}}{(4\tts \pi)^2\ts m_{S_3}^{2}}\,,
  \label{eq:Legammaij}
\\
\big[ L_{\tts dd}^{V,LL}(m_b) \big]_{2323}^{\mathrm{NP}}
  &=
  - \frac{5}{8}
  \frac{
  \big( \bar{Y}_{3}^{QL} \bar{Y}_{3}^{QL\dagger}\big)_{32}\ts
  \big( \bar{Y}_{3}^{QL} \bar{Y}_{3}^{QL\dagger}\big)_{32}
  }{(4\tts\pi)^2\ts m_{S_3}^2}\,,
  \label{eq:LddVLL2323}
\end{align}
where $Q_d=-1/3$, $Q_e=-1$, $I_{d_L}^3=I_{e_L}^3=-1/2$, 
and $I_{\nu d}(x)$ is the loop function defined by 
\begin{align}
I_{\nu d}(x)
&=
  -
  \log\bigg(\frac{m_{S_3}^2}{m_W^2}\bigg)
  -
  \frac{3(x + 1 )}{2(x - 1)}
  +
  \frac{x^2 + 10\tts x - 8}{(x - 1 )^2}
  \log x\,.
\end{align}
Similar one-loop expressions for the low-energy coefficients can also be found in Refs.~\cite{Crivellin:2019dwb,Gherardi:2020qhc,Bordone:2020lnb,Crivellin:2020mjs}.

\section{\texorpdfstring{$\bm{Z\to\mu^\mp\tau^\pm}$}{ZtoMuTau}}
\label{sec:ZtoMuTau}

The $S_3$ affects the $Z$-boson effective couplings with charged
leptons which are defined as 
\begin{align}
\mathcal{L}
&=
\frac{e}{s_W c_W}\,
Z_\mu\ts 
\Big[
\bar{e}_{Li}\ts \gamma_\mu
\big(g_{L}^e\big)_{ij} 
e_{Lj}
+
\bar{e}_{Ri}\ts \gamma_\mu
\big(g_{R}^e\big)_{ij} 
e_{Rj}
\Big]
\,,
\label{eq:L_Zffbar}
\end{align}
where 
$(g_{L}^e)_{ij} = g_{L}^{\tts e,\mathrm{SM}}\ts \delta_{ij} + (g_{L}^e)^{\mathrm{NP}}_{ij}$ 
and 
$(g_{R}^e)_{ij} = g_{R}^{\tts e,\mathrm{SM}}\ts \delta_{ij} + (g_{R}^e)^{\mathrm{NP}}_{ij}$ 
with the SM tree-level couplings 
$g_{L}^{\tts e,\mathrm{SM}} = I^3_{e_L} - Q_e s_W^2$ 
and $g_{R}^{\tts e,\mathrm{SM}} = - Q_e s_W^2$. 
According to Ref.~\cite{Arnan:2019olv} (see also Refs.~\cite{Feruglio:2016gvd,Feruglio:2017rjo,Crivellin:2020mjs}), the $S_3$ contribution to the left-handed coupling reads as
\begin{align}
\big( g_L^{\tts e} \big)^{\mathrm{NP}}_{ij}
&=
\frac{N_c}{(4\tts\pi)^2}
\bigg[
\big( g_{L}^{\tts u,\mathrm{SM}} - g_{R}^{\tts u,\mathrm{SM}} \big)
\frac{ x_t(x_t-1-\log x_t)}{(x_t-1)^2}
+
\frac{x_Z}{12}\tts
F(x_t)
+
\mathcal{O}(x_Z^2)
\bigg]
\big(Y_{3}^{QL*} \big)_{3i}
\big(Y_{3}^{QL} \big)_{3j}
\nonumber\\
&\hspace{5mm}
+
\frac{N_c\ts x_Z}{3(4\tts\pi)^2}
\bigg[
g_{L}^{\tts u,\mathrm{SM}}
\bigg(
\log x_Z
-
i\pi
- 
\frac{1}{6}
\bigg)
+
\frac{g_{L}^{\tts e,\mathrm{SM}}}{6}
\bigg]
\sum_{w=1}^{2}
\big(Y_{3}^{QL*}\big)_{wi} 
\big(Y_{3}^{QL} \big)_{wj}
\nonumber\\
&\hspace{5mm}
+
\frac{2\ts N_c\ts x_Z}{3(4\tts\pi)^2}
\bigg[
g_{L}^{\tts d,\mathrm{SM}}
\bigg(
\log x_Z
-
i\pi
- 
\frac{1}{6}
\bigg)
+
\frac{g_{L}^{\tts e,\mathrm{SM}}}{6}
\bigg]
\sum_{w=1}^{3}
\big(\bar{Y}_{3}^{QL*}\big)_{wi} 
\big(\bar{Y}_{3}^{QL} \big)_{wj}
\,,
\end{align}
where $x_Z=m_Z^2/m_{S_3}^2$, $x_t=m_t^2/m_{S_3}^2$, 
$g_{L}^{\tts u,\mathrm{SM}}$, $g_{R}^{\tts u,\mathrm{SM}}$ 
and $g_{L}^{\tts d,\mathrm{SM}}$ are the SM couplings for up-type
and down-type quarks defined analogous to those for charged
leptons, 
and the function $F(x)$ is defined as 
\begin{align}
F(x)
&=
- g_{L}^{\tts u,\mathrm{SM}}
\frac{(x - 1) (5\tts x^2 - 7 x + 8) - 2 (x^3 + 2) \log x}{(x - 1)^4}
- g_{R}^{\tts u,\mathrm{SM}}
\frac{(x - 1) (x^2 - 5\tts x - 2) + 6\tts x \log x}{(x - 1)^4}
\nonumber\\
&\hspace{5mm}
+ g_{L}^{\tts e,\mathrm{SM}}
\frac{(x - 1) (-11 x^2 + 7\tts x - 2) + 6\tts x^3 \log x}{3 (x - 1)^4}\,.
\end{align}
Using the above effective coupling, 
the branching ratio for $Z\to\mu^\mp\tau^\pm$ is given by 
\begin{align}
\mathcal{B}(Z\to\mu^\mp\tau^\pm)
=
\mathcal{B}(Z\to\mu^-\tau^+)
+
\mathcal{B}(Z\to\mu^+\tau^-)
=
\frac{G_Fm_Z^3}{3\tts\pi\sqrt{2}\,\Gamma_Z}
\Big(\,
\Big| \big(g_L^e\big)^{\mathrm{NP}}_{23} \Big|^2
+
\Big| \big(g_L^e\big)^{\mathrm{NP}}_{32} \Big|^2
\,\Big)
\,,
\end{align}
where $\Gamma_Z$ is the total decay width of $Z$ boson.

\bibliography{ref}

\begin{thebibliography}{143}%
\makeatletter
\providecommand \@ifxundefined [1]{%
 \@ifx{#1\undefined}
}%
\providecommand \@ifnum [1]{%
 \ifnum #1\expandafter \@firstoftwo
 \else \expandafter \@secondoftwo
 \fi
}%
\providecommand \@ifx [1]{%
 \ifx #1\expandafter \@firstoftwo
 \else \expandafter \@secondoftwo
 \fi
}%
\providecommand \natexlab [1]{#1}%
\providecommand \enquote  [1]{``#1''}%
\providecommand \bibnamefont  [1]{#1}%
\providecommand \bibfnamefont [1]{#1}%
\providecommand \citenamefont [1]{#1}%
\providecommand \href@noop [0]{\@secondoftwo}%
\providecommand \href [0]{\begingroup \@sanitize@url \@href}%
\providecommand \@href[1]{\@@startlink{#1}\@@href}%
\providecommand \@@href[1]{\endgroup#1\@@endlink}%
\providecommand \@sanitize@url [0]{\catcode `\\12\catcode `\$12\catcode
  `\&12\catcode `\#12\catcode `\^12\catcode `\_12\catcode `\%12\relax}%
\providecommand \@@startlink[1]{}%
\providecommand \@@endlink[0]{}%
\providecommand \url  [0]{\begingroup\@sanitize@url \@url }%
\providecommand \@url [1]{\endgroup\@href {#1}{\urlprefix }}%
\providecommand \urlprefix  [0]{URL }%
\providecommand \Eprint [0]{\href }%
\providecommand \doibase [0]{https://doi.org/}%
\providecommand \selectlanguage [0]{\@gobble}%
\providecommand \bibinfo  [0]{\@secondoftwo}%
\providecommand \bibfield  [0]{\@secondoftwo}%
\providecommand \translation [1]{[#1]}%
\providecommand \BibitemOpen [0]{}%
\providecommand \bibitemStop [0]{}%
\providecommand \bibitemNoStop [0]{.\EOS\space}%
\providecommand \EOS [0]{\spacefactor3000\relax}%
\providecommand \BibitemShut  [1]{\csname bibitem#1\endcsname}%
\let\auto@bib@innerbib\@empty
\bibitem [{\citenamefont {Pati}\ and\ \citenamefont
  {Salam}(1973)}]{Pati:1973rp}%
  \BibitemOpen
  \bibfield  {author} {\bibinfo {author} {\bibfnamefont {J.~C.}\ \bibnamefont
  {Pati}}\ and\ \bibinfo {author} {\bibfnamefont {A.}~\bibnamefont {Salam}},\
  }\href {https://doi.org/10.1103/PhysRevLett.31.661} {\bibfield  {journal}
  {\bibinfo  {journal} {Phys. Rev. Lett.}\ }\textbf {\bibinfo {volume} {31}},\
  \bibinfo {pages} {661} (\bibinfo {year} {1973})}\BibitemShut {NoStop}%
\bibitem [{\citenamefont {Pati}\ and\ \citenamefont
  {Salam}(1974)}]{Pati:1974yy}%
  \BibitemOpen
  \bibfield  {author} {\bibinfo {author} {\bibfnamefont {J.~C.}\ \bibnamefont
  {Pati}}\ and\ \bibinfo {author} {\bibfnamefont {A.}~\bibnamefont {Salam}},\
  }\href {https://doi.org/10.1103/PhysRevD.10.275} {\bibfield  {journal}
  {\bibinfo  {journal} {Phys. Rev. D}\ }\textbf {\bibinfo {volume} {10}},\
  \bibinfo {pages} {275} (\bibinfo {year} {1974})},\ \bibinfo {note} {[Erratum:
  Phys. Rev. D 11, 703 (1975)]}\BibitemShut {NoStop}%
\bibitem [{\citenamefont {Georgi}\ and\ \citenamefont
  {Glashow}(1974)}]{Georgi:1974sy}%
  \BibitemOpen
  \bibfield  {author} {\bibinfo {author} {\bibfnamefont {H.}~\bibnamefont
  {Georgi}}\ and\ \bibinfo {author} {\bibfnamefont {S.~L.}\ \bibnamefont
  {Glashow}},\ }\href {https://doi.org/10.1103/PhysRevLett.32.438} {\bibfield
  {journal} {\bibinfo  {journal} {Phys. Rev. Lett.}\ }\textbf {\bibinfo
  {volume} {32}},\ \bibinfo {pages} {438} (\bibinfo {year} {1974})}\BibitemShut
  {NoStop}%
\bibitem [{\citenamefont {Georgi}\ \emph {et~al.}(1974)\citenamefont {Georgi},
  \citenamefont {Quinn},\ and\ \citenamefont {Weinberg}}]{Georgi:1974yf}%
  \BibitemOpen
  \bibfield  {author} {\bibinfo {author} {\bibfnamefont {H.}~\bibnamefont
  {Georgi}}, \bibinfo {author} {\bibfnamefont {H.~R.}\ \bibnamefont {Quinn}},\
  and\ \bibinfo {author} {\bibfnamefont {S.}~\bibnamefont {Weinberg}},\ }\href
  {https://doi.org/10.1103/PhysRevLett.33.451} {\bibfield  {journal} {\bibinfo
  {journal} {Phys. Rev. Lett.}\ }\textbf {\bibinfo {volume} {33}},\ \bibinfo
  {pages} {451} (\bibinfo {year} {1974})}\BibitemShut {NoStop}%
\bibitem [{\citenamefont {Georgi}(1975)}]{Georgi:1974my}%
  \BibitemOpen
  \bibfield  {author} {\bibinfo {author} {\bibfnamefont {H.}~\bibnamefont
  {Georgi}},\ }\href {https://doi.org/10.1063/1.2947450} {\bibfield  {journal}
  {\bibinfo  {journal} {AIP Conf. Proc.}\ }\textbf {\bibinfo {volume} {23}},\
  \bibinfo {pages} {575} (\bibinfo {year} {1975})}\BibitemShut {NoStop}%
\bibitem [{\citenamefont {Fritzsch}\ and\ \citenamefont
  {Minkowski}(1975)}]{Fritzsch:1974nn}%
  \BibitemOpen
  \bibfield  {author} {\bibinfo {author} {\bibfnamefont {H.}~\bibnamefont
  {Fritzsch}}\ and\ \bibinfo {author} {\bibfnamefont {P.}~\bibnamefont
  {Minkowski}},\ }\href {https://doi.org/10.1016/0003-4916(75)90211-0}
  {\bibfield  {journal} {\bibinfo  {journal} {Annals Phys.}\ }\textbf {\bibinfo
  {volume} {93}},\ \bibinfo {pages} {193} (\bibinfo {year} {1975})}\BibitemShut
  {NoStop}%
\bibitem [{\citenamefont {Langacker}(1981)}]{Langacker:1980js}%
  \BibitemOpen
  \bibfield  {author} {\bibinfo {author} {\bibfnamefont {P.}~\bibnamefont
  {Langacker}},\ }\href {https://doi.org/10.1016/0370-1573(81)90059-4}
  {\bibfield  {journal} {\bibinfo  {journal} {Phys. Rept.}\ }\textbf {\bibinfo
  {volume} {72}},\ \bibinfo {pages} {185} (\bibinfo {year} {1981})}\BibitemShut
  {NoStop}%
\bibitem [{\citenamefont {Langacker}(1990)}]{Langacker:1990jh}%
  \BibitemOpen
  \bibfield  {author} {\bibinfo {author} {\bibfnamefont {P.}~\bibnamefont
  {Langacker}},\ }in\ \href@noop {} {\emph {\bibinfo {booktitle} {{Prodceedings
  of the 1st International Symposium on Particles, Strings and Cosmology}}}}\
  (\bibinfo  {publisher} {World Scientific},\ \bibinfo {year} {1990})\ pp.\
  \bibinfo {pages} {237--269}\BibitemShut {NoStop}%
\bibitem [{\citenamefont {Ellis}\ \emph {et~al.}(1991)\citenamefont {Ellis},
  \citenamefont {Kelley},\ and\ \citenamefont {Nanopoulos}}]{Ellis:1990wk}%
  \BibitemOpen
  \bibfield  {author} {\bibinfo {author} {\bibfnamefont {J.~R.}\ \bibnamefont
  {Ellis}}, \bibinfo {author} {\bibfnamefont {S.}~\bibnamefont {Kelley}},\ and\
  \bibinfo {author} {\bibfnamefont {D.~V.}\ \bibnamefont {Nanopoulos}},\ }\href
  {https://doi.org/10.1016/0370-2693(91)90980-5} {\bibfield  {journal}
  {\bibinfo  {journal} {Phys. Lett. B}\ }\textbf {\bibinfo {volume} {260}},\
  \bibinfo {pages} {131} (\bibinfo {year} {1991})}\BibitemShut {NoStop}%
\bibitem [{\citenamefont {Amaldi}\ \emph {et~al.}(1991)\citenamefont {Amaldi},
  \citenamefont {de~Boer},\ and\ \citenamefont {Furstenau}}]{Amaldi:1991cn}%
  \BibitemOpen
  \bibfield  {author} {\bibinfo {author} {\bibfnamefont {U.}~\bibnamefont
  {Amaldi}}, \bibinfo {author} {\bibfnamefont {W.}~\bibnamefont {de~Boer}},\
  and\ \bibinfo {author} {\bibfnamefont {H.}~\bibnamefont {Furstenau}},\ }\href
  {https://doi.org/10.1016/0370-2693(91)91641-8} {\bibfield  {journal}
  {\bibinfo  {journal} {Phys. Lett. B}\ }\textbf {\bibinfo {volume} {260}},\
  \bibinfo {pages} {447} (\bibinfo {year} {1991})}\BibitemShut {NoStop}%
\bibitem [{\citenamefont {Langacker}\ and\ \citenamefont
  {Luo}(1991)}]{Langacker:1991an}%
  \BibitemOpen
  \bibfield  {author} {\bibinfo {author} {\bibfnamefont {P.}~\bibnamefont
  {Langacker}}\ and\ \bibinfo {author} {\bibfnamefont {M.-x.}\ \bibnamefont
  {Luo}},\ }\href {https://doi.org/10.1103/PhysRevD.44.817} {\bibfield
  {journal} {\bibinfo  {journal} {Phys. Rev. D}\ }\textbf {\bibinfo {volume}
  {44}},\ \bibinfo {pages} {817} (\bibinfo {year} {1991})}\BibitemShut
  {NoStop}%
\bibitem [{\citenamefont {Giunti}\ \emph {et~al.}(1991)\citenamefont {Giunti},
  \citenamefont {Kim},\ and\ \citenamefont {Lee}}]{Giunti:1991ta}%
  \BibitemOpen
  \bibfield  {author} {\bibinfo {author} {\bibfnamefont {C.}~\bibnamefont
  {Giunti}}, \bibinfo {author} {\bibfnamefont {C.~W.}\ \bibnamefont {Kim}},\
  and\ \bibinfo {author} {\bibfnamefont {U.~W.}\ \bibnamefont {Lee}},\ }\href
  {https://doi.org/10.1142/S0217732391001883} {\bibfield  {journal} {\bibinfo
  {journal} {Mod. Phys. Lett. A}\ }\textbf {\bibinfo {volume} {06}},\ \bibinfo
  {pages} {1745} (\bibinfo {year} {1991})}\BibitemShut {NoStop}%
\bibitem [{\citenamefont {Babu}\ and\ \citenamefont {Ma}(1984)}]{Babu:1984vx}%
  \BibitemOpen
  \bibfield  {author} {\bibinfo {author} {\bibfnamefont {K.~S.}\ \bibnamefont
  {Babu}}\ and\ \bibinfo {author} {\bibfnamefont {E.}~\bibnamefont {Ma}},\
  }\href {https://doi.org/10.1016/0370-2693(84)91283-8} {\bibfield  {journal}
  {\bibinfo  {journal} {Phys. Lett. B}\ }\textbf {\bibinfo {volume} {144}},\
  \bibinfo {pages} {381} (\bibinfo {year} {1984})}\BibitemShut {NoStop}%
\bibitem [{\citenamefont {Murayama}\ and\ \citenamefont
  {Yanagida}(1992)}]{Murayama:1991ah}%
  \BibitemOpen
  \bibfield  {author} {\bibinfo {author} {\bibfnamefont {H.}~\bibnamefont
  {Murayama}}\ and\ \bibinfo {author} {\bibfnamefont {T.}~\bibnamefont
  {Yanagida}},\ }\href {https://doi.org/10.1142/S0217732392000070} {\bibfield
  {journal} {\bibinfo  {journal} {Mod. Phys. Lett. A}\ }\textbf {\bibinfo
  {volume} {07}},\ \bibinfo {pages} {147} (\bibinfo {year} {1992})}\BibitemShut
  {NoStop}%
\bibitem [{\citenamefont {Giveon}\ \emph {et~al.}(1991)\citenamefont {Giveon},
  \citenamefont {Hall},\ and\ \citenamefont {Sarid}}]{Giveon:1991zm}%
  \BibitemOpen
  \bibfield  {author} {\bibinfo {author} {\bibfnamefont {A.}~\bibnamefont
  {Giveon}}, \bibinfo {author} {\bibfnamefont {L.~J.}\ \bibnamefont {Hall}},\
  and\ \bibinfo {author} {\bibfnamefont {U.}~\bibnamefont {Sarid}},\ }\href
  {https://doi.org/10.1016/0370-2693(91)91289-8} {\bibfield  {journal}
  {\bibinfo  {journal} {Phys. Lett. B}\ }\textbf {\bibinfo {volume} {271}},\
  \bibinfo {pages} {138} (\bibinfo {year} {1991})}\BibitemShut {NoStop}%
\bibitem [{\citenamefont {Dorsner}\ and\ \citenamefont
  {Fileviez~Perez}(2005)}]{Dorsner:2005fq}%
  \BibitemOpen
  \bibfield  {author} {\bibinfo {author} {\bibfnamefont {I.}~\bibnamefont
  {Dorsner}}\ and\ \bibinfo {author} {\bibfnamefont {P.}~\bibnamefont
  {Fileviez~Perez}},\ }\href {https://doi.org/10.1016/j.nuclphysb.2005.06.016}
  {\bibfield  {journal} {\bibinfo  {journal} {Nucl. Phys. B}\ }\textbf
  {\bibinfo {volume} {723}},\ \bibinfo {pages} {53} (\bibinfo {year} {2005})},\
  \Eprint {https://arxiv.org/abs/hep-ph/0504276} {arXiv:hep-ph/0504276}
  \BibitemShut {NoStop}%
\bibitem [{\citenamefont {Dorsner}\ \emph {et~al.}(2006)\citenamefont
  {Dorsner}, \citenamefont {Fileviez~Perez},\ and\ \citenamefont
  {Gonzalez~Felipe}}]{Dorsner:2005ii}%
  \BibitemOpen
  \bibfield  {author} {\bibinfo {author} {\bibfnamefont {I.}~\bibnamefont
  {Dorsner}}, \bibinfo {author} {\bibfnamefont {P.}~\bibnamefont
  {Fileviez~Perez}},\ and\ \bibinfo {author} {\bibfnamefont {R.}~\bibnamefont
  {Gonzalez~Felipe}},\ }\href {https://doi.org/10.1016/j.nuclphysb.2006.05.006}
  {\bibfield  {journal} {\bibinfo  {journal} {Nucl. Phys. B}\ }\textbf
  {\bibinfo {volume} {747}},\ \bibinfo {pages} {312} (\bibinfo {year}
  {2006})},\ \Eprint {https://arxiv.org/abs/hep-ph/0512068}
  {arXiv:hep-ph/0512068} \BibitemShut {NoStop}%
\bibitem [{\citenamefont {Dorsner}\ and\ \citenamefont
  {Fileviez~Perez}(2006)}]{Dorsner:2006dj}%
  \BibitemOpen
  \bibfield  {author} {\bibinfo {author} {\bibfnamefont {I.}~\bibnamefont
  {Dorsner}}\ and\ \bibinfo {author} {\bibfnamefont {P.}~\bibnamefont
  {Fileviez~Perez}},\ }\href {https://doi.org/10.1016/j.physletb.2006.09.034}
  {\bibfield  {journal} {\bibinfo  {journal} {Phys. Lett. B}\ }\textbf
  {\bibinfo {volume} {642}},\ \bibinfo {pages} {248} (\bibinfo {year}
  {2006})},\ \Eprint {https://arxiv.org/abs/hep-ph/0606062}
  {arXiv:hep-ph/0606062} \BibitemShut {NoStop}%
\bibitem [{\citenamefont {Dorsner}\ \emph {et~al.}(2007)\citenamefont
  {Dorsner}, \citenamefont {Fileviez~Perez},\ and\ \citenamefont
  {Rodrigo}}]{Dorsner:2006hw}%
  \BibitemOpen
  \bibfield  {author} {\bibinfo {author} {\bibfnamefont {I.}~\bibnamefont
  {Dorsner}}, \bibinfo {author} {\bibfnamefont {P.}~\bibnamefont
  {Fileviez~Perez}},\ and\ \bibinfo {author} {\bibfnamefont {G.}~\bibnamefont
  {Rodrigo}},\ }\href {https://doi.org/10.1103/PhysRevD.75.125007} {\bibfield
  {journal} {\bibinfo  {journal} {Phys. Rev. D}\ }\textbf {\bibinfo {volume}
  {75}},\ \bibinfo {pages} {125007} (\bibinfo {year} {2007})},\ \Eprint
  {https://arxiv.org/abs/hep-ph/0607208} {arXiv:hep-ph/0607208} \BibitemShut
  {NoStop}%
\bibitem [{\citenamefont {Bajc}\ and\ \citenamefont
  {Senjanovic}(2007)}]{Bajc:2006ia}%
  \BibitemOpen
  \bibfield  {author} {\bibinfo {author} {\bibfnamefont {B.}~\bibnamefont
  {Bajc}}\ and\ \bibinfo {author} {\bibfnamefont {G.}~\bibnamefont
  {Senjanovic}},\ }\href {https://doi.org/10.1088/1126-6708/2007/08/014}
  {\bibfield  {journal} {\bibinfo  {journal} {J. High Energy Phys.}\ }\textbf
  {\bibinfo {volume} {08}},\ \bibinfo {pages} {014 (2007)}},\ \Eprint
  {https://arxiv.org/abs/hep-ph/0612029} {arXiv:hep-ph/0612029} \BibitemShut
  {NoStop}%
\bibitem [{\citenamefont {Fileviez~Perez}(2007)}]{FileviezPerez:2007bcw}%
  \BibitemOpen
  \bibfield  {author} {\bibinfo {author} {\bibfnamefont {P.}~\bibnamefont
  {Fileviez~Perez}},\ }\href {https://doi.org/10.1016/j.physletb.2007.07.075}
  {\bibfield  {journal} {\bibinfo  {journal} {Phys. Lett. B}\ }\textbf
  {\bibinfo {volume} {654}},\ \bibinfo {pages} {189} (\bibinfo {year}
  {2007})},\ \Eprint {https://arxiv.org/abs/hep-ph/0702287}
  {arXiv:hep-ph/0702287} \BibitemShut {NoStop}%
\bibitem [{\citenamefont {Dorsner}\ and\ \citenamefont
  {Mocioiu}(2008)}]{Dorsner:2007fy}%
  \BibitemOpen
  \bibfield  {author} {\bibinfo {author} {\bibfnamefont {I.}~\bibnamefont
  {Dorsner}}\ and\ \bibinfo {author} {\bibfnamefont {I.}~\bibnamefont
  {Mocioiu}},\ }\href {https://doi.org/10.1016/j.nuclphysb.2007.12.004}
  {\bibfield  {journal} {\bibinfo  {journal} {Nucl. Phys. B}\ }\textbf
  {\bibinfo {volume} {796}},\ \bibinfo {pages} {123} (\bibinfo {year}
  {2008})},\ \Eprint {https://arxiv.org/abs/0708.3332} {arXiv:0708.3332
  [hep-ph]} \BibitemShut {NoStop}%
\bibitem [{\citenamefont {Fileviez~Perez}\ \emph {et~al.}(2008)\citenamefont
  {Fileviez~Perez}, \citenamefont {Iminniyaz},\ and\ \citenamefont
  {Rodrigo}}]{FileviezPerez:2008afb}%
  \BibitemOpen
  \bibfield  {author} {\bibinfo {author} {\bibfnamefont {P.}~\bibnamefont
  {Fileviez~Perez}}, \bibinfo {author} {\bibfnamefont {H.}~\bibnamefont
  {Iminniyaz}},\ and\ \bibinfo {author} {\bibfnamefont {G.}~\bibnamefont
  {Rodrigo}},\ }\href {https://doi.org/10.1103/PhysRevD.78.015013} {\bibfield
  {journal} {\bibinfo  {journal} {Phys. Rev. D}\ }\textbf {\bibinfo {volume}
  {78}},\ \bibinfo {pages} {015013} (\bibinfo {year} {2008})},\ \Eprint
  {https://arxiv.org/abs/0803.4156} {arXiv:0803.4156 [hep-ph]} \BibitemShut
  {NoStop}%
\bibitem [{\citenamefont {Dorsner}\ \emph {et~al.}(2010)\citenamefont
  {Dorsner}, \citenamefont {Fajfer}, \citenamefont {Kamenik},\ and\
  \citenamefont {Kosnik}}]{Dorsner:2009mq}%
  \BibitemOpen
  \bibfield  {author} {\bibinfo {author} {\bibfnamefont {I.}~\bibnamefont
  {Dorsner}}, \bibinfo {author} {\bibfnamefont {S.}~\bibnamefont {Fajfer}},
  \bibinfo {author} {\bibfnamefont {J.~F.}\ \bibnamefont {Kamenik}},\ and\
  \bibinfo {author} {\bibfnamefont {N.}~\bibnamefont {Kosnik}},\ }\href
  {https://doi.org/10.1103/PhysRevD.81.055009} {\bibfield  {journal} {\bibinfo
  {journal} {Phys. Rev. D}\ }\textbf {\bibinfo {volume} {81}},\ \bibinfo
  {pages} {055009} (\bibinfo {year} {2010})},\ \Eprint
  {https://arxiv.org/abs/0912.0972} {arXiv:0912.0972 [hep-ph]} \BibitemShut
  {NoStop}%
\bibitem [{\citenamefont {Fileviez~Perez}\ and\ \citenamefont
  {Murgui}(2016)}]{FileviezPerez:2016sal}%
  \BibitemOpen
  \bibfield  {author} {\bibinfo {author} {\bibfnamefont {P.}~\bibnamefont
  {Fileviez~Perez}}\ and\ \bibinfo {author} {\bibfnamefont {C.}~\bibnamefont
  {Murgui}},\ }\href {https://doi.org/10.1103/PhysRevD.94.075014} {\bibfield
  {journal} {\bibinfo  {journal} {Phys. Rev. D}\ }\textbf {\bibinfo {volume}
  {94}},\ \bibinfo {pages} {075014} (\bibinfo {year} {2016})},\ \Eprint
  {https://arxiv.org/abs/1604.03377} {arXiv:1604.03377 [hep-ph]} \BibitemShut
  {NoStop}%
\bibitem [{\citenamefont {Cox}\ \emph {et~al.}(2017)\citenamefont {Cox},
  \citenamefont {Kusenko}, \citenamefont {Sumensari},\ and\ \citenamefont
  {Yanagida}}]{Cox:2016epl}%
  \BibitemOpen
  \bibfield  {author} {\bibinfo {author} {\bibfnamefont {P.}~\bibnamefont
  {Cox}}, \bibinfo {author} {\bibfnamefont {A.}~\bibnamefont {Kusenko}},
  \bibinfo {author} {\bibfnamefont {O.}~\bibnamefont {Sumensari}},\ and\
  \bibinfo {author} {\bibfnamefont {T.~T.}\ \bibnamefont {Yanagida}},\ }\href
  {https://doi.org/10.1007/JHEP03(2017)035} {\bibfield  {journal} {\bibinfo
  {journal} {J. High Energy Phys.}\ }\textbf {\bibinfo {volume} {03}},\
  \bibinfo {pages} {035 (2017)}},\ \Eprint {https://arxiv.org/abs/1612.03923}
  {arXiv:1612.03923 [hep-ph]} \BibitemShut {NoStop}%
\bibitem [{\citenamefont {Dor\v{s}ner}\ \emph
  {et~al.}(2017{\natexlab{a}})\citenamefont {Dor\v{s}ner}, \citenamefont
  {Fajfer},\ and\ \citenamefont {Ko\v{s}nik}}]{Dorsner:2017wwn}%
  \BibitemOpen
  \bibfield  {author} {\bibinfo {author} {\bibfnamefont {I.}~\bibnamefont
  {Dor\v{s}ner}}, \bibinfo {author} {\bibfnamefont {S.}~\bibnamefont
  {Fajfer}},\ and\ \bibinfo {author} {\bibfnamefont {N.}~\bibnamefont
  {Ko\v{s}nik}},\ }\href {https://doi.org/10.1140/epjc/s10052-017-4987-2}
  {\bibfield  {journal} {\bibinfo  {journal} {Eur. Phys. J. C}\ }\textbf
  {\bibinfo {volume} {77}},\ \bibinfo {pages} {417} (\bibinfo {year}
  {2017}{\natexlab{a}})},\ \Eprint {https://arxiv.org/abs/1701.08322}
  {arXiv:1701.08322 [hep-ph]} \BibitemShut {NoStop}%
\bibitem [{\citenamefont {Be\v{c}irevi\'c}\ \emph {et~al.}(2018)\citenamefont
  {Be\v{c}irevi\'c}, \citenamefont {Dor\v{s}ner}, \citenamefont {Fajfer},
  \citenamefont {Faroughy}, \citenamefont {Ko\v{s}nik},\ and\ \citenamefont
  {Sumensari}}]{Becirevic:2018afm}%
  \BibitemOpen
  \bibfield  {author} {\bibinfo {author} {\bibfnamefont {D.}~\bibnamefont
  {Be\v{c}irevi\'c}}, \bibinfo {author} {\bibfnamefont {I.}~\bibnamefont
  {Dor\v{s}ner}}, \bibinfo {author} {\bibfnamefont {S.}~\bibnamefont {Fajfer}},
  \bibinfo {author} {\bibfnamefont {D.~A.}\ \bibnamefont {Faroughy}}, \bibinfo
  {author} {\bibfnamefont {N.}~\bibnamefont {Ko\v{s}nik}},\ and\ \bibinfo
  {author} {\bibfnamefont {O.}~\bibnamefont {Sumensari}},\ }\href
  {https://doi.org/10.1103/PhysRevD.98.055003} {\bibfield  {journal} {\bibinfo
  {journal} {Phys. Rev. D}\ }\textbf {\bibinfo {volume} {98}},\ \bibinfo
  {pages} {055003} (\bibinfo {year} {2018})},\ \Eprint
  {https://arxiv.org/abs/1806.05689} {arXiv:1806.05689 [hep-ph]} \BibitemShut
  {NoStop}%
\bibitem [{\citenamefont {Schwichtenberg}(2019)}]{Schwichtenberg:2018cka}%
  \BibitemOpen
  \bibfield  {author} {\bibinfo {author} {\bibfnamefont {J.}~\bibnamefont
  {Schwichtenberg}},\ }\href {https://doi.org/10.1140/epjc/s10052-019-6878-1}
  {\bibfield  {journal} {\bibinfo  {journal} {Eur. Phys. J. C}\ }\textbf
  {\bibinfo {volume} {79}},\ \bibinfo {pages} {351} (\bibinfo {year} {2019})},\
  \Eprint {https://arxiv.org/abs/1808.10329} {arXiv:1808.10329 [hep-ph]}
  \BibitemShut {NoStop}%
\bibitem [{\citenamefont {Haba}\ \emph {et~al.}(2019)\citenamefont {Haba},
  \citenamefont {Mimura},\ and\ \citenamefont {Yamada}}]{Haba:2018vvu}%
  \BibitemOpen
  \bibfield  {author} {\bibinfo {author} {\bibfnamefont {N.}~\bibnamefont
  {Haba}}, \bibinfo {author} {\bibfnamefont {Y.}~\bibnamefont {Mimura}},\ and\
  \bibinfo {author} {\bibfnamefont {T.}~\bibnamefont {Yamada}},\ }\href
  {https://doi.org/10.1103/PhysRevD.99.075018} {\bibfield  {journal} {\bibinfo
  {journal} {Phys. Rev. D}\ }\textbf {\bibinfo {volume} {99}},\ \bibinfo
  {pages} {075018} (\bibinfo {year} {2019})},\ \Eprint
  {https://arxiv.org/abs/1812.08521} {arXiv:1812.08521 [hep-ph]} \BibitemShut
  {NoStop}%
\bibitem [{\citenamefont {Georgi}\ and\ \citenamefont
  {Jarlskog}(1979)}]{Georgi:1979df}%
  \BibitemOpen
  \bibfield  {author} {\bibinfo {author} {\bibfnamefont {H.}~\bibnamefont
  {Georgi}}\ and\ \bibinfo {author} {\bibfnamefont {C.}~\bibnamefont
  {Jarlskog}},\ }\href {https://doi.org/10.1016/0370-2693(79)90842-6}
  {\bibfield  {journal} {\bibinfo  {journal} {Phys. Lett. B}\ }\textbf
  {\bibinfo {volume} {86}},\ \bibinfo {pages} {297} (\bibinfo {year}
  {1979})}\BibitemShut {NoStop}%
\bibitem [{\citenamefont {Rahat}\ \emph {et~al.}(2018)\citenamefont {Rahat},
  \citenamefont {Ramond},\ and\ \citenamefont {Xu}}]{Rahat:2018sgs}%
  \BibitemOpen
  \bibfield  {author} {\bibinfo {author} {\bibfnamefont {M.~H.}\ \bibnamefont
  {Rahat}}, \bibinfo {author} {\bibfnamefont {P.}~\bibnamefont {Ramond}},\ and\
  \bibinfo {author} {\bibfnamefont {B.}~\bibnamefont {Xu}},\ }\href
  {https://doi.org/10.1103/PhysRevD.98.055030} {\bibfield  {journal} {\bibinfo
  {journal} {Phys. Rev. D}\ }\textbf {\bibinfo {volume} {98}},\ \bibinfo
  {pages} {055030} (\bibinfo {year} {2018})},\ \Eprint
  {https://arxiv.org/abs/1805.10684} {arXiv:1805.10684 [hep-ph]} \BibitemShut
  {NoStop}%
\bibitem [{\citenamefont {P\'erez}\ \emph {et~al.}(2019)\citenamefont
  {P\'erez}, \citenamefont {Rahat}, \citenamefont {Ramond}, \citenamefont
  {Stuart},\ and\ \citenamefont {Xu}}]{Perez:2019aqq}%
  \BibitemOpen
  \bibfield  {author} {\bibinfo {author} {\bibfnamefont {M.~J.}\ \bibnamefont
  {P\'erez}}, \bibinfo {author} {\bibfnamefont {M.~H.}\ \bibnamefont {Rahat}},
  \bibinfo {author} {\bibfnamefont {P.}~\bibnamefont {Ramond}}, \bibinfo
  {author} {\bibfnamefont {A.~J.}\ \bibnamefont {Stuart}},\ and\ \bibinfo
  {author} {\bibfnamefont {B.}~\bibnamefont {Xu}},\ }\href
  {https://doi.org/10.1103/PhysRevD.100.075008} {\bibfield  {journal} {\bibinfo
   {journal} {Phys. Rev. D}\ }\textbf {\bibinfo {volume} {100}},\ \bibinfo
  {pages} {075008} (\bibinfo {year} {2019})},\ \Eprint
  {https://arxiv.org/abs/1907.10698} {arXiv:1907.10698 [hep-ph]} \BibitemShut
  {NoStop}%
\bibitem [{\citenamefont {P\'erez}\ \emph {et~al.}(2020)\citenamefont
  {P\'erez}, \citenamefont {Rahat}, \citenamefont {Ramond}, \citenamefont
  {Stuart},\ and\ \citenamefont {Xu}}]{Perez:2020nqq}%
  \BibitemOpen
  \bibfield  {author} {\bibinfo {author} {\bibfnamefont {M.~J.}\ \bibnamefont
  {P\'erez}}, \bibinfo {author} {\bibfnamefont {M.~H.}\ \bibnamefont {Rahat}},
  \bibinfo {author} {\bibfnamefont {P.}~\bibnamefont {Ramond}}, \bibinfo
  {author} {\bibfnamefont {A.~J.}\ \bibnamefont {Stuart}},\ and\ \bibinfo
  {author} {\bibfnamefont {B.}~\bibnamefont {Xu}},\ }\href
  {https://doi.org/10.1103/PhysRevD.101.075018} {\bibfield  {journal} {\bibinfo
   {journal} {Phys. Rev. D}\ }\textbf {\bibinfo {volume} {101}},\ \bibinfo
  {pages} {075018} (\bibinfo {year} {2020})},\ \Eprint
  {https://arxiv.org/abs/2001.04019} {arXiv:2001.04019 [hep-ph]} \BibitemShut
  {NoStop}%
\bibitem [{\citenamefont {Buchmuller}\ \emph {et~al.}(1987)\citenamefont
  {Buchmuller}, \citenamefont {Ruckl},\ and\ \citenamefont
  {Wyler}}]{Buchmuller:1986zs}%
  \BibitemOpen
  \bibfield  {author} {\bibinfo {author} {\bibfnamefont {W.}~\bibnamefont
  {Buchmuller}}, \bibinfo {author} {\bibfnamefont {R.}~\bibnamefont {Ruckl}},\
  and\ \bibinfo {author} {\bibfnamefont {D.}~\bibnamefont {Wyler}},\ }\href
  {https://doi.org/10.1016/0370-2693(87)90637-X} {\bibfield  {journal}
  {\bibinfo  {journal} {Phys. Lett. B}\ }\textbf {\bibinfo {volume} {191}},\
  \bibinfo {pages} {442} (\bibinfo {year} {1987})},\ \bibinfo {note} {[Erratum:
  Phys. Lett. B 448, 320 (1999)]}\BibitemShut {NoStop}%
\bibitem [{\citenamefont {Dor\v{s}ner}\ \emph {et~al.}(2016)\citenamefont
  {Dor\v{s}ner}, \citenamefont {Fajfer}, \citenamefont {Greljo}, \citenamefont
  {Kamenik},\ and\ \citenamefont {Ko\v{s}nik}}]{Dorsner:2016wpm}%
  \BibitemOpen
  \bibfield  {author} {\bibinfo {author} {\bibfnamefont {I.}~\bibnamefont
  {Dor\v{s}ner}}, \bibinfo {author} {\bibfnamefont {S.}~\bibnamefont {Fajfer}},
  \bibinfo {author} {\bibfnamefont {A.}~\bibnamefont {Greljo}}, \bibinfo
  {author} {\bibfnamefont {J.~F.}\ \bibnamefont {Kamenik}},\ and\ \bibinfo
  {author} {\bibfnamefont {N.}~\bibnamefont {Ko\v{s}nik}},\ }\href
  {https://doi.org/10.1016/j.physrep.2016.06.001} {\bibfield  {journal}
  {\bibinfo  {journal} {Phys. Rept.}\ }\textbf {\bibinfo {volume} {641}},\
  \bibinfo {pages} {1} (\bibinfo {year} {2016})},\ \Eprint
  {https://arxiv.org/abs/1603.04993} {arXiv:1603.04993 [hep-ph]} \BibitemShut
  {NoStop}%
\bibitem [{\citenamefont {Sakaki}\ \emph {et~al.}(2013)\citenamefont {Sakaki},
  \citenamefont {Watanabe}, \citenamefont {Tanaka},\ and\ \citenamefont
  {Tayduganov}}]{Sakaki:2013bfa}%
  \BibitemOpen
  \bibfield  {author} {\bibinfo {author} {\bibfnamefont {Y.}~\bibnamefont
  {Sakaki}}, \bibinfo {author} {\bibfnamefont {R.}~\bibnamefont {Watanabe}},
  \bibinfo {author} {\bibfnamefont {M.}~\bibnamefont {Tanaka}},\ and\ \bibinfo
  {author} {\bibfnamefont {A.}~\bibnamefont {Tayduganov}},\ }\href
  {https://doi.org/10.1103/PhysRevD.88.094012} {\bibfield  {journal} {\bibinfo
  {journal} {Phys. Rev. D}\ }\textbf {\bibinfo {volume} {88}},\ \bibinfo
  {pages} {094012} (\bibinfo {year} {2013})},\ \Eprint
  {https://arxiv.org/abs/1309.0301} {arXiv:1309.0301 [hep-ph]} \BibitemShut
  {NoStop}%
\bibitem [{\citenamefont {Hiller}\ and\ \citenamefont
  {Schmaltz}(2014)}]{Hiller:2014yaa}%
  \BibitemOpen
  \bibfield  {author} {\bibinfo {author} {\bibfnamefont {G.}~\bibnamefont
  {Hiller}}\ and\ \bibinfo {author} {\bibfnamefont {M.}~\bibnamefont
  {Schmaltz}},\ }\href {https://doi.org/10.1103/PhysRevD.90.054014} {\bibfield
  {journal} {\bibinfo  {journal} {Phys. Rev. D}\ }\textbf {\bibinfo {volume}
  {90}},\ \bibinfo {pages} {054014} (\bibinfo {year} {2014})},\ \Eprint
  {https://arxiv.org/abs/1408.1627} {arXiv:1408.1627 [hep-ph]} \BibitemShut
  {NoStop}%
\bibitem [{\citenamefont {Alok}\ \emph {et~al.}(2017)\citenamefont {Alok},
  \citenamefont {Bhattacharya}, \citenamefont {Kumar}, \citenamefont {Kumar},
  \citenamefont {London},\ and\ \citenamefont {Sankar}}]{Alok:2017jgr}%
  \BibitemOpen
  \bibfield  {author} {\bibinfo {author} {\bibfnamefont {A.~K.}\ \bibnamefont
  {Alok}}, \bibinfo {author} {\bibfnamefont {B.}~\bibnamefont {Bhattacharya}},
  \bibinfo {author} {\bibfnamefont {D.}~\bibnamefont {Kumar}}, \bibinfo
  {author} {\bibfnamefont {J.}~\bibnamefont {Kumar}}, \bibinfo {author}
  {\bibfnamefont {D.}~\bibnamefont {London}},\ and\ \bibinfo {author}
  {\bibfnamefont {S.~U.}\ \bibnamefont {Sankar}},\ }\href
  {https://doi.org/10.1103/PhysRevD.96.015034} {\bibfield  {journal} {\bibinfo
  {journal} {Phys. Rev. D}\ }\textbf {\bibinfo {volume} {96}},\ \bibinfo
  {pages} {015034} (\bibinfo {year} {2017})},\ \Eprint
  {https://arxiv.org/abs/1703.09247} {arXiv:1703.09247 [hep-ph]} \BibitemShut
  {NoStop}%
\bibitem [{\citenamefont {Dor\v{s}ner}\ \emph
  {et~al.}(2017{\natexlab{b}})\citenamefont {Dor\v{s}ner}, \citenamefont
  {Fajfer}, \citenamefont {Faroughy},\ and\ \citenamefont
  {Ko\v{s}nik}}]{Dorsner:2017ufx}%
  \BibitemOpen
  \bibfield  {author} {\bibinfo {author} {\bibfnamefont {I.}~\bibnamefont
  {Dor\v{s}ner}}, \bibinfo {author} {\bibfnamefont {S.}~\bibnamefont {Fajfer}},
  \bibinfo {author} {\bibfnamefont {D.~A.}\ \bibnamefont {Faroughy}},\ and\
  \bibinfo {author} {\bibfnamefont {N.}~\bibnamefont {Ko\v{s}nik}},\ }\href
  {https://doi.org/10.1007/JHEP10(2017)188} {\bibfield  {journal} {\bibinfo
  {journal} {J. High Energy Phys.}\ }\textbf {\bibinfo {volume} {10}},\
  \bibinfo {pages} {188 (2017)}},\ \Eprint {https://arxiv.org/abs/1706.07779}
  {arXiv:1706.07779 [hep-ph]} \BibitemShut {NoStop}%
\bibitem [{\citenamefont {Di~Luzio}\ \emph {et~al.}(2018)\citenamefont
  {Di~Luzio}, \citenamefont {Kirk},\ and\ \citenamefont
  {Lenz}}]{DiLuzio:2017fdq}%
  \BibitemOpen
  \bibfield  {author} {\bibinfo {author} {\bibfnamefont {L.}~\bibnamefont
  {Di~Luzio}}, \bibinfo {author} {\bibfnamefont {M.}~\bibnamefont {Kirk}},\
  and\ \bibinfo {author} {\bibfnamefont {A.}~\bibnamefont {Lenz}},\ }\href
  {https://doi.org/10.1103/PhysRevD.97.095035} {\bibfield  {journal} {\bibinfo
  {journal} {Phys. Rev. D}\ }\textbf {\bibinfo {volume} {97}},\ \bibinfo
  {pages} {095035} (\bibinfo {year} {2018})},\ \Eprint
  {https://arxiv.org/abs/1712.06572} {arXiv:1712.06572 [hep-ph]} \BibitemShut
  {NoStop}%
\bibitem [{\citenamefont {Fajfer}\ \emph {et~al.}(2018)\citenamefont {Fajfer},
  \citenamefont {Ko\v{s}nik},\ and\ \citenamefont
  {Vale~Silva}}]{Fajfer:2018bfj}%
  \BibitemOpen
  \bibfield  {author} {\bibinfo {author} {\bibfnamefont {S.}~\bibnamefont
  {Fajfer}}, \bibinfo {author} {\bibfnamefont {N.}~\bibnamefont {Ko\v{s}nik}},\
  and\ \bibinfo {author} {\bibfnamefont {L.}~\bibnamefont {Vale~Silva}},\
  }\href {https://doi.org/10.1140/epjc/s10052-018-5757-5} {\bibfield  {journal}
  {\bibinfo  {journal} {Eur. Phys. J. C}\ }\textbf {\bibinfo {volume} {78}},\
  \bibinfo {pages} {275} (\bibinfo {year} {2018})},\ \Eprint
  {https://arxiv.org/abs/1802.00786} {arXiv:1802.00786 [hep-ph]} \BibitemShut
  {NoStop}%
\bibitem [{\citenamefont {Alda}\ \emph {et~al.}(2019)\citenamefont {Alda},
  \citenamefont {Guasch},\ and\ \citenamefont {Penaranda}}]{Alda:2018mfy}%
  \BibitemOpen
  \bibfield  {author} {\bibinfo {author} {\bibfnamefont {J.}~\bibnamefont
  {Alda}}, \bibinfo {author} {\bibfnamefont {J.}~\bibnamefont {Guasch}},\ and\
  \bibinfo {author} {\bibfnamefont {S.}~\bibnamefont {Penaranda}},\ }\href
  {https://doi.org/10.1140/epjc/s10052-019-7092-x} {\bibfield  {journal}
  {\bibinfo  {journal} {Eur. Phys. J. C}\ }\textbf {\bibinfo {volume} {79}},\
  \bibinfo {pages} {588} (\bibinfo {year} {2019})},\ \Eprint
  {https://arxiv.org/abs/1805.03636} {arXiv:1805.03636 [hep-ph]} \BibitemShut
  {NoStop}%
\bibitem [{\citenamefont {Mandal}\ and\ \citenamefont
  {Pich}(2019)}]{Mandal:2019gff}%
  \BibitemOpen
  \bibfield  {author} {\bibinfo {author} {\bibfnamefont {R.}~\bibnamefont
  {Mandal}}\ and\ \bibinfo {author} {\bibfnamefont {A.}~\bibnamefont {Pich}},\
  }\href {https://doi.org/10.1007/JHEP12(2019)089} {\bibfield  {journal}
  {\bibinfo  {journal} {J. High Energy Phys.}\ }\textbf {\bibinfo {volume}
  {12}},\ \bibinfo {pages} {089 (2019)}},\ \Eprint
  {https://arxiv.org/abs/1908.11155} {arXiv:1908.11155 [hep-ph]} \BibitemShut
  {NoStop}%
\bibitem [{\citenamefont {Di~Luzio}\ \emph {et~al.}(2019)\citenamefont
  {Di~Luzio}, \citenamefont {Kirk}, \citenamefont {Lenz},\ and\ \citenamefont
  {Rauh}}]{DiLuzio:2019jyq}%
  \BibitemOpen
  \bibfield  {author} {\bibinfo {author} {\bibfnamefont {L.}~\bibnamefont
  {Di~Luzio}}, \bibinfo {author} {\bibfnamefont {M.}~\bibnamefont {Kirk}},
  \bibinfo {author} {\bibfnamefont {A.}~\bibnamefont {Lenz}},\ and\ \bibinfo
  {author} {\bibfnamefont {T.}~\bibnamefont {Rauh}},\ }\href
  {https://doi.org/10.1007/JHEP12(2019)009} {\bibfield  {journal} {\bibinfo
  {journal} {J. High Energy Phys.}\ }\textbf {\bibinfo {volume} {12}},\
  \bibinfo {pages} {009 (2019)}},\ \Eprint {https://arxiv.org/abs/1909.11087}
  {arXiv:1909.11087 [hep-ph]} \BibitemShut {NoStop}%
\bibitem [{\citenamefont {Angelescu}\ \emph {et~al.}(2021)\citenamefont
  {Angelescu}, \citenamefont {Be\v{c}irevi\'c}, \citenamefont {Faroughy},
  \citenamefont {Jaffredo},\ and\ \citenamefont
  {Sumensari}}]{Angelescu:2021lln}%
  \BibitemOpen
  \bibfield  {author} {\bibinfo {author} {\bibfnamefont {A.}~\bibnamefont
  {Angelescu}}, \bibinfo {author} {\bibfnamefont {D.}~\bibnamefont
  {Be\v{c}irevi\'c}}, \bibinfo {author} {\bibfnamefont {D.~A.}\ \bibnamefont
  {Faroughy}}, \bibinfo {author} {\bibfnamefont {F.}~\bibnamefont {Jaffredo}},\
  and\ \bibinfo {author} {\bibfnamefont {O.}~\bibnamefont {Sumensari}},\ }\href
  {https://doi.org/10.1103/PhysRevD.104.055017} {\bibfield  {journal} {\bibinfo
   {journal} {Phys. Rev. D}\ }\textbf {\bibinfo {volume} {104}},\ \bibinfo
  {pages} {055017} (\bibinfo {year} {2021})},\ \Eprint
  {https://arxiv.org/abs/2103.12504} {arXiv:2103.12504 [hep-ph]} \BibitemShut
  {NoStop}%
\bibitem [{\citenamefont {Crivellin}\ \emph
  {et~al.}(2021{\natexlab{a}})\citenamefont {Crivellin}, \citenamefont
  {M\"uller},\ and\ \citenamefont {Schnell}}]{Crivellin:2021egp}%
  \BibitemOpen
  \bibfield  {author} {\bibinfo {author} {\bibfnamefont {A.}~\bibnamefont
  {Crivellin}}, \bibinfo {author} {\bibfnamefont {D.}~\bibnamefont
  {M\"uller}},\ and\ \bibinfo {author} {\bibfnamefont {L.}~\bibnamefont
  {Schnell}},\ }\href {https://doi.org/10.1103/PhysRevD.103.115023} {\bibfield
  {journal} {\bibinfo  {journal} {Phys. Rev. D}\ }\textbf {\bibinfo {volume}
  {103}},\ \bibinfo {pages} {115023} (\bibinfo {year} {2021}{\natexlab{a}})},\
  \bibinfo {note} {[Addendum: Phys. Rev. D 104, 055020 (2021),
  arXiv:2104.06417]},\ \Eprint {https://arxiv.org/abs/2101.07811}
  {arXiv:2101.07811 [hep-ph]} \BibitemShut {NoStop}%
\bibitem [{\citenamefont {Ko\v{s}nik}\ and\ \citenamefont
  {Smolkovi\v{c}}(2021)}]{Kosnik:2021wyp}%
  \BibitemOpen
  \bibfield  {author} {\bibinfo {author} {\bibfnamefont {N.}~\bibnamefont
  {Ko\v{s}nik}}\ and\ \bibinfo {author} {\bibfnamefont {A.}~\bibnamefont
  {Smolkovi\v{c}}},\ }\href {https://doi.org/10.1103/PhysRevD.104.115004}
  {\bibfield  {journal} {\bibinfo  {journal} {Phys. Rev. D}\ }\textbf {\bibinfo
  {volume} {104}},\ \bibinfo {pages} {115004} (\bibinfo {year} {2021})},\
  \Eprint {https://arxiv.org/abs/2108.11929} {arXiv:2108.11929 [hep-ph]}
  \BibitemShut {NoStop}%
\bibitem [{\citenamefont {Chau}\ and\ \citenamefont
  {Keung}(1984)}]{Chau:1984fp}%
  \BibitemOpen
  \bibfield  {author} {\bibinfo {author} {\bibfnamefont {L.-L.}\ \bibnamefont
  {Chau}}\ and\ \bibinfo {author} {\bibfnamefont {W.-Y.}\ \bibnamefont
  {Keung}},\ }\href {https://doi.org/10.1103/PhysRevLett.53.1802} {\bibfield
  {journal} {\bibinfo  {journal} {Phys. Rev. Lett.}\ }\textbf {\bibinfo
  {volume} {53}},\ \bibinfo {pages} {1802} (\bibinfo {year}
  {1984})}\BibitemShut {NoStop}%
\bibitem [{\citenamefont {Zyla}\ \emph {et~al.}(2020)\citenamefont {Zyla} \emph
  {et~al.}}]{ParticleDataGroup:2020ssz}%
  \BibitemOpen
  \bibfield  {author} {\bibinfo {author} {\bibfnamefont {P.~A.}\ \bibnamefont
  {Zyla}} \emph {et~al.} (\bibinfo {collaboration} {Particle Data Group}),\
  }\href {https://doi.org/10.1093/ptep/ptaa104} {\bibfield  {journal} {\bibinfo
   {journal} {Prog. Theor. Exp. Phys.}\ }\textbf {\bibinfo {volume} {2020}},\
  \bibinfo {pages} {083C01} (\bibinfo {year} {2020})}\BibitemShut {NoStop}%
\bibitem [{\citenamefont {Hisano}\ \emph {et~al.}(1993)\citenamefont {Hisano},
  \citenamefont {Murayama},\ and\ \citenamefont {Yanagida}}]{Hisano:1992jj}%
  \BibitemOpen
  \bibfield  {author} {\bibinfo {author} {\bibfnamefont {J.}~\bibnamefont
  {Hisano}}, \bibinfo {author} {\bibfnamefont {H.}~\bibnamefont {Murayama}},\
  and\ \bibinfo {author} {\bibfnamefont {T.}~\bibnamefont {Yanagida}},\ }\href
  {https://doi.org/10.1016/0550-3213(93)90636-4} {\bibfield  {journal}
  {\bibinfo  {journal} {Nucl. Phys. B}\ }\textbf {\bibinfo {volume} {402}},\
  \bibinfo {pages} {46} (\bibinfo {year} {1993})},\ \Eprint
  {https://arxiv.org/abs/hep-ph/9207279} {arXiv:hep-ph/9207279} \BibitemShut
  {NoStop}%
\bibitem [{\citenamefont {Arason}\ \emph {et~al.}(1992)\citenamefont {Arason},
  \citenamefont {Castano}, \citenamefont {Kesthelyi}, \citenamefont
  {Mikaelian}, \citenamefont {Piard}, \citenamefont {Ramond},\ and\
  \citenamefont {Wright}}]{Arason:1991ic}%
  \BibitemOpen
  \bibfield  {author} {\bibinfo {author} {\bibfnamefont {H.}~\bibnamefont
  {Arason}}, \bibinfo {author} {\bibfnamefont {D.~J.}\ \bibnamefont {Castano}},
  \bibinfo {author} {\bibfnamefont {B.}~\bibnamefont {Kesthelyi}}, \bibinfo
  {author} {\bibfnamefont {S.}~\bibnamefont {Mikaelian}}, \bibinfo {author}
  {\bibfnamefont {E.~J.}\ \bibnamefont {Piard}}, \bibinfo {author}
  {\bibfnamefont {P.}~\bibnamefont {Ramond}},\ and\ \bibinfo {author}
  {\bibfnamefont {B.~D.}\ \bibnamefont {Wright}},\ }\href
  {https://doi.org/10.1103/PhysRevD.46.3945} {\bibfield  {journal} {\bibinfo
  {journal} {Phys. Rev. D}\ }\textbf {\bibinfo {volume} {46}},\ \bibinfo
  {pages} {3945} (\bibinfo {year} {1992})}\BibitemShut {NoStop}%
\bibitem [{\citenamefont {Nath}\ and\ \citenamefont
  {Fileviez~Perez}(2007)}]{Nath:2006ut}%
  \BibitemOpen
  \bibfield  {author} {\bibinfo {author} {\bibfnamefont {P.}~\bibnamefont
  {Nath}}\ and\ \bibinfo {author} {\bibfnamefont {P.}~\bibnamefont
  {Fileviez~Perez}},\ }\href {https://doi.org/10.1016/j.physrep.2007.02.010}
  {\bibfield  {journal} {\bibinfo  {journal} {Phys. Rept.}\ }\textbf {\bibinfo
  {volume} {441}},\ \bibinfo {pages} {191} (\bibinfo {year} {2007})},\ \Eprint
  {https://arxiv.org/abs/hep-ph/0601023} {arXiv:hep-ph/0601023 [hep-ph]}
  \BibitemShut {NoStop}%
\bibitem [{\citenamefont {Takenaka}\ \emph {et~al.}(2020)\citenamefont
  {Takenaka} \emph {et~al.}}]{Super-Kamiokande:2020wjk}%
  \BibitemOpen
  \bibfield  {author} {\bibinfo {author} {\bibfnamefont {A.}~\bibnamefont
  {Takenaka}} \emph {et~al.} (\bibinfo {collaboration} {Super-Kamiokande
  Collaboration}),\ }\href {https://doi.org/10.1103/PhysRevD.102.112011}
  {\bibfield  {journal} {\bibinfo  {journal} {Phys. Rev. D}\ }\textbf {\bibinfo
  {volume} {102}},\ \bibinfo {pages} {112011} (\bibinfo {year} {2020})},\
  \Eprint {https://arxiv.org/abs/2010.16098} {arXiv:2010.16098 [hep-ex]}
  \BibitemShut {NoStop}%
\bibitem [{\citenamefont {Chetyrkin}\ \emph {et~al.}(2000)\citenamefont
  {Chetyrkin}, \citenamefont {Kuhn},\ and\ \citenamefont
  {Steinhauser}}]{Chetyrkin:2000yt}%
  \BibitemOpen
  \bibfield  {author} {\bibinfo {author} {\bibfnamefont {K.~G.}\ \bibnamefont
  {Chetyrkin}}, \bibinfo {author} {\bibfnamefont {J.~H.}\ \bibnamefont
  {Kuhn}},\ and\ \bibinfo {author} {\bibfnamefont {M.}~\bibnamefont
  {Steinhauser}},\ }\href {https://doi.org/10.1016/S0010-4655(00)00155-7}
  {\bibfield  {journal} {\bibinfo  {journal} {Comput. Phys. Commun.}\ }\textbf
  {\bibinfo {volume} {133}},\ \bibinfo {pages} {43} (\bibinfo {year} {2000})},\
  \Eprint {https://arxiv.org/abs/hep-ph/0004189} {arXiv:hep-ph/0004189}
  \BibitemShut {NoStop}%
\bibitem [{\citenamefont {Herren}\ and\ \citenamefont
  {Steinhauser}(2018)}]{Herren:2017osy}%
  \BibitemOpen
  \bibfield  {author} {\bibinfo {author} {\bibfnamefont {F.}~\bibnamefont
  {Herren}}\ and\ \bibinfo {author} {\bibfnamefont {M.}~\bibnamefont
  {Steinhauser}},\ }\href {https://doi.org/10.1016/j.cpc.2017.11.014}
  {\bibfield  {journal} {\bibinfo  {journal} {Comput. Phys. Commun.}\ }\textbf
  {\bibinfo {volume} {224}},\ \bibinfo {pages} {333} (\bibinfo {year}
  {2018})},\ \Eprint {https://arxiv.org/abs/1703.03751} {arXiv:1703.03751
  [hep-ph]} \BibitemShut {NoStop}%
\bibitem [{\citenamefont {Juste~Rozas}(2023)}]{JusteRozas:2853694}%
  \BibitemOpen
  \bibfield  {author} {\bibinfo {author} {\bibfnamefont {A.}~\bibnamefont
  {Juste~Rozas}} (\bibinfo {collaboration} {on behalf of the ATLAS and CMS
  Collaborations}),\ }\href {https://cds.cern.ch/record/2853694} {\bibfield
  {journal} {\bibinfo  {journal} {ATL-PHYS-SLIDE-2023-034}\ } (\bibinfo {year}
  {2023})},\ \bibinfo {note} {talk at the 57th Rencontres de Moriond on
  Electroweak Interactions and Unified Theories, La Thuile, Mar.
  2023}\BibitemShut {NoStop}%
\bibitem [{\citenamefont {de~Blas}\ \emph {et~al.}(2015)\citenamefont
  {de~Blas}, \citenamefont {Chala}, \citenamefont {Perez-Victoria},\ and\
  \citenamefont {Santiago}}]{deBlas:2014mba}%
  \BibitemOpen
  \bibfield  {author} {\bibinfo {author} {\bibfnamefont {J.}~\bibnamefont
  {de~Blas}}, \bibinfo {author} {\bibfnamefont {M.}~\bibnamefont {Chala}},
  \bibinfo {author} {\bibfnamefont {M.}~\bibnamefont {Perez-Victoria}},\ and\
  \bibinfo {author} {\bibfnamefont {J.}~\bibnamefont {Santiago}},\ }\href
  {https://doi.org/10.1007/JHEP04(2015)078} {\bibfield  {journal} {\bibinfo
  {journal} {J. High Energy Phys.}\ }\textbf {\bibinfo {volume} {04}},\
  \bibinfo {pages} {078 (2015)}},\ \Eprint {https://arxiv.org/abs/1412.8480}
  {arXiv:1412.8480 [hep-ph]} \BibitemShut {NoStop}%
\bibitem [{\citenamefont {de~Blas}\ \emph {et~al.}(2018)\citenamefont
  {de~Blas}, \citenamefont {Criado}, \citenamefont {Perez-Victoria},\ and\
  \citenamefont {Santiago}}]{deBlas:2017xtg}%
  \BibitemOpen
  \bibfield  {author} {\bibinfo {author} {\bibfnamefont {J.}~\bibnamefont
  {de~Blas}}, \bibinfo {author} {\bibfnamefont {J.}~\bibnamefont {Criado}},
  \bibinfo {author} {\bibfnamefont {M.}~\bibnamefont {Perez-Victoria}},\ and\
  \bibinfo {author} {\bibfnamefont {J.}~\bibnamefont {Santiago}},\ }\href
  {https://doi.org/10.1007/JHEP03(2018)109} {\bibfield  {journal} {\bibinfo
  {journal} {J. High Energy Phys.}\ }\textbf {\bibinfo {volume} {03}},\
  \bibinfo {pages} {109 (2018)}},\ \Eprint {https://arxiv.org/abs/1711.10391}
  {arXiv:1711.10391 [hep-ph]} \BibitemShut {NoStop}%
\bibitem [{\citenamefont {Gherardi}\ \emph {et~al.}(2020)\citenamefont
  {Gherardi}, \citenamefont {Marzocca},\ and\ \citenamefont
  {Venturini}}]{Gherardi:2020det}%
  \BibitemOpen
  \bibfield  {author} {\bibinfo {author} {\bibfnamefont {V.}~\bibnamefont
  {Gherardi}}, \bibinfo {author} {\bibfnamefont {D.}~\bibnamefont {Marzocca}},\
  and\ \bibinfo {author} {\bibfnamefont {E.}~\bibnamefont {Venturini}},\ }\href
  {https://doi.org/10.1007/JHEP07(2020)225} {\bibfield  {journal} {\bibinfo
  {journal} {J. High Energy Phys.}\ }\textbf {\bibinfo {volume} {07}},\
  \bibinfo {pages} {225 (2020)}},\ \bibinfo {note} {[Erratum: J. High Energy
  Phys. 01, 006 (2021)]},\ \Eprint {https://arxiv.org/abs/2003.12525}
  {arXiv:2003.12525 [hep-ph]} \BibitemShut {NoStop}%
\bibitem [{\citenamefont {Jenkins}\ \emph {et~al.}(2013)\citenamefont
  {Jenkins}, \citenamefont {Manohar},\ and\ \citenamefont
  {Trott}}]{Jenkins:2013zja}%
  \BibitemOpen
  \bibfield  {author} {\bibinfo {author} {\bibfnamefont {E.~E.}\ \bibnamefont
  {Jenkins}}, \bibinfo {author} {\bibfnamefont {A.~V.}\ \bibnamefont
  {Manohar}},\ and\ \bibinfo {author} {\bibfnamefont {M.}~\bibnamefont
  {Trott}},\ }\href {https://doi.org/10.1007/JHEP10(2013)087} {\bibfield
  {journal} {\bibinfo  {journal} {J. High Energy Phys.}\ }\textbf {\bibinfo
  {volume} {10}},\ \bibinfo {pages} {087 (2013)}},\ \Eprint
  {https://arxiv.org/abs/1308.2627} {arXiv:1308.2627 [hep-ph]} \BibitemShut
  {NoStop}%
\bibitem [{\citenamefont {Jenkins}\ \emph {et~al.}(2014)\citenamefont
  {Jenkins}, \citenamefont {Manohar},\ and\ \citenamefont
  {Trott}}]{Jenkins:2013wua}%
  \BibitemOpen
  \bibfield  {author} {\bibinfo {author} {\bibfnamefont {E.~E.}\ \bibnamefont
  {Jenkins}}, \bibinfo {author} {\bibfnamefont {A.~V.}\ \bibnamefont
  {Manohar}},\ and\ \bibinfo {author} {\bibfnamefont {M.}~\bibnamefont
  {Trott}},\ }\href {https://doi.org/10.1007/JHEP01(2014)035} {\bibfield
  {journal} {\bibinfo  {journal} {J. High Energy Phys.}\ }\textbf {\bibinfo
  {volume} {01}},\ \bibinfo {pages} {035 (2014)}},\ \Eprint
  {https://arxiv.org/abs/1310.4838} {arXiv:1310.4838 [hep-ph]} \BibitemShut
  {NoStop}%
\bibitem [{\citenamefont {Alonso}\ \emph {et~al.}(2014)\citenamefont {Alonso},
  \citenamefont {Jenkins}, \citenamefont {Manohar},\ and\ \citenamefont
  {Trott}}]{Alonso:2013hga}%
  \BibitemOpen
  \bibfield  {author} {\bibinfo {author} {\bibfnamefont {R.}~\bibnamefont
  {Alonso}}, \bibinfo {author} {\bibfnamefont {E.~E.}\ \bibnamefont {Jenkins}},
  \bibinfo {author} {\bibfnamefont {A.~V.}\ \bibnamefont {Manohar}},\ and\
  \bibinfo {author} {\bibfnamefont {M.}~\bibnamefont {Trott}},\ }\href
  {https://doi.org/10.1007/JHEP04(2014)159} {\bibfield  {journal} {\bibinfo
  {journal} {J. High Energy Phys.}\ }\textbf {\bibinfo {volume} {04}},\
  \bibinfo {pages} {159 (2014)}},\ \Eprint {https://arxiv.org/abs/1312.2014}
  {arXiv:1312.2014 [hep-ph]} \BibitemShut {NoStop}%
\bibitem [{\citenamefont {Grzadkowski}\ \emph {et~al.}(2010)\citenamefont
  {Grzadkowski}, \citenamefont {Iskrzynski}, \citenamefont {Misiak},\ and\
  \citenamefont {Rosiek}}]{Grzadkowski:2010es}%
  \BibitemOpen
  \bibfield  {author} {\bibinfo {author} {\bibfnamefont {B.}~\bibnamefont
  {Grzadkowski}}, \bibinfo {author} {\bibfnamefont {M.}~\bibnamefont
  {Iskrzynski}}, \bibinfo {author} {\bibfnamefont {M.}~\bibnamefont {Misiak}},\
  and\ \bibinfo {author} {\bibfnamefont {J.}~\bibnamefont {Rosiek}},\ }\href
  {https://doi.org/10.1007/JHEP10(2010)085} {\bibfield  {journal} {\bibinfo
  {journal} {J. High Energy Phys.}\ }\textbf {\bibinfo {volume} {10}},\
  \bibinfo {pages} {085 (2010)}},\ \Eprint {https://arxiv.org/abs/1008.4884}
  {arXiv:1008.4884 [hep-ph]} \BibitemShut {NoStop}%
\bibitem [{\citenamefont {Jenkins}\ \emph
  {et~al.}(2018{\natexlab{a}})\citenamefont {Jenkins}, \citenamefont
  {Manohar},\ and\ \citenamefont {Stoffer}}]{Jenkins:2017jig}%
  \BibitemOpen
  \bibfield  {author} {\bibinfo {author} {\bibfnamefont {E.~E.}\ \bibnamefont
  {Jenkins}}, \bibinfo {author} {\bibfnamefont {A.~V.}\ \bibnamefont
  {Manohar}},\ and\ \bibinfo {author} {\bibfnamefont {P.}~\bibnamefont
  {Stoffer}},\ }\href {https://doi.org/10.1007/JHEP03(2018)016} {\bibfield
  {journal} {\bibinfo  {journal} {J. High Energy Phys.}\ }\textbf {\bibinfo
  {volume} {03}},\ \bibinfo {pages} {016 (2018)}},\ \Eprint
  {https://arxiv.org/abs/1709.04486} {arXiv:1709.04486 [hep-ph]} \BibitemShut
  {NoStop}%
\bibitem [{\citenamefont {Aebischer}\ \emph {et~al.}(2016)\citenamefont
  {Aebischer}, \citenamefont {Crivellin}, \citenamefont {Fael},\ and\
  \citenamefont {Greub}}]{Aebischer:2015fzz}%
  \BibitemOpen
  \bibfield  {author} {\bibinfo {author} {\bibfnamefont {J.}~\bibnamefont
  {Aebischer}}, \bibinfo {author} {\bibfnamefont {A.}~\bibnamefont
  {Crivellin}}, \bibinfo {author} {\bibfnamefont {M.}~\bibnamefont {Fael}},\
  and\ \bibinfo {author} {\bibfnamefont {C.}~\bibnamefont {Greub}},\ }\href
  {https://doi.org/10.1007/JHEP05(2016)037} {\bibfield  {journal} {\bibinfo
  {journal} {J. High Energy Phys.}\ }\textbf {\bibinfo {volume} {05}},\
  \bibinfo {pages} {037 (2016)}},\ \Eprint {https://arxiv.org/abs/1512.02830}
  {arXiv:1512.02830 [hep-ph]} \BibitemShut {NoStop}%
\bibitem [{\citenamefont {Dekens}\ and\ \citenamefont
  {Stoffer}(2019)}]{Dekens:2019ept}%
  \BibitemOpen
  \bibfield  {author} {\bibinfo {author} {\bibfnamefont {W.}~\bibnamefont
  {Dekens}}\ and\ \bibinfo {author} {\bibfnamefont {P.}~\bibnamefont
  {Stoffer}},\ }\href {https://doi.org/10.1007/JHEP10(2019)197} {\bibfield
  {journal} {\bibinfo  {journal} {J. High Energy Phys.}\ }\textbf {\bibinfo
  {volume} {10}},\ \bibinfo {pages} {197 (2019)}},\ \Eprint
  {https://arxiv.org/abs/1908.05295} {arXiv:1908.05295 [hep-ph]} \BibitemShut
  {NoStop}%
\bibitem [{\citenamefont {Aebischer}\ \emph {et~al.}(2017)\citenamefont
  {Aebischer}, \citenamefont {Fael}, \citenamefont {Greub},\ and\ \citenamefont
  {Virto}}]{Aebischer:2017gaw}%
  \BibitemOpen
  \bibfield  {author} {\bibinfo {author} {\bibfnamefont {J.}~\bibnamefont
  {Aebischer}}, \bibinfo {author} {\bibfnamefont {M.}~\bibnamefont {Fael}},
  \bibinfo {author} {\bibfnamefont {C.}~\bibnamefont {Greub}},\ and\ \bibinfo
  {author} {\bibfnamefont {J.}~\bibnamefont {Virto}},\ }\href
  {https://doi.org/10.1007/JHEP09(2017)158} {\bibfield  {journal} {\bibinfo
  {journal} {J. High Energy Phys.}\ }\textbf {\bibinfo {volume} {09}},\
  \bibinfo {pages} {158 (2017)}},\ \Eprint {https://arxiv.org/abs/1704.06639}
  {arXiv:1704.06639 [hep-ph]} \BibitemShut {NoStop}%
\bibitem [{\citenamefont {Jenkins}\ \emph
  {et~al.}(2018{\natexlab{b}})\citenamefont {Jenkins}, \citenamefont
  {Manohar},\ and\ \citenamefont {Stoffer}}]{Jenkins:2017dyc}%
  \BibitemOpen
  \bibfield  {author} {\bibinfo {author} {\bibfnamefont {E.~E.}\ \bibnamefont
  {Jenkins}}, \bibinfo {author} {\bibfnamefont {A.~V.}\ \bibnamefont
  {Manohar}},\ and\ \bibinfo {author} {\bibfnamefont {P.}~\bibnamefont
  {Stoffer}},\ }\href {https://doi.org/10.1007/JHEP01(2018)084} {\bibfield
  {journal} {\bibinfo  {journal} {J. High Energy Phys.}\ }\textbf {\bibinfo
  {volume} {01}},\ \bibinfo {pages} {084 (2018)}},\ \Eprint
  {https://arxiv.org/abs/1711.05270} {arXiv:1711.05270 [hep-ph]} \BibitemShut
  {NoStop}%
\bibitem [{\citenamefont {Buchalla}\ \emph {et~al.}(1996)\citenamefont
  {Buchalla}, \citenamefont {Buras},\ and\ \citenamefont
  {Lautenbacher}}]{Buchalla:1995vs}%
  \BibitemOpen
  \bibfield  {author} {\bibinfo {author} {\bibfnamefont {G.}~\bibnamefont
  {Buchalla}}, \bibinfo {author} {\bibfnamefont {A.~J.}\ \bibnamefont
  {Buras}},\ and\ \bibinfo {author} {\bibfnamefont {M.~E.}\ \bibnamefont
  {Lautenbacher}},\ }\href {https://doi.org/10.1103/RevModPhys.68.1125}
  {\bibfield  {journal} {\bibinfo  {journal} {Rev. Mod. Phys.}\ }\textbf
  {\bibinfo {volume} {68}},\ \bibinfo {pages} {1125} (\bibinfo {year}
  {1996})},\ \Eprint {https://arxiv.org/abs/hep-ph/9512380}
  {arXiv:hep-ph/9512380} \BibitemShut {NoStop}%
\bibitem [{\citenamefont {Aaij}\ \emph
  {et~al.}(2023{\natexlab{a}})\citenamefont {Aaij} \emph
  {et~al.}}]{LHCb:2022qnv}%
  \BibitemOpen
  \bibfield  {author} {\bibinfo {author} {\bibfnamefont {R.}~\bibnamefont
  {Aaij}} \emph {et~al.} (\bibinfo {collaboration} {LHCb Collaboration}),\
  }\href {https://doi.org/10.1103/PhysRevLett.131.051803} {\bibfield  {journal}
  {\bibinfo  {journal} {Phys. Rev. Lett.}\ }\textbf {\bibinfo {volume} {131}},\
  \bibinfo {pages} {051803} (\bibinfo {year} {2023}{\natexlab{a}})},\ \Eprint
  {https://arxiv.org/abs/2212.09152} {arXiv:2212.09152 [hep-ex]} \BibitemShut
  {NoStop}%
\bibitem [{\citenamefont {Aaij}\ \emph
  {et~al.}(2023{\natexlab{b}})\citenamefont {Aaij} \emph
  {et~al.}}]{LHCb:2022vje}%
  \BibitemOpen
  \bibfield  {author} {\bibinfo {author} {\bibfnamefont {R.}~\bibnamefont
  {Aaij}} \emph {et~al.} (\bibinfo {collaboration} {LHCb Collaboration}),\
  }\href {https://doi.org/10.1103/PhysRevD.108.032002} {\bibfield  {journal}
  {\bibinfo  {journal} {Phys. Rev. D}\ }\textbf {\bibinfo {volume} {108}},\
  \bibinfo {pages} {032002} (\bibinfo {year} {2023}{\natexlab{b}})},\ \Eprint
  {https://arxiv.org/abs/2212.09153} {arXiv:2212.09153 [hep-ex]} \BibitemShut
  {NoStop}%
\bibitem [{\citenamefont {Aaij}\ \emph {et~al.}(2018)\citenamefont {Aaij} \emph
  {et~al.}}]{LHCb:2018roe}%
  \BibitemOpen
  \bibfield  {author} {\bibinfo {author} {\bibfnamefont {R.}~\bibnamefont
  {Aaij}} \emph {et~al.} (\bibinfo {collaboration} {LHCb Collaboration}),\
  }\Eprint {https://arxiv.org/abs/1808.08865} {arXiv:1808.08865 [hep-ex]}
  (\bibinfo {year} {2018})\BibitemShut {NoStop}%
\bibitem [{\citenamefont {Workman}\ \emph {et~al.}(2022)\citenamefont {Workman}
  \emph {et~al.}}]{ParticleDataGroup:2022pth}%
  \BibitemOpen
  \bibfield  {author} {\bibinfo {author} {\bibfnamefont {R.~L.}\ \bibnamefont
  {Workman}} \emph {et~al.} (\bibinfo {collaboration} {Particle Data Group}),\
  }\href {https://doi.org/10.1093/ptep/ptac097} {\bibfield  {journal} {\bibinfo
   {journal} {Prog. Theor. Exp. Phys.}\ }\textbf {\bibinfo {volume} {2022}},\
  \bibinfo {pages} {083C01} (\bibinfo {year} {2022})}\BibitemShut {NoStop}%
\bibitem [{\citenamefont {Lees}\ \emph {et~al.}(2013)\citenamefont {Lees} \emph
  {et~al.}}]{Lees:2013kla}%
  \BibitemOpen
  \bibfield  {author} {\bibinfo {author} {\bibfnamefont {J.~P.}\ \bibnamefont
  {Lees}} \emph {et~al.} (\bibinfo {collaboration} {BaBar Collaboration}),\
  }\href {https://doi.org/10.1103/PhysRevD.87.112005} {\bibfield  {journal}
  {\bibinfo  {journal} {Phys. Rev. D}\ }\textbf {\bibinfo {volume} {87}},\
  \bibinfo {pages} {112005} (\bibinfo {year} {2013})},\ \Eprint
  {https://arxiv.org/abs/1303.7465} {arXiv:1303.7465 [hep-ex]} \BibitemShut
  {NoStop}%
\bibitem [{\citenamefont {Altmannshofer}\ \emph {et~al.}(2019)\citenamefont
  {Altmannshofer} \emph {et~al.}}]{Kou:2018nap}%
  \BibitemOpen
  \bibfield  {author} {\bibinfo {author} {\bibfnamefont {W.}~\bibnamefont
  {Altmannshofer}} \emph {et~al.} (\bibinfo {collaboration} {Belle-II
  Collaboration}),\ }\href {https://doi.org/10.1093/ptep/ptz106} {\bibfield
  {journal} {\bibinfo  {journal} {Prog. Theor. Exp. Phys.}\ }\textbf {\bibinfo
  {volume} {2019}},\ \bibinfo {pages} {123C01} (\bibinfo {year} {2019})},\
  \bibinfo {note} {[Erratum: Prog. Theor. Exp. Phys. 2020, 029201 (2020)]},\
  \Eprint {https://arxiv.org/abs/1808.10567} {arXiv:1808.10567 [hep-ex]}
  \BibitemShut {NoStop}%
\bibitem [{\citenamefont {Grygier}\ \emph {et~al.}(2017)\citenamefont {Grygier}
  \emph {et~al.}}]{Grygier:2017tzo}%
  \BibitemOpen
  \bibfield  {author} {\bibinfo {author} {\bibfnamefont {J.}~\bibnamefont
  {Grygier}} \emph {et~al.} (\bibinfo {collaboration} {Belle Collaboration}),\
  }\href {https://doi.org/10.1103/PhysRevD.96.091101} {\bibfield  {journal}
  {\bibinfo  {journal} {Phys. Rev. D}\ }\textbf {\bibinfo {volume} {96}},\
  \bibinfo {pages} {091101} (\bibinfo {year} {2017})},\ \bibinfo {note}
  {[Addendum: Phys. Rev. D 97, 099902 (2018)]},\ \Eprint
  {https://arxiv.org/abs/1702.03224} {arXiv:1702.03224 [hep-ex]} \BibitemShut
  {NoStop}%
\bibitem [{\citenamefont {Lutz}\ \emph {et~al.}(2013)\citenamefont {Lutz} \emph
  {et~al.}}]{Belle:2013tnz}%
  \BibitemOpen
  \bibfield  {author} {\bibinfo {author} {\bibfnamefont {O.}~\bibnamefont
  {Lutz}} \emph {et~al.} (\bibinfo {collaboration} {Belle Collaboration}),\
  }\href {https://doi.org/10.1103/PhysRevD.87.111103} {\bibfield  {journal}
  {\bibinfo  {journal} {Phys. Rev. D}\ }\textbf {\bibinfo {volume} {87}},\
  \bibinfo {pages} {111103} (\bibinfo {year} {2013})},\ \Eprint
  {https://arxiv.org/abs/1303.3719} {arXiv:1303.3719 [hep-ex]} \BibitemShut
  {NoStop}%
\bibitem [{\citenamefont {Amhis}\ \emph {et~al.}(2023)\citenamefont {Amhis}
  \emph {et~al.}}]{HeavyFlavorAveragingGroup:2022wzx}%
  \BibitemOpen
  \bibfield  {author} {\bibinfo {author} {\bibfnamefont {Y.~S.}\ \bibnamefont
  {Amhis}} \emph {et~al.} (\bibinfo {collaboration} {Heavy Flavor Averaging
  Group}),\ }\href {https://doi.org/10.1103/PhysRevD.107.052008} {\bibfield
  {journal} {\bibinfo  {journal} {Phys. Rev. D}\ }\textbf {\bibinfo {volume}
  {107}},\ \bibinfo {pages} {052008} (\bibinfo {year} {2023})},\ \bibinfo
  {note} {and online updates at https://hflav.web.cern.ch},\ \Eprint
  {https://arxiv.org/abs/2206.07501} {arXiv:2206.07501 [hep-ex]} \BibitemShut
  {NoStop}%
\bibitem [{\citenamefont {Aaij}\ \emph {et~al.}(2017)\citenamefont {Aaij} \emph
  {et~al.}}]{Aaij:2017xqt}%
  \BibitemOpen
  \bibfield  {author} {\bibinfo {author} {\bibfnamefont {R.}~\bibnamefont
  {Aaij}} \emph {et~al.} (\bibinfo {collaboration} {LHCb Collaboration}),\
  }\href {https://doi.org/10.1103/PhysRevLett.118.251802} {\bibfield  {journal}
  {\bibinfo  {journal} {Phys. Rev. Lett.}\ }\textbf {\bibinfo {volume} {118}},\
  \bibinfo {pages} {251802} (\bibinfo {year} {2017})},\ \Eprint
  {https://arxiv.org/abs/1703.02508} {arXiv:1703.02508 [hep-ex]} \BibitemShut
  {NoStop}%
\bibitem [{\citenamefont {Lees}\ \emph {et~al.}(2017)\citenamefont {Lees} \emph
  {et~al.}}]{TheBaBar:2016xwe}%
  \BibitemOpen
  \bibfield  {author} {\bibinfo {author} {\bibfnamefont {J.}~\bibnamefont
  {Lees}} \emph {et~al.} (\bibinfo {collaboration} {BaBar Collaboration}),\
  }\href {https://doi.org/10.1103/PhysRevLett.118.031802} {\bibfield  {journal}
  {\bibinfo  {journal} {Phys. Rev. Lett.}\ }\textbf {\bibinfo {volume} {118}},\
  \bibinfo {pages} {031802} (\bibinfo {year} {2017})},\ \Eprint
  {https://arxiv.org/abs/1605.09637} {arXiv:1605.09637 [hep-ex]} \BibitemShut
  {NoStop}%
\bibitem [{\citenamefont {Dong}\ \emph {et~al.}(2023)\citenamefont {Dong} \emph
  {et~al.}}]{Belle:2021ecr}%
  \BibitemOpen
  \bibfield  {author} {\bibinfo {author} {\bibfnamefont {T.~V.}\ \bibnamefont
  {Dong}} \emph {et~al.} (\bibinfo {collaboration} {Belle Collaboration}),\
  }\href {https://doi.org/10.1103/PhysRevD.108.L011102} {\bibfield  {journal}
  {\bibinfo  {journal} {Phys. Rev. D}\ }\textbf {\bibinfo {volume} {108}},\
  \bibinfo {pages} {L011102} (\bibinfo {year} {2023})},\ \Eprint
  {https://arxiv.org/abs/2110.03871} {arXiv:2110.03871 [hep-ex]} \BibitemShut
  {NoStop}%
\bibitem [{\citenamefont {Aggarwal}\ \emph {et~al.}(2022)\citenamefont
  {Aggarwal} \emph {et~al.}}]{Belle-II:2022cgf}%
  \BibitemOpen
  \bibfield  {author} {\bibinfo {author} {\bibfnamefont {L.}~\bibnamefont
  {Aggarwal}} \emph {et~al.} (\bibinfo {collaboration} {Belle-II
  Collaboration}),\ }\Eprint {https://arxiv.org/abs/2207.06307}
  {arXiv:2207.06307 [hep-ex]}  (\bibinfo {year} {2022})\BibitemShut {NoStop}%
\bibitem [{\citenamefont {Aaij}\ \emph {et~al.}(2019)\citenamefont {Aaij} \emph
  {et~al.}}]{Aaij:2019okb}%
  \BibitemOpen
  \bibfield  {author} {\bibinfo {author} {\bibfnamefont {R.}~\bibnamefont
  {Aaij}} \emph {et~al.} (\bibinfo {collaboration} {LHCb Collaboration}),\
  }\href {https://doi.org/10.1103/PhysRevLett.123.211801} {\bibfield  {journal}
  {\bibinfo  {journal} {Phys. Rev. Lett.}\ }\textbf {\bibinfo {volume} {123}},\
  \bibinfo {pages} {211801} (\bibinfo {year} {2019})},\ \Eprint
  {https://arxiv.org/abs/1905.06614} {arXiv:1905.06614 [hep-ex]} \BibitemShut
  {NoStop}%
\bibitem [{\citenamefont {Watanuki}\ \emph {et~al.}(2023)\citenamefont
  {Watanuki} \emph {et~al.}}]{Belle:2022pcr}%
  \BibitemOpen
  \bibfield  {author} {\bibinfo {author} {\bibfnamefont {S.}~\bibnamefont
  {Watanuki}} \emph {et~al.} (\bibinfo {collaboration} {Belle Collaboration}),\
  }\href {https://doi.org/10.1103/PhysRevLett.130.261802} {\bibfield  {journal}
  {\bibinfo  {journal} {Phys. Rev. Lett.}\ }\textbf {\bibinfo {volume} {130}},\
  \bibinfo {pages} {261802} (\bibinfo {year} {2023})},\ \Eprint
  {https://arxiv.org/abs/2212.04128} {arXiv:2212.04128 [hep-ex]} \BibitemShut
  {NoStop}%
\bibitem [{\citenamefont {Aaij}\ \emph
  {et~al.}(2023{\natexlab{c}})\citenamefont {Aaij} \emph
  {et~al.}}]{LHCb:2022wrs}%
  \BibitemOpen
  \bibfield  {author} {\bibinfo {author} {\bibfnamefont {R.}~\bibnamefont
  {Aaij}} \emph {et~al.} (\bibinfo {collaboration} {LHCb Collaboration}),\
  }\href {https://doi.org/10.1007/JHEP06(2023)143} {\bibfield  {journal}
  {\bibinfo  {journal} {J. High Energy Phys.}\ }\textbf {\bibinfo {volume}
  {06}},\ \bibinfo {pages} {143 (2023)}},\ \Eprint
  {https://arxiv.org/abs/2209.09846} {arXiv:2209.09846 [hep-ex]} \BibitemShut
  {NoStop}%
\bibitem [{\citenamefont {Tsuzuki}\ \emph {et~al.}(2023)\citenamefont {Tsuzuki}
  \emph {et~al.}}]{Belle:2023ziz}%
  \BibitemOpen
  \bibfield  {author} {\bibinfo {author} {\bibfnamefont {N.}~\bibnamefont
  {Tsuzuki}} \emph {et~al.} (\bibinfo {collaboration} {Belle Collaboration}),\
  }\href {https://doi.org/10.1007/JHEP06(2023)118} {\bibfield  {journal}
  {\bibinfo  {journal} {J. High Energy Phys.}\ }\textbf {\bibinfo {volume}
  {06}},\ \bibinfo {pages} {118}},\ \Eprint {https://arxiv.org/abs/2301.03768}
  {arXiv:2301.03768 [hep-ex]} \BibitemShut {NoStop}%
\bibitem [{\citenamefont {Banerjee}\ \emph {et~al.}(2022)\citenamefont
  {Banerjee} \emph {et~al.}}]{Banerjee:2022xuw}%
  \BibitemOpen
  \bibfield  {author} {\bibinfo {author} {\bibfnamefont {S.}~\bibnamefont
  {Banerjee}} \emph {et~al.},\ }\Eprint {https://arxiv.org/abs/2203.14919}
  {arXiv:2203.14919 [hep-ph]}  (\bibinfo {year} {2022})\BibitemShut {NoStop}%
\bibitem [{\citenamefont {Patra}\ \emph {et~al.}(2022)\citenamefont {Patra}
  \emph {et~al.}}]{Belle:2022cce}%
  \BibitemOpen
  \bibfield  {author} {\bibinfo {author} {\bibfnamefont {S.}~\bibnamefont
  {Patra}} \emph {et~al.} (\bibinfo {collaboration} {Belle Collaboration}),\
  }\href {https://doi.org/10.1007/JHEP05(2022)095} {\bibfield  {journal}
  {\bibinfo  {journal} {J. High Energy Phys.}\ }\textbf {\bibinfo {volume}
  {05}},\ \bibinfo {pages} {095}},\ \Eprint {https://arxiv.org/abs/2201.09620}
  {arXiv:2201.09620 [hep-ex]} \BibitemShut {NoStop}%
\bibitem [{\citenamefont {Lees}\ \emph {et~al.}(2010)\citenamefont {Lees} \emph
  {et~al.}}]{BaBar:2010vxb}%
  \BibitemOpen
  \bibfield  {author} {\bibinfo {author} {\bibfnamefont {J.~P.}\ \bibnamefont
  {Lees}} \emph {et~al.} (\bibinfo {collaboration} {BaBar Collaboration}),\
  }\href {https://doi.org/10.1103/PhysRevLett.104.151802} {\bibfield  {journal}
  {\bibinfo  {journal} {Phys. Rev. Lett.}\ }\textbf {\bibinfo {volume} {104}},\
  \bibinfo {pages} {151802} (\bibinfo {year} {2010})},\ \Eprint
  {https://arxiv.org/abs/1001.1883} {arXiv:1001.1883 [hep-ex]} \BibitemShut
  {NoStop}%
\bibitem [{\citenamefont {Abdesselam}\ \emph {et~al.}(2021)\citenamefont
  {Abdesselam} \emph {et~al.}}]{Belle:2021ysv}%
  \BibitemOpen
  \bibfield  {author} {\bibinfo {author} {\bibfnamefont {A.}~\bibnamefont
  {Abdesselam}} \emph {et~al.} (\bibinfo {collaboration} {Belle
  Collaboration}),\ }\href {https://doi.org/10.1007/JHEP10(2021)019} {\bibfield
   {journal} {\bibinfo  {journal} {J. High Energy Phys.}\ }\textbf {\bibinfo
  {volume} {10}},\ \bibinfo {pages} {19 (2021)}},\ \Eprint
  {https://arxiv.org/abs/2103.12994} {arXiv:2103.12994 [hep-ex]} \BibitemShut
  {NoStop}%
\bibitem [{\citenamefont {Hayasaka}\ \emph {et~al.}(2010)\citenamefont
  {Hayasaka} \emph {et~al.}}]{Hayasaka:2010np}%
  \BibitemOpen
  \bibfield  {author} {\bibinfo {author} {\bibfnamefont {K.}~\bibnamefont
  {Hayasaka}} \emph {et~al.},\ }\href
  {https://doi.org/10.1016/j.physletb.2010.03.037} {\bibfield  {journal}
  {\bibinfo  {journal} {Phys. Lett. B}\ }\textbf {\bibinfo {volume} {687}},\
  \bibinfo {pages} {139} (\bibinfo {year} {2010})},\ \Eprint
  {https://arxiv.org/abs/1001.3221} {arXiv:1001.3221 [hep-ex]} \BibitemShut
  {NoStop}%
\bibitem [{\citenamefont {Aad}\ \emph {et~al.}(2022)\citenamefont {Aad} \emph
  {et~al.}}]{ATLAS:2021bdj}%
  \BibitemOpen
  \bibfield  {author} {\bibinfo {author} {\bibfnamefont {G.}~\bibnamefont
  {Aad}} \emph {et~al.} (\bibinfo {collaboration} {ATLAS Collaboration}),\
  }\href {https://doi.org/10.1103/PhysRevLett.127.271801} {\bibfield  {journal}
  {\bibinfo  {journal} {Phys. Rev. Lett.}\ }\textbf {\bibinfo {volume} {127}},\
  \bibinfo {pages} {271801} (\bibinfo {year} {2022})},\ \Eprint
  {https://arxiv.org/abs/2105.12491} {arXiv:2105.12491 [hep-ex]} \BibitemShut
  {NoStop}%
\bibitem [{\citenamefont {Dam}(2019)}]{Dam:2018rfz}%
  \BibitemOpen
  \bibfield  {author} {\bibinfo {author} {\bibfnamefont {M.}~\bibnamefont
  {Dam}},\ }\href {https://doi.org/10.21468/SciPostPhysProc.1.041} {\bibfield
  {journal} {\bibinfo  {journal} {SciPost Phys. Proc.}\ }\textbf {\bibinfo
  {volume} {1}},\ \bibinfo {pages} {041} (\bibinfo {year} {2019})},\ \Eprint
  {https://arxiv.org/abs/1811.09408} {arXiv:1811.09408 [hep-ex]} \BibitemShut
  {NoStop}%
\bibitem [{\citenamefont {Gubernari}\ \emph {et~al.}(2022)\citenamefont
  {Gubernari}, \citenamefont {Reboud}, \citenamefont {van Dyk},\ and\
  \citenamefont {Virto}}]{Gubernari:2022hxn}%
  \BibitemOpen
  \bibfield  {author} {\bibinfo {author} {\bibfnamefont {N.}~\bibnamefont
  {Gubernari}}, \bibinfo {author} {\bibfnamefont {M.}~\bibnamefont {Reboud}},
  \bibinfo {author} {\bibfnamefont {D.}~\bibnamefont {van Dyk}},\ and\ \bibinfo
  {author} {\bibfnamefont {J.}~\bibnamefont {Virto}},\ }\href
  {https://doi.org/10.1007/JHEP09(2022)133} {\bibfield  {journal} {\bibinfo
  {journal} {J. High Energy Phys.}\ }\textbf {\bibinfo {volume} {09}},\
  \bibinfo {pages} {133 (2022)}},\ \Eprint {https://arxiv.org/abs/2206.03797}
  {arXiv:2206.03797 [hep-ph]} \BibitemShut {NoStop}%
\bibitem [{\citenamefont {J\"ager}\ and\ \citenamefont
  {Martin~Camalich}(2013)}]{Jager:2012uw}%
  \BibitemOpen
  \bibfield  {author} {\bibinfo {author} {\bibfnamefont {S.}~\bibnamefont
  {J\"ager}}\ and\ \bibinfo {author} {\bibfnamefont {J.}~\bibnamefont
  {Martin~Camalich}},\ }\href {https://doi.org/10.1007/JHEP05(2013)043}
  {\bibfield  {journal} {\bibinfo  {journal} {J. High Energy Phys.}\ }\textbf
  {\bibinfo {volume} {05}},\ \bibinfo {pages} {043 (2013)}},\ \Eprint
  {https://arxiv.org/abs/1212.2263} {arXiv:1212.2263 [hep-ph]} \BibitemShut
  {NoStop}%
\bibitem [{\citenamefont {Lyon}\ and\ \citenamefont
  {Zwicky}(2014)}]{Lyon:2014hpa}%
  \BibitemOpen
  \bibfield  {author} {\bibinfo {author} {\bibfnamefont {J.}~\bibnamefont
  {Lyon}}\ and\ \bibinfo {author} {\bibfnamefont {R.}~\bibnamefont {Zwicky}},\
  }\Eprint {https://arxiv.org/abs/1406.0566} {arXiv:1406.0566 [hep-ph]}
  (\bibinfo {year} {2014})\BibitemShut {NoStop}%
\bibitem [{\citenamefont {Descotes-Genon}\ \emph {et~al.}(2014)\citenamefont
  {Descotes-Genon}, \citenamefont {Hofer}, \citenamefont {Matias},\ and\
  \citenamefont {Virto}}]{Descotes-Genon:2014uoa}%
  \BibitemOpen
  \bibfield  {author} {\bibinfo {author} {\bibfnamefont {S.}~\bibnamefont
  {Descotes-Genon}}, \bibinfo {author} {\bibfnamefont {L.}~\bibnamefont
  {Hofer}}, \bibinfo {author} {\bibfnamefont {J.}~\bibnamefont {Matias}},\ and\
  \bibinfo {author} {\bibfnamefont {J.}~\bibnamefont {Virto}},\ }\href
  {https://doi.org/10.1007/JHEP12(2014)125} {\bibfield  {journal} {\bibinfo
  {journal} {J. High Energy Phys.}\ }\textbf {\bibinfo {volume} {12}},\
  \bibinfo {pages} {125 (2014)}},\ \Eprint {https://arxiv.org/abs/1407.8526}
  {arXiv:1407.8526 [hep-ph]} \BibitemShut {NoStop}%
\bibitem [{\citenamefont {J\"ager}\ and\ \citenamefont
  {Martin~Camalich}(2016)}]{Jager:2014rwa}%
  \BibitemOpen
  \bibfield  {author} {\bibinfo {author} {\bibfnamefont {S.}~\bibnamefont
  {J\"ager}}\ and\ \bibinfo {author} {\bibfnamefont {J.}~\bibnamefont
  {Martin~Camalich}},\ }\href {https://doi.org/10.1103/PhysRevD.93.014028}
  {\bibfield  {journal} {\bibinfo  {journal} {Phys. Rev. D}\ }\textbf {\bibinfo
  {volume} {93}},\ \bibinfo {pages} {014028} (\bibinfo {year} {2016})},\
  \Eprint {https://arxiv.org/abs/1412.3183} {arXiv:1412.3183 [hep-ph]}
  \BibitemShut {NoStop}%
\bibitem [{\citenamefont {Ciuchini}\ \emph {et~al.}(2016)\citenamefont
  {Ciuchini}, \citenamefont {Fedele}, \citenamefont {Franco}, \citenamefont
  {Mishima}, \citenamefont {Paul}, \citenamefont {Silvestrini},\ and\
  \citenamefont {Valli}}]{Ciuchini:2015qxb}%
  \BibitemOpen
  \bibfield  {author} {\bibinfo {author} {\bibfnamefont {M.}~\bibnamefont
  {Ciuchini}}, \bibinfo {author} {\bibfnamefont {M.}~\bibnamefont {Fedele}},
  \bibinfo {author} {\bibfnamefont {E.}~\bibnamefont {Franco}}, \bibinfo
  {author} {\bibfnamefont {S.}~\bibnamefont {Mishima}}, \bibinfo {author}
  {\bibfnamefont {A.}~\bibnamefont {Paul}}, \bibinfo {author} {\bibfnamefont
  {L.}~\bibnamefont {Silvestrini}},\ and\ \bibinfo {author} {\bibfnamefont
  {M.}~\bibnamefont {Valli}},\ }\href {https://doi.org/10.1007/JHEP06(2016)116}
  {\bibfield  {journal} {\bibinfo  {journal} {J. High Energy Phys.}\ }\textbf
  {\bibinfo {volume} {06}},\ \bibinfo {pages} {116 (2016)}},\ \Eprint
  {https://arxiv.org/abs/1512.07157} {arXiv:1512.07157 [hep-ph]} \BibitemShut
  {NoStop}%
\bibitem [{\citenamefont {Ciuchini}\ \emph {et~al.}(2023)\citenamefont
  {Ciuchini}, \citenamefont {Fedele}, \citenamefont {Franco}, \citenamefont
  {Paul}, \citenamefont {Silvestrini},\ and\ \citenamefont
  {Valli}}]{Ciuchini:2022wbq}%
  \BibitemOpen
  \bibfield  {author} {\bibinfo {author} {\bibfnamefont {M.}~\bibnamefont
  {Ciuchini}}, \bibinfo {author} {\bibfnamefont {M.}~\bibnamefont {Fedele}},
  \bibinfo {author} {\bibfnamefont {E.}~\bibnamefont {Franco}}, \bibinfo
  {author} {\bibfnamefont {A.}~\bibnamefont {Paul}}, \bibinfo {author}
  {\bibfnamefont {L.}~\bibnamefont {Silvestrini}},\ and\ \bibinfo {author}
  {\bibfnamefont {M.}~\bibnamefont {Valli}},\ }\href
  {https://doi.org/10.1103/PhysRevD.107.055036} {\bibfield  {journal} {\bibinfo
   {journal} {Phys. Rev. D}\ }\textbf {\bibinfo {volume} {107}},\ \bibinfo
  {pages} {055036} (\bibinfo {year} {2023})},\ \Eprint
  {https://arxiv.org/abs/2212.10516} {arXiv:2212.10516 [hep-ph]} \BibitemShut
  {NoStop}%
\bibitem [{\citenamefont {Buras}\ \emph {et~al.}(2015)\citenamefont {Buras},
  \citenamefont {Girrbach-Noe}, \citenamefont {Niehoff},\ and\ \citenamefont
  {Straub}}]{Buras:2014fpa}%
  \BibitemOpen
  \bibfield  {author} {\bibinfo {author} {\bibfnamefont {A.~J.}\ \bibnamefont
  {Buras}}, \bibinfo {author} {\bibfnamefont {J.}~\bibnamefont {Girrbach-Noe}},
  \bibinfo {author} {\bibfnamefont {C.}~\bibnamefont {Niehoff}},\ and\ \bibinfo
  {author} {\bibfnamefont {D.~M.}\ \bibnamefont {Straub}},\ }\href
  {https://doi.org/10.1007/JHEP02(2015)184} {\bibfield  {journal} {\bibinfo
  {journal} {J. High Energy Phys.}\ }\textbf {\bibinfo {volume} {02}},\
  \bibinfo {pages} {184 (2015)}},\ \Eprint {https://arxiv.org/abs/1409.4557}
  {arXiv:1409.4557 [hep-ph]} \BibitemShut {NoStop}%
\bibitem [{\citenamefont {Ganiev}()}]{Belle2:BtoKnunu}%
  \BibitemOpen
  \bibfield  {author} {\bibinfo {author} {\bibfnamefont {E.}~\bibnamefont
  {Ganiev}} (\bibinfo {collaboration} {Belle-II Collaboration}),\ }\bibinfo
  {note} {talk at the European Physical Society Conference on High Energy
  Physics (EPS-HEP), Hamburg, Aug. 2023}\BibitemShut {NoStop}%
\bibitem [{\citenamefont {Celis}\ \emph {et~al.}(2017)\citenamefont {Celis},
  \citenamefont {Fuentes-Martin}, \citenamefont {Vicente},\ and\ \citenamefont
  {Virto}}]{Celis:2017doq}%
  \BibitemOpen
  \bibfield  {author} {\bibinfo {author} {\bibfnamefont {A.}~\bibnamefont
  {Celis}}, \bibinfo {author} {\bibfnamefont {J.}~\bibnamefont
  {Fuentes-Martin}}, \bibinfo {author} {\bibfnamefont {A.}~\bibnamefont
  {Vicente}},\ and\ \bibinfo {author} {\bibfnamefont {J.}~\bibnamefont
  {Virto}},\ }\href {https://doi.org/10.1103/PhysRevD.96.035026} {\bibfield
  {journal} {\bibinfo  {journal} {Phys. Rev. D}\ }\textbf {\bibinfo {volume}
  {96}},\ \bibinfo {pages} {035026} (\bibinfo {year} {2017})},\ \Eprint
  {https://arxiv.org/abs/1704.05672} {arXiv:1704.05672 [hep-ph]} \BibitemShut
  {NoStop}%
\bibitem [{\citenamefont {Capdevila}\ \emph {et~al.}(2017)\citenamefont
  {Capdevila}, \citenamefont {Descotes-Genon}, \citenamefont {Hofer},\ and\
  \citenamefont {Matias}}]{Capdevila:2017ert}%
  \BibitemOpen
  \bibfield  {author} {\bibinfo {author} {\bibfnamefont {B.}~\bibnamefont
  {Capdevila}}, \bibinfo {author} {\bibfnamefont {S.}~\bibnamefont
  {Descotes-Genon}}, \bibinfo {author} {\bibfnamefont {L.}~\bibnamefont
  {Hofer}},\ and\ \bibinfo {author} {\bibfnamefont {J.}~\bibnamefont
  {Matias}},\ }\href {https://doi.org/10.1007/JHEP04(2017)016} {\bibfield
  {journal} {\bibinfo  {journal} {J. High Energy Phys.}\ }\textbf {\bibinfo
  {volume} {04}},\ \bibinfo {pages} {016 (2017)}},\ \Eprint
  {https://arxiv.org/abs/1701.08672} {arXiv:1701.08672 [hep-ph]} \BibitemShut
  {NoStop}%
\bibitem [{\citenamefont {Bordone}\ \emph {et~al.}(2016)\citenamefont
  {Bordone}, \citenamefont {Isidori},\ and\ \citenamefont
  {Pattori}}]{Bordone:2016gaq}%
  \BibitemOpen
  \bibfield  {author} {\bibinfo {author} {\bibfnamefont {M.}~\bibnamefont
  {Bordone}}, \bibinfo {author} {\bibfnamefont {G.}~\bibnamefont {Isidori}},\
  and\ \bibinfo {author} {\bibfnamefont {A.}~\bibnamefont {Pattori}},\ }\href
  {https://doi.org/10.1140/epjc/s10052-016-4274-7} {\bibfield  {journal}
  {\bibinfo  {journal} {Eur. Phys. J. C}\ ,\ \bibinfo {pages} {440}} (\bibinfo
  {year} {2016})},\ \Eprint {https://arxiv.org/abs/1605.07633}
  {arXiv:1605.07633 [hep-ph]} \BibitemShut {NoStop}%
\bibitem [{\citenamefont {Isidori}\ \emph {et~al.}(2020)\citenamefont
  {Isidori}, \citenamefont {Nabeebaccus},\ and\ \citenamefont
  {Zwicky}}]{Isidori:2020acz}%
  \BibitemOpen
  \bibfield  {author} {\bibinfo {author} {\bibfnamefont {G.}~\bibnamefont
  {Isidori}}, \bibinfo {author} {\bibfnamefont {S.}~\bibnamefont
  {Nabeebaccus}},\ and\ \bibinfo {author} {\bibfnamefont {R.}~\bibnamefont
  {Zwicky}},\ }\href {https://doi.org/10.1007/JHEP12(2020)104} {\bibfield
  {journal} {\bibinfo  {journal} {J. High Energy Phys.}\ }\textbf {\bibinfo
  {volume} {12}},\ \bibinfo {pages} {104 (2020)}},\ \Eprint
  {https://arxiv.org/abs/2009.00929} {arXiv:2009.00929 [hep-ph]} \BibitemShut
  {NoStop}%
\bibitem [{\citenamefont {Bobeth}\ \emph {et~al.}(2014)\citenamefont {Bobeth},
  \citenamefont {Gorbahn}, \citenamefont {Hermann}, \citenamefont {Misiak},
  \citenamefont {Stamou},\ and\ \citenamefont {Steinhauser}}]{Bobeth:2013uxa}%
  \BibitemOpen
  \bibfield  {author} {\bibinfo {author} {\bibfnamefont {C.}~\bibnamefont
  {Bobeth}}, \bibinfo {author} {\bibfnamefont {M.}~\bibnamefont {Gorbahn}},
  \bibinfo {author} {\bibfnamefont {T.}~\bibnamefont {Hermann}}, \bibinfo
  {author} {\bibfnamefont {M.}~\bibnamefont {Misiak}}, \bibinfo {author}
  {\bibfnamefont {E.}~\bibnamefont {Stamou}},\ and\ \bibinfo {author}
  {\bibfnamefont {M.}~\bibnamefont {Steinhauser}},\ }\href
  {https://doi.org/10.1103/PhysRevLett.112.101801} {\bibfield  {journal}
  {\bibinfo  {journal} {Phys. Rev. Lett.}\ }\textbf {\bibinfo {volume} {112}},\
  \bibinfo {pages} {101801} (\bibinfo {year} {2014})},\ \Eprint
  {https://arxiv.org/abs/1311.0903} {arXiv:1311.0903 [hep-ph]} \BibitemShut
  {NoStop}%
\bibitem [{\citenamefont {Blake}\ \emph {et~al.}(2017)\citenamefont {Blake},
  \citenamefont {Lanfranchi},\ and\ \citenamefont {Straub}}]{Blake:2016olu}%
  \BibitemOpen
  \bibfield  {author} {\bibinfo {author} {\bibfnamefont {T.}~\bibnamefont
  {Blake}}, \bibinfo {author} {\bibfnamefont {G.}~\bibnamefont {Lanfranchi}},\
  and\ \bibinfo {author} {\bibfnamefont {D.~M.}\ \bibnamefont {Straub}},\
  }\href {https://doi.org/10.1016/j.ppnp.2016.10.001} {\bibfield  {journal}
  {\bibinfo  {journal} {Prog. Part. Nucl. Phys.}\ }\textbf {\bibinfo {volume}
  {92}},\ \bibinfo {pages} {50} (\bibinfo {year} {2017})},\ \Eprint
  {https://arxiv.org/abs/1606.00916} {arXiv:1606.00916 [hep-ph]} \BibitemShut
  {NoStop}%
\bibitem [{\citenamefont {De~Bruyn}\ \emph
  {et~al.}(2012{\natexlab{a}})\citenamefont {De~Bruyn}, \citenamefont
  {Fleischer}, \citenamefont {Knegjens}, \citenamefont {Koppenburg},
  \citenamefont {Merk},\ and\ \citenamefont {Tuning}}]{DeBruyn:2012wj}%
  \BibitemOpen
  \bibfield  {author} {\bibinfo {author} {\bibfnamefont {K.}~\bibnamefont
  {De~Bruyn}}, \bibinfo {author} {\bibfnamefont {R.}~\bibnamefont {Fleischer}},
  \bibinfo {author} {\bibfnamefont {R.}~\bibnamefont {Knegjens}}, \bibinfo
  {author} {\bibfnamefont {P.}~\bibnamefont {Koppenburg}}, \bibinfo {author}
  {\bibfnamefont {M.}~\bibnamefont {Merk}},\ and\ \bibinfo {author}
  {\bibfnamefont {N.}~\bibnamefont {Tuning}},\ }\href
  {https://doi.org/10.1103/PhysRevD.86.014027} {\bibfield  {journal} {\bibinfo
  {journal} {Phys. Rev. D}\ }\textbf {\bibinfo {volume} {86}},\ \bibinfo
  {pages} {014027} (\bibinfo {year} {2012}{\natexlab{a}})},\ \Eprint
  {https://arxiv.org/abs/1204.1735} {arXiv:1204.1735 [hep-ph]} \BibitemShut
  {NoStop}%
\bibitem [{\citenamefont {De~Bruyn}\ \emph
  {et~al.}(2012{\natexlab{b}})\citenamefont {De~Bruyn}, \citenamefont
  {Fleischer}, \citenamefont {Knegjens}, \citenamefont {Koppenburg},
  \citenamefont {Merk}, \citenamefont {Pellegrino},\ and\ \citenamefont
  {Tuning}}]{DeBruyn:2012wk}%
  \BibitemOpen
  \bibfield  {author} {\bibinfo {author} {\bibfnamefont {K.}~\bibnamefont
  {De~Bruyn}}, \bibinfo {author} {\bibfnamefont {R.}~\bibnamefont {Fleischer}},
  \bibinfo {author} {\bibfnamefont {R.}~\bibnamefont {Knegjens}}, \bibinfo
  {author} {\bibfnamefont {P.}~\bibnamefont {Koppenburg}}, \bibinfo {author}
  {\bibfnamefont {M.}~\bibnamefont {Merk}}, \bibinfo {author} {\bibfnamefont
  {A.}~\bibnamefont {Pellegrino}},\ and\ \bibinfo {author} {\bibfnamefont
  {N.}~\bibnamefont {Tuning}},\ }\href
  {https://doi.org/10.1103/PhysRevLett.109.041801} {\bibfield  {journal}
  {\bibinfo  {journal} {Phys. Rev. Lett.}\ }\textbf {\bibinfo {volume} {109}},\
  \bibinfo {pages} {041801} (\bibinfo {year} {2012}{\natexlab{b}})},\ \Eprint
  {https://arxiv.org/abs/1204.1737} {arXiv:1204.1737 [hep-ph]} \BibitemShut
  {NoStop}%
\bibitem [{\citenamefont {Crivellin}\ \emph {et~al.}(2017)\citenamefont
  {Crivellin}, \citenamefont {M\"uller},\ and\ \citenamefont
  {Ota}}]{Crivellin:2017zlb}%
  \BibitemOpen
  \bibfield  {author} {\bibinfo {author} {\bibfnamefont {A.}~\bibnamefont
  {Crivellin}}, \bibinfo {author} {\bibfnamefont {D.}~\bibnamefont
  {M\"uller}},\ and\ \bibinfo {author} {\bibfnamefont {T.}~\bibnamefont
  {Ota}},\ }\href {https://doi.org/10.1007/JHEP09(2017)040} {\bibfield
  {journal} {\bibinfo  {journal} {J. High Energy Phys.}\ }\textbf {\bibinfo
  {volume} {09}},\ \bibinfo {pages} {040 (2017)}},\ \Eprint
  {https://arxiv.org/abs/1703.09226} {arXiv:1703.09226 [hep-ph]} \BibitemShut
  {NoStop}%
\bibitem [{\citenamefont {Buttazzo}\ \emph {et~al.}(2017)\citenamefont
  {Buttazzo}, \citenamefont {Greljo}, \citenamefont {Isidori},\ and\
  \citenamefont {Marzocca}}]{Buttazzo:2017ixm}%
  \BibitemOpen
  \bibfield  {author} {\bibinfo {author} {\bibfnamefont {D.}~\bibnamefont
  {Buttazzo}}, \bibinfo {author} {\bibfnamefont {A.}~\bibnamefont {Greljo}},
  \bibinfo {author} {\bibfnamefont {G.}~\bibnamefont {Isidori}},\ and\ \bibinfo
  {author} {\bibfnamefont {D.}~\bibnamefont {Marzocca}},\ }\href
  {https://doi.org/10.1007/JHEP11(2017)044} {\bibfield  {journal} {\bibinfo
  {journal} {J. High Energy Phys.}\ }\textbf {\bibinfo {volume} {11}},\
  \bibinfo {pages} {044 (2017)}},\ \Eprint {https://arxiv.org/abs/1706.07808}
  {arXiv:1706.07808 [hep-ph]} \BibitemShut {NoStop}%
\bibitem [{\citenamefont {Bobeth}\ and\ \citenamefont
  {Haisch}(2013)}]{Bobeth:2011st}%
  \BibitemOpen
  \bibfield  {author} {\bibinfo {author} {\bibfnamefont {C.}~\bibnamefont
  {Bobeth}}\ and\ \bibinfo {author} {\bibfnamefont {U.}~\bibnamefont
  {Haisch}},\ }\href {https://doi.org/10.5506/APhysPolB.44.127} {\bibfield
  {journal} {\bibinfo  {journal} {Acta Phys. Polon. B}\ }\textbf {\bibinfo
  {volume} {44}},\ \bibinfo {pages} {127} (\bibinfo {year} {2013})},\ \Eprint
  {https://arxiv.org/abs/1109.1826} {arXiv:1109.1826 [hep-ph]} \BibitemShut
  {NoStop}%
\bibitem [{\citenamefont {Capdevila}\ \emph {et~al.}(2018)\citenamefont
  {Capdevila}, \citenamefont {Crivellin}, \citenamefont {Descotes-Genon},
  \citenamefont {Hofer},\ and\ \citenamefont {Matias}}]{Capdevila:2017iqn}%
  \BibitemOpen
  \bibfield  {author} {\bibinfo {author} {\bibfnamefont {B.}~\bibnamefont
  {Capdevila}}, \bibinfo {author} {\bibfnamefont {A.}~\bibnamefont
  {Crivellin}}, \bibinfo {author} {\bibfnamefont {S.}~\bibnamefont
  {Descotes-Genon}}, \bibinfo {author} {\bibfnamefont {L.}~\bibnamefont
  {Hofer}},\ and\ \bibinfo {author} {\bibfnamefont {J.}~\bibnamefont
  {Matias}},\ }\href {https://doi.org/10.1103/PhysRevLett.120.181802}
  {\bibfield  {journal} {\bibinfo  {journal} {Phys. Rev. Lett.}\ }\textbf
  {\bibinfo {volume} {120}},\ \bibinfo {pages} {181802} (\bibinfo {year}
  {2018})},\ \Eprint {https://arxiv.org/abs/1712.01919} {arXiv:1712.01919
  [hep-ph]} \BibitemShut {NoStop}%
\bibitem [{\citenamefont {Dedes}\ \emph {et~al.}(2009)\citenamefont {Dedes},
  \citenamefont {Rosiek},\ and\ \citenamefont {Tanedo}}]{Dedes:2008iw}%
  \BibitemOpen
  \bibfield  {author} {\bibinfo {author} {\bibfnamefont {A.}~\bibnamefont
  {Dedes}}, \bibinfo {author} {\bibfnamefont {J.}~\bibnamefont {Rosiek}},\ and\
  \bibinfo {author} {\bibfnamefont {P.}~\bibnamefont {Tanedo}},\ }\href
  {https://doi.org/10.1103/PhysRevD.79.055006} {\bibfield  {journal} {\bibinfo
  {journal} {Phys. Rev. D}\ }\textbf {\bibinfo {volume} {79}},\ \bibinfo
  {pages} {055006} (\bibinfo {year} {2009})},\ \Eprint
  {https://arxiv.org/abs/0812.4320} {arXiv:0812.4320 [hep-ph]} \BibitemShut
  {NoStop}%
\bibitem [{\citenamefont {Be\v{c}irevi\'c}\ \emph {et~al.}(2016)\citenamefont
  {Be\v{c}irevi\'c}, \citenamefont {Sumensari},\ and\ \citenamefont
  {Zukanovich~Funchal}}]{Becirevic:2016zri}%
  \BibitemOpen
  \bibfield  {author} {\bibinfo {author} {\bibfnamefont {D.}~\bibnamefont
  {Be\v{c}irevi\'c}}, \bibinfo {author} {\bibfnamefont {O.}~\bibnamefont
  {Sumensari}},\ and\ \bibinfo {author} {\bibfnamefont {R.}~\bibnamefont
  {Zukanovich~Funchal}},\ }\href
  {https://doi.org/10.1140/epjc/s10052-016-3985-0} {\bibfield  {journal}
  {\bibinfo  {journal} {Eur. Phys. J. C}\ }\textbf {\bibinfo {volume} {76}},\
  \bibinfo {pages} {134} (\bibinfo {year} {2016})},\ \Eprint
  {https://arxiv.org/abs/1602.00881} {arXiv:1602.00881 [hep-ph]} \BibitemShut
  {NoStop}%
\bibitem [{\citenamefont {Abada}\ \emph {et~al.}(2015)\citenamefont {Abada},
  \citenamefont {Be\v{c}irevi\'c}, \citenamefont {Lucente},\ and\ \citenamefont
  {Sumensari}}]{Abada:2015zea}%
  \BibitemOpen
  \bibfield  {author} {\bibinfo {author} {\bibfnamefont {A.}~\bibnamefont
  {Abada}}, \bibinfo {author} {\bibfnamefont {D.}~\bibnamefont
  {Be\v{c}irevi\'c}}, \bibinfo {author} {\bibfnamefont {M.}~\bibnamefont
  {Lucente}},\ and\ \bibinfo {author} {\bibfnamefont {O.}~\bibnamefont
  {Sumensari}},\ }\href {https://doi.org/10.1103/PhysRevD.91.113013} {\bibfield
   {journal} {\bibinfo  {journal} {Phys. Rev. D}\ }\textbf {\bibinfo {volume}
  {91}},\ \bibinfo {pages} {113013} (\bibinfo {year} {2015})},\ \Eprint
  {https://arxiv.org/abs/1503.04159} {arXiv:1503.04159 [hep-ph]} \BibitemShut
  {NoStop}%
\bibitem [{\citenamefont {Hazard}\ and\ \citenamefont
  {Petrov}(2016)}]{Hazard:2016fnc}%
  \BibitemOpen
  \bibfield  {author} {\bibinfo {author} {\bibfnamefont {D.~E.}\ \bibnamefont
  {Hazard}}\ and\ \bibinfo {author} {\bibfnamefont {A.~A.}\ \bibnamefont
  {Petrov}},\ }\href {https://doi.org/10.1103/PhysRevD.94.074023} {\bibfield
  {journal} {\bibinfo  {journal} {Phys. Rev. D}\ }\textbf {\bibinfo {volume}
  {94}},\ \bibinfo {pages} {074023} (\bibinfo {year} {2016})},\ \Eprint
  {https://arxiv.org/abs/1607.00815} {arXiv:1607.00815 [hep-ph]} \BibitemShut
  {NoStop}%
\bibitem [{\citenamefont {Calibbi}\ \emph {et~al.}(2022)\citenamefont
  {Calibbi}, \citenamefont {Li}, \citenamefont {Marcano},\ and\ \citenamefont
  {Schmidt}}]{Calibbi:2022ddo}%
  \BibitemOpen
  \bibfield  {author} {\bibinfo {author} {\bibfnamefont {L.}~\bibnamefont
  {Calibbi}}, \bibinfo {author} {\bibfnamefont {T.}~\bibnamefont {Li}},
  \bibinfo {author} {\bibfnamefont {X.}~\bibnamefont {Marcano}},\ and\ \bibinfo
  {author} {\bibfnamefont {M.~A.}\ \bibnamefont {Schmidt}},\ }\href
  {https://doi.org/10.1103/PhysRevD.106.115039} {\bibfield  {journal} {\bibinfo
   {journal} {Phys. Rev. D}\ }\textbf {\bibinfo {volume} {106}},\ \bibinfo
  {pages} {115039} (\bibinfo {year} {2022})},\ \Eprint
  {https://arxiv.org/abs/2207.10913} {arXiv:2207.10913 [hep-ph]} \BibitemShut
  {NoStop}%
\bibitem [{\citenamefont {Goto}\ \emph {et~al.}(2011)\citenamefont {Goto},
  \citenamefont {Okada},\ and\ \citenamefont {Yamamoto}}]{Goto:2010sn}%
  \BibitemOpen
  \bibfield  {author} {\bibinfo {author} {\bibfnamefont {T.}~\bibnamefont
  {Goto}}, \bibinfo {author} {\bibfnamefont {Y.}~\bibnamefont {Okada}},\ and\
  \bibinfo {author} {\bibfnamefont {Y.}~\bibnamefont {Yamamoto}},\ }\href
  {https://doi.org/10.1103/PhysRevD.83.053011} {\bibfield  {journal} {\bibinfo
  {journal} {Phys. Rev. D}\ }\textbf {\bibinfo {volume} {83}},\ \bibinfo
  {pages} {053011} (\bibinfo {year} {2011})},\ \Eprint
  {https://arxiv.org/abs/1012.4385} {arXiv:1012.4385 [hep-ph]} \BibitemShut
  {NoStop}%
\bibitem [{\citenamefont {Okada}\ \emph {et~al.}(2000)\citenamefont {Okada},
  \citenamefont {Okumura},\ and\ \citenamefont {Shimizu}}]{Okada:1999zk}%
  \BibitemOpen
  \bibfield  {author} {\bibinfo {author} {\bibfnamefont {Y.}~\bibnamefont
  {Okada}}, \bibinfo {author} {\bibfnamefont {K.-i.}\ \bibnamefont {Okumura}},\
  and\ \bibinfo {author} {\bibfnamefont {Y.}~\bibnamefont {Shimizu}},\ }\href
  {https://doi.org/10.1103/PhysRevD.61.094001} {\bibfield  {journal} {\bibinfo
  {journal} {Phys. Rev. D}\ }\textbf {\bibinfo {volume} {61}},\ \bibinfo
  {pages} {094001} (\bibinfo {year} {2000})},\ \Eprint
  {https://arxiv.org/abs/hep-ph/9906446} {arXiv:hep-ph/9906446} \BibitemShut
  {NoStop}%
\bibitem [{\citenamefont {Kuno}\ and\ \citenamefont
  {Okada}(2001)}]{Kuno:1999jp}%
  \BibitemOpen
  \bibfield  {author} {\bibinfo {author} {\bibfnamefont {Y.}~\bibnamefont
  {Kuno}}\ and\ \bibinfo {author} {\bibfnamefont {Y.}~\bibnamefont {Okada}},\
  }\href {https://doi.org/10.1103/RevModPhys.73.151} {\bibfield  {journal}
  {\bibinfo  {journal} {Rev. Mod. Phys.}\ }\textbf {\bibinfo {volume} {73}},\
  \bibinfo {pages} {151} (\bibinfo {year} {2001})},\ \Eprint
  {https://arxiv.org/abs/hep-ph/9909265} {arXiv:hep-ph/9909265} \BibitemShut
  {NoStop}%
\bibitem [{\citenamefont {Aoyama}\ \emph {et~al.}(2020)\citenamefont {Aoyama}
  \emph {et~al.}}]{Aoyama:2020ynm}%
  \BibitemOpen
  \bibfield  {author} {\bibinfo {author} {\bibfnamefont {T.}~\bibnamefont
  {Aoyama}} \emph {et~al.},\ }\href
  {https://doi.org/10.1016/j.physrep.2020.07.006} {\bibfield  {journal}
  {\bibinfo  {journal} {Phys. Rept.}\ }\textbf {\bibinfo {volume} {887}},\
  \bibinfo {pages} {1} (\bibinfo {year} {2020})},\ \Eprint
  {https://arxiv.org/abs/2006.04822} {arXiv:2006.04822 [hep-ph]} \BibitemShut
  {NoStop}%
\bibitem [{\citenamefont {Aguillard}\ \emph {et~al.}(2023)\citenamefont
  {Aguillard} \emph {et~al.}}]{Muong-2:2023cdq}%
  \BibitemOpen
  \bibfield  {author} {\bibinfo {author} {\bibfnamefont {D.~P.}\ \bibnamefont
  {Aguillard}} \emph {et~al.} (\bibinfo {collaboration} {Muon g-2
  Collaboration}),\ }\href {https://doi.org/10.1103/PhysRevLett.131.161802}
  {\bibfield  {journal} {\bibinfo  {journal} {Phys. Rev. Lett.}\ }\textbf
  {\bibinfo {volume} {131}},\ \bibinfo {pages} {161802} (\bibinfo {year}
  {2023})},\ \Eprint {https://arxiv.org/abs/2308.06230} {arXiv:2308.06230
  [hep-ex]} \BibitemShut {NoStop}%
\bibitem [{\citenamefont {Arnan}\ \emph {et~al.}(2019)\citenamefont {Arnan},
  \citenamefont {Becirevic}, \citenamefont {Mescia},\ and\ \citenamefont
  {Sumensari}}]{Arnan:2019olv}%
  \BibitemOpen
  \bibfield  {author} {\bibinfo {author} {\bibfnamefont {P.}~\bibnamefont
  {Arnan}}, \bibinfo {author} {\bibfnamefont {D.}~\bibnamefont {Becirevic}},
  \bibinfo {author} {\bibfnamefont {F.}~\bibnamefont {Mescia}},\ and\ \bibinfo
  {author} {\bibfnamefont {O.}~\bibnamefont {Sumensari}},\ }\href
  {https://doi.org/10.1007/JHEP02(2019)109} {\bibfield  {journal} {\bibinfo
  {journal} {J. High Energy Phys.}\ }\textbf {\bibinfo {volume} {02}},\
  \bibinfo {pages} {109 (2019)}},\ \Eprint {https://arxiv.org/abs/1901.06315}
  {arXiv:1901.06315 [hep-ph]} \BibitemShut {NoStop}%
\bibitem [{\citenamefont {Abreu}\ \emph {et~al.}(1997)\citenamefont {Abreu}
  \emph {et~al.}}]{DELPHI:1996iox}%
  \BibitemOpen
  \bibfield  {author} {\bibinfo {author} {\bibfnamefont {P.}~\bibnamefont
  {Abreu}} \emph {et~al.} (\bibinfo {collaboration} {DELPHI Collaboration}),\
  }\href {https://doi.org/10.1007/s002880050313} {\bibfield  {journal}
  {\bibinfo  {journal} {Z. Phys. C}\ }\textbf {\bibinfo {volume} {73}},\
  \bibinfo {pages} {243} (\bibinfo {year} {1997})}\BibitemShut {NoStop}%
\bibitem [{\citenamefont {Buras}\ \emph {et~al.}(1978)\citenamefont {Buras},
  \citenamefont {Ellis}, \citenamefont {Gaillard},\ and\ \citenamefont
  {Nanopoulos}}]{Buras:1977yy}%
  \BibitemOpen
  \bibfield  {author} {\bibinfo {author} {\bibfnamefont {A.~J.}\ \bibnamefont
  {Buras}}, \bibinfo {author} {\bibfnamefont {J.~R.}\ \bibnamefont {Ellis}},
  \bibinfo {author} {\bibfnamefont {M.~K.}\ \bibnamefont {Gaillard}},\ and\
  \bibinfo {author} {\bibfnamefont {D.~V.}\ \bibnamefont {Nanopoulos}},\ }\href
  {https://doi.org/10.1016/0550-3213(78)90214-6} {\bibfield  {journal}
  {\bibinfo  {journal} {Nucl. Phys. B}\ }\textbf {\bibinfo {volume} {135}},\
  \bibinfo {pages} {66} (\bibinfo {year} {1978})}\BibitemShut {NoStop}%
\bibitem [{\citenamefont {Frampton}\ \emph {et~al.}(1979)\citenamefont
  {Frampton}, \citenamefont {Nandi},\ and\ \citenamefont
  {Scanio}}]{Frampton:1979wf}%
  \BibitemOpen
  \bibfield  {author} {\bibinfo {author} {\bibfnamefont {P.~H.}\ \bibnamefont
  {Frampton}}, \bibinfo {author} {\bibfnamefont {S.}~\bibnamefont {Nandi}},\
  and\ \bibinfo {author} {\bibfnamefont {J.~J.~G.}\ \bibnamefont {Scanio}},\
  }\href {https://doi.org/10.1016/0370-2693(79)90584-7} {\bibfield  {journal}
  {\bibinfo  {journal} {Phys. Lett. B}\ }\textbf {\bibinfo {volume} {85}},\
  \bibinfo {pages} {225} (\bibinfo {year} {1979})}\BibitemShut {NoStop}%
\bibitem [{\citenamefont {Kalyniak}\ and\ \citenamefont
  {Ng}(1982)}]{Kalyniak:1982pt}%
  \BibitemOpen
  \bibfield  {author} {\bibinfo {author} {\bibfnamefont {P.}~\bibnamefont
  {Kalyniak}}\ and\ \bibinfo {author} {\bibfnamefont {J.~N.}\ \bibnamefont
  {Ng}},\ }\href {https://doi.org/10.1103/PhysRevD.26.890} {\bibfield
  {journal} {\bibinfo  {journal} {Phys. Rev. D}\ }\textbf {\bibinfo {volume}
  {26}},\ \bibinfo {pages} {890} (\bibinfo {year} {1982})}\BibitemShut
  {NoStop}%
\bibitem [{\citenamefont {Eckert}\ \emph {et~al.}(1983)\citenamefont {Eckert},
  \citenamefont {Gerard}, \citenamefont {Ruegg},\ and\ \citenamefont
  {Schucker}}]{Eckert:1983bn}%
  \BibitemOpen
  \bibfield  {author} {\bibinfo {author} {\bibfnamefont {P.}~\bibnamefont
  {Eckert}}, \bibinfo {author} {\bibfnamefont {J.~M.}\ \bibnamefont {Gerard}},
  \bibinfo {author} {\bibfnamefont {H.}~\bibnamefont {Ruegg}},\ and\ \bibinfo
  {author} {\bibfnamefont {T.}~\bibnamefont {Schucker}},\ }\href
  {https://doi.org/10.1016/0370-2693(83)91308-4} {\bibfield  {journal}
  {\bibinfo  {journal} {Phys. Lett. B}\ }\textbf {\bibinfo {volume} {125}},\
  \bibinfo {pages} {385} (\bibinfo {year} {1983})}\BibitemShut {NoStop}%
\bibitem [{\citenamefont {Machacek}\ and\ \citenamefont
  {Vaughn}(1983)}]{Machacek:1983tz}%
  \BibitemOpen
  \bibfield  {author} {\bibinfo {author} {\bibfnamefont {M.~E.}\ \bibnamefont
  {Machacek}}\ and\ \bibinfo {author} {\bibfnamefont {M.~T.}\ \bibnamefont
  {Vaughn}},\ }\href {https://doi.org/10.1016/0550-3213(83)90610-7} {\bibfield
  {journal} {\bibinfo  {journal} {Nucl. Phys. B}\ }\textbf {\bibinfo {volume}
  {222}},\ \bibinfo {pages} {83} (\bibinfo {year} {1983})}\BibitemShut
  {NoStop}%
\bibitem [{\citenamefont {Machacek}\ and\ \citenamefont
  {Vaughn}(1984)}]{Machacek:1983fi}%
  \BibitemOpen
  \bibfield  {author} {\bibinfo {author} {\bibfnamefont {M.~E.}\ \bibnamefont
  {Machacek}}\ and\ \bibinfo {author} {\bibfnamefont {M.~T.}\ \bibnamefont
  {Vaughn}},\ }\href {https://doi.org/10.1016/0550-3213(84)90533-9} {\bibfield
  {journal} {\bibinfo  {journal} {Nucl. Phys. B}\ }\textbf {\bibinfo {volume}
  {236}},\ \bibinfo {pages} {221} (\bibinfo {year} {1984})}\BibitemShut
  {NoStop}%
\bibitem [{\citenamefont {Machacek}\ and\ \citenamefont
  {Vaughn}(1985)}]{Machacek:1984zw}%
  \BibitemOpen
  \bibfield  {author} {\bibinfo {author} {\bibfnamefont {M.~E.}\ \bibnamefont
  {Machacek}}\ and\ \bibinfo {author} {\bibfnamefont {M.~T.}\ \bibnamefont
  {Vaughn}},\ }\href {https://doi.org/10.1016/0550-3213(85)90040-9} {\bibfield
  {journal} {\bibinfo  {journal} {Nucl. Phys. B}\ }\textbf {\bibinfo {volume}
  {249}},\ \bibinfo {pages} {70} (\bibinfo {year} {1985})}\BibitemShut
  {NoStop}%
\bibitem [{\citenamefont {Luo}\ \emph {et~al.}(2003)\citenamefont {Luo},
  \citenamefont {Wang},\ and\ \citenamefont {Xiao}}]{Luo:2002ti}%
  \BibitemOpen
  \bibfield  {author} {\bibinfo {author} {\bibfnamefont {M.-x.}\ \bibnamefont
  {Luo}}, \bibinfo {author} {\bibfnamefont {H.-w.}\ \bibnamefont {Wang}},\ and\
  \bibinfo {author} {\bibfnamefont {Y.}~\bibnamefont {Xiao}},\ }\href
  {https://doi.org/10.1103/PhysRevD.67.065019} {\bibfield  {journal} {\bibinfo
  {journal} {Phys. Rev. D}\ }\textbf {\bibinfo {volume} {67}},\ \bibinfo
  {pages} {065019} (\bibinfo {year} {2003})},\ \Eprint
  {https://arxiv.org/abs/hep-ph/0211440} {arXiv:hep-ph/0211440} \BibitemShut
  {NoStop}%
\bibitem [{\citenamefont {Kowalska}\ \emph {et~al.}(2021)\citenamefont
  {Kowalska}, \citenamefont {Sessolo},\ and\ \citenamefont
  {Yamamoto}}]{Kowalska:2020gie}%
  \BibitemOpen
  \bibfield  {author} {\bibinfo {author} {\bibfnamefont {K.}~\bibnamefont
  {Kowalska}}, \bibinfo {author} {\bibfnamefont {E.~M.}\ \bibnamefont
  {Sessolo}},\ and\ \bibinfo {author} {\bibfnamefont {Y.}~\bibnamefont
  {Yamamoto}},\ }\href {https://doi.org/10.1140/epjc/s10052-021-09072-1}
  {\bibfield  {journal} {\bibinfo  {journal} {Eur. Phys. J. C}\ }\textbf
  {\bibinfo {volume} {81}},\ \bibinfo {pages} {272} (\bibinfo {year} {2021})},\
  \Eprint {https://arxiv.org/abs/2007.03567} {arXiv:2007.03567 [hep-ph]}
  \BibitemShut {NoStop}%
\bibitem [{\citenamefont {Fedele}\ \emph {et~al.}(2023)\citenamefont {Fedele},
  \citenamefont {Wuest},\ and\ \citenamefont {Nierste}}]{Fedele:2023rxb}%
  \BibitemOpen
  \bibfield  {author} {\bibinfo {author} {\bibfnamefont {M.}~\bibnamefont
  {Fedele}}, \bibinfo {author} {\bibfnamefont {F.}~\bibnamefont {Wuest}},\ and\
  \bibinfo {author} {\bibfnamefont {U.}~\bibnamefont {Nierste}},\ }\Eprint
  {https://arxiv.org/abs/2307.15117} {arXiv:2307.15117 [hep-ph]}  (\bibinfo
  {year} {2023})\BibitemShut {NoStop}%
\bibitem [{\citenamefont {Crivellin}\ \emph {et~al.}(2020)\citenamefont
  {Crivellin}, \citenamefont {M\"uller},\ and\ \citenamefont
  {Saturnino}}]{Crivellin:2019dwb}%
  \BibitemOpen
  \bibfield  {author} {\bibinfo {author} {\bibfnamefont {A.}~\bibnamefont
  {Crivellin}}, \bibinfo {author} {\bibfnamefont {D.}~\bibnamefont
  {M\"uller}},\ and\ \bibinfo {author} {\bibfnamefont {F.}~\bibnamefont
  {Saturnino}},\ }\href {https://doi.org/10.1007/JHEP06(2020)020} {\bibfield
  {journal} {\bibinfo  {journal} {J. High Energy Phys.}\ }\textbf {\bibinfo
  {volume} {06}},\ \bibinfo {pages} {020 (2020)}},\ \Eprint
  {https://arxiv.org/abs/1912.04224} {arXiv:1912.04224 [hep-ph]} \BibitemShut
  {NoStop}%
\bibitem [{\citenamefont {Gherardi}\ \emph {et~al.}(2021)\citenamefont
  {Gherardi}, \citenamefont {Marzocca},\ and\ \citenamefont
  {Venturini}}]{Gherardi:2020qhc}%
  \BibitemOpen
  \bibfield  {author} {\bibinfo {author} {\bibfnamefont {V.}~\bibnamefont
  {Gherardi}}, \bibinfo {author} {\bibfnamefont {D.}~\bibnamefont {Marzocca}},\
  and\ \bibinfo {author} {\bibfnamefont {E.}~\bibnamefont {Venturini}},\ }\href
  {https://doi.org/10.1007/JHEP01(2021)138} {\bibfield  {journal} {\bibinfo
  {journal} {J. High Energy Phys.}\ }\textbf {\bibinfo {volume} {01}},\
  \bibinfo {pages} {138 (2021)}},\ \Eprint {https://arxiv.org/abs/2008.09548}
  {arXiv:2008.09548 [hep-ph]} \BibitemShut {NoStop}%
\bibitem [{\citenamefont {Bordone}\ \emph {et~al.}(2021)\citenamefont
  {Bordone}, \citenamefont {Cat\`a}, \citenamefont {Feldmann},\ and\
  \citenamefont {Mandal}}]{Bordone:2020lnb}%
  \BibitemOpen
  \bibfield  {author} {\bibinfo {author} {\bibfnamefont {M.}~\bibnamefont
  {Bordone}}, \bibinfo {author} {\bibfnamefont {O.}~\bibnamefont {Cat\`a}},
  \bibinfo {author} {\bibfnamefont {T.}~\bibnamefont {Feldmann}},\ and\
  \bibinfo {author} {\bibfnamefont {R.}~\bibnamefont {Mandal}},\ }\href
  {https://doi.org/10.1007/JHEP03(2021)122} {\bibfield  {journal} {\bibinfo
  {journal} {J. High Energy Phys.}\ }\textbf {\bibinfo {volume} {03}},\
  \bibinfo {pages} {122 (2021)}},\ \Eprint {https://arxiv.org/abs/2010.03297}
  {arXiv:2010.03297 [hep-ph]} \BibitemShut {NoStop}%
\bibitem [{\citenamefont {Crivellin}\ \emph
  {et~al.}(2021{\natexlab{b}})\citenamefont {Crivellin}, \citenamefont {Greub},
  \citenamefont {M\"uller},\ and\ \citenamefont
  {Saturnino}}]{Crivellin:2020mjs}%
  \BibitemOpen
  \bibfield  {author} {\bibinfo {author} {\bibfnamefont {A.}~\bibnamefont
  {Crivellin}}, \bibinfo {author} {\bibfnamefont {C.}~\bibnamefont {Greub}},
  \bibinfo {author} {\bibfnamefont {D.}~\bibnamefont {M\"uller}},\ and\
  \bibinfo {author} {\bibfnamefont {F.}~\bibnamefont {Saturnino}},\ }\href
  {https://doi.org/10.1007/JHEP02(2021)182} {\bibfield  {journal} {\bibinfo
  {journal} {J. High Energy Phys.}\ }\textbf {\bibinfo {volume} {02}},\
  \bibinfo {pages} {182 (2021)}},\ \Eprint {https://arxiv.org/abs/2010.06593}
  {arXiv:2010.06593 [hep-ph]} \BibitemShut {NoStop}%
\bibitem [{\citenamefont {Feruglio}\ \emph
  {et~al.}(2017{\natexlab{a}})\citenamefont {Feruglio}, \citenamefont
  {Paradisi},\ and\ \citenamefont {Pattori}}]{Feruglio:2016gvd}%
  \BibitemOpen
  \bibfield  {author} {\bibinfo {author} {\bibfnamefont {F.}~\bibnamefont
  {Feruglio}}, \bibinfo {author} {\bibfnamefont {P.}~\bibnamefont {Paradisi}},\
  and\ \bibinfo {author} {\bibfnamefont {A.}~\bibnamefont {Pattori}},\ }\href
  {https://doi.org/10.1103/PhysRevLett.118.011801} {\bibfield  {journal}
  {\bibinfo  {journal} {Phys. Rev. Lett.}\ }\textbf {\bibinfo {volume} {118}},\
  \bibinfo {pages} {011801} (\bibinfo {year} {2017}{\natexlab{a}})},\ \Eprint
  {https://arxiv.org/abs/1606.00524} {arXiv:1606.00524 [hep-ph]} \BibitemShut
  {NoStop}%
\bibitem [{\citenamefont {Feruglio}\ \emph
  {et~al.}(2017{\natexlab{b}})\citenamefont {Feruglio}, \citenamefont
  {Paradisi},\ and\ \citenamefont {Pattori}}]{Feruglio:2017rjo}%
  \BibitemOpen
  \bibfield  {author} {\bibinfo {author} {\bibfnamefont {F.}~\bibnamefont
  {Feruglio}}, \bibinfo {author} {\bibfnamefont {P.}~\bibnamefont {Paradisi}},\
  and\ \bibinfo {author} {\bibfnamefont {A.}~\bibnamefont {Pattori}},\ }\href
  {https://doi.org/10.1007/JHEP09(2017)061} {\bibfield  {journal} {\bibinfo
  {journal} {J. High Energy Phys.}\ }\textbf {\bibinfo {volume} {09}},\
  \bibinfo {pages} {061 (2017)}},\ \Eprint {https://arxiv.org/abs/1705.00929}
  {arXiv:1705.00929 [hep-ph]} \BibitemShut {NoStop}%
\end{thebibliography}%
\bibliographystyle{apsrev4-2}
\end{document}